\documentclass[12pt]{article}
\usepackage{authblk}
\usepackage{graphicx} %Loading the package
\graphicspath{{}}

\usepackage[myheadings]{fullpage}
\usepackage[margin=1in]{geometry}
\usepackage[natbibapa]{apacite}
\usepackage[dvipsnames]{xcolor}
\usepackage{subfigure}
\usepackage{amssymb}
\usepackage{amsmath,amssymb,amsfonts} % Math packages
\usepackage{mathtools} % enahance amsmath
\usepackage{adjustbox}

\usepackage{hhline}
\usepackage{float} 
\usepackage{cellspace}
\usepackage{multirow}% http://ctan.org/pkg/multirow

\usepackage{makecell}

\usepackage{graphicx}
\usepackage{hyperref}
\usepackage[dvipsnames]{xcolor}
\hypersetup{
    colorlinks,
    linkcolor={red!50!black},
    citecolor=NavyBlue,
    urlcolor={blue!80!black}
}

\newenvironment{revised}{\par\color{black}}{\par}
\newcommand{\inlineRevised}[1]{\textcolor{black}{#1}}

\begin{document}

%\title{Computational methods for Bayesian semiparametric Item Response Theory models}

\date{}

\title{Computational strategies and estimation performance with Bayesian semiparametric Item Response Theory models}

\author[1]{Sally Paganin \thanks{\href{mailto:spaganin@hsph.harvard.edu}{spaganin@hsph.harvard.edu}}}
\author[1]{Christopher J. Paciorek}
\author[2]{Claudia Wehrhahn}
\author[3]{Abel Rodr{\'i}guez}
\author[1]{Sophia Rabe-Hesketh}
\author[1]{Perry de Valpine}
\affil[1]{University of California, Berkeley}
\affil[2]{University of California, Santa Cruz}
\affil[3]{University of Washington, Seattle}

\maketitle

\begin{abstract}

Item response theory (IRT) models typically rely on a normality assumption for subject-specific latent traits, which is often unrealistic in practice. Semiparametric extensions based on Dirichlet process mixtures offer a more flexible representation of the unknown distribution of the latent trait. However, the use of such models in the IRT literature has been extremely limited, in good part because of the lack of comprehensive studies and accessible software tools. 
This paper provides guidance for practitioners on  semiparametric IRT models and their implementation. In particular, we rely on NIMBLE, a flexible software system for hierarchical models that enables the use of Dirichlet process mixtures. We highlight efficient sampling strategies for model estimation and compare inferential results under parametric and semiparametric models. 

\textit{Keywords}: binary IRT models, Dirichlet process mixture, MCMC strategies, NIMBLE. 

\end{abstract}\vspace{\fill}\pagebreak

\section{Introduction}

Traditional approaches in item response theory (IRT) modeling rely on the assumption that subject-specific latent traits follow a normal distribution. This assumption is often considered for computational convenience, but there are many situations in which it may be unrealistic \citep{samejima1997departure}. For example, \cite{micceri1989unicorn} gives a comprehensive review of many psychometric datasets where the distribution of latent individual trait does not respect the normality assumption and presents instead asymmetries, heavy-tails or multimodality. In addition, estimation of IRT parameters in the presence of non-normal latent traits has been shown to produce biased estimates of the parameters \citep[see, for example][]{seong1990sensitivity,kirisci2001robustness,schmitt2006semi,finch2016rasch}.

Different proposals have been made in the general IRT literature for relaxing this normality assumption, using either Markov chain Monte Carlo (MCMC) or Marginal Maximum Likelihood (MML) estimation methods. One option is to rely on more general parametric assumptions. For example, \cite{azevedo2011bayesian} considered a skew-normal distribution~\citep{azzalini1985class}, while others have suggested finite mixtures of normal distributions~\citep{bolt2001mixture,bambirra2018bayesian}. 
Alternatively, one can refrain from making distributional assumption on the latent abilities by using nonparametric maximum likelihood estimation \citep{laird78,mislevy1984estimating}, B-splines \citep{woods2006item,johnson2007modeling} or empirical histograms \citep{woods2007empirical}.  

This paper considers a Bayesian nonparametric approach that uses a Dirichlet process mixture \citep{ferguson1973,lo1984class,escobar1995bayesian} as a nonparametric prior on the distribution of the subject-specific latent trait.
\inlineRevised{Dirichlet process mixtures are often used as flexible models to describe the unknown distribution of an heterogeneous population of interest. These models are sometimes interpreted as mixture models with an infinite number of components. In practice these model treat the number of groups as an unknown parameter and estimate it from the data, so that the model can easily account for multi-modality, asymmetries or outliers in the latent trait distribution.} We focus in particular on semiparametric extensions of logistic IRT models for binary responses.  Such models are semiparametric because they retain other, parametric, assumptions of binomial mixed models, such as the functional form of the link function. 

Even though some Bayesian nonparametric extensions of binary IRT models have been presented in the literature, they have been given only limited consideration. Within this approach, the semiparametric 1PL model has been the focus of more effort as well as software \citep[\texttt{DPpackage}, no longer actively maintained]{jara2011dppackage}.
\cite{sanmartin2011} investigated a semiparametric generalization of the 1PL model from a theoretical perspective, while \cite{finch2016rasch} provided results from simulation studies. An example using the semiparametric 2PL model is given in \cite{duncan2008nonparametric}. 
\inlineRevised{However, such semiparametric models have not received much attention in applied IRT modeling, in good part because of the lack of comprehensive studies and accessible software tools.}

\inlineRevised{The goal of this paper is to provide a practical guide to semiparametric IRT models for both (i) applied researchers interested in using Dirichlet process mixtures, and (ii) those familiar with Bayesian nonparametrics concepts who are interested in IRT models. To achieve these goals, we fill three major gaps that hinder widespread application of semiparametric Bayesian IRT models.}

First, we implement the semiparametric 1PL, 2PL and 3PL models in NIMBLE \citep{devalpine2017nimble} (\textsc{R} package \texttt{nimble}, \citealp{nimble-software2020}), a flexible \textsc{R}-based system for hierarchical modeling. In particular, NIMBLE provides functionality for fitting hierarchical models that involve Dirichlet process priors either via a Chinese Restaurant Process (CRP) \citep{aldous1985exchangeability,pitman1996some,blackwell1973ferguson} or a truncated stick-breaking (SB) \citep{sethuraman1994} representation of the prior. Hence, NIMBLE supports a much wider class of models than those that are implemented in standard software packages. Code is provided for all examples in a publicly accessible GitHub repository (\url{https://github.com/salleuska/IRT_nimble_code}). 

Second, focusing on the 2PL model, we study the efficiency of several MCMC sampling strategies in both simulated and real-data scenarios. \inlineRevised{We define sampling strategies as the combination of model parameterization, identifiability constraints and sampling algorithms, focusing on general MCMC algorithms available in easy-to-access software tools for Bayesian hierarchical models.} 
We find that some choices of parameterization and identifiability constraints can yield order-of-magnitude differences in sampling efficiency compared to others. This approach also allows us to compare various random walk Metropolis-Hastings MCMC strategies to the Hamiltonian Monte Carlo (HMC) strategy implemented in the widely used Stan package \citep{stan2018}. 
\inlineRevised{Although there is Stan support for many parametric IRT models \citep{burker2021bayesian,edstan2017}}, HMC algorithms are not readily available for Dirichlet process prior models, since HMC cannot sample discrete parameters (the component indicators), which cannot be easily integrated out in infinite mixture models.

Finally, we present a comparison of inferential results for item and subject parameters under parametric and semiparametric specifications. To make these comparisons fair, we carefully  elicit prior distributions for the models by matching the prior predictive distribution of the data to a common distribution \citep{berger1996intrinsic}. We also illustrate how to estimate the entire distribution of latent traits and its functionals under the two specifications.
\inlineRevised{As expected, we find that the semiparametric model improves recovery of item and individual latent trait parameters in the case of non-normal latent traits.} More suprisingly, there seems to be little inferential penalty in using a semiparametric model when a parametric model would be correct, supporting the benefit of greater robustness to mis-specification. \inlineRevised{These conclusions are based on analyses carried out on simulated data as well as two real datasets related to education and medical assessments: the 2007 Trends in International Mathematics and Science Study (TIMSS) and the 1996 Health Survey for England. For both the real data examples, the semiparametric model performs better than the parametric counterpart, with the semiparametric model identifying distinct modes in the distribution of the latent trait missed by the parametric model.}

We note that other authors have considered nonparametric IRT models that rely on  a general monotonic function in place of the logistic/probit link function.  These models are sometimes referred as NIRT models. Some work using the Dirichlet process falls in this class of models \citep{qin1998nonparametric, miyazaki2009bayesian,karabastos2017}. While we do not pursue this direction in this paper, focusing instead on nonparametric modeling of the latent trait distribution, such an extension is relatively straightforward.

\inlineRevised{The remainder of the paper is organized as follows. In Section~\ref{sec:IRT} we present the standard IRT model and the Bayesian semiparametric extension along with considerations for identifiability. 
We then present different potential sampling strategies (Section~\ref{sec:sampling_strategies}) and discuss the goals of our experiments. To fairly compare the different strategies, we give guidance on selecting prior distributions in Section~\ref{sec:prior_distributions}. We introduce simulated and real-world data in Section~\ref{sec:data}. Comparison of the results in terms of MCMC efficiency and statistical inference is presented in Sections~\ref{sec:results_efficiency} and \ref{sec:results_inference}. 
In Section~\ref{sec:discussion}, we conclude that having access to semiparametric models can be broadly useful, as it allows inference on the entire underlying latent trait distribution and its functionals, with NIMBLE being a flexible framework for estimation of such models.}

\section{IRT models and background}\label{sec:IRT}

IRT models are widely used in various social science disciplines to scale binary responses into continuous constructs. For conciseness, in this section we introduce model notation in the context of educational assessment, where typically data are answers to exam questions from a set of individuals \inlineRevised{and the latent trait is interpreted as an individual's ability.}  
In particular, let $y_{ij}$ denote the answer of individual $j$ to item $i$ for $j = 1, \ldots, N$ and $i = 1, \ldots, I$, with $y_{ij} = 1$ when the answer is correct and $0$ otherwise. 
Responses from different individuals are assumed to be independent, while responses from the same individual are assumed independent conditional on the latent trait (this is sometimes called the \emph{local independence assumption} in the psychometric literature). 

\subsection{Binary logistic IRT models}\label{subsec:2PL}

Let $\pi_{ij}$ denote the probability that individual $j$ answers item $i$ correctly, given the model parameters $\eta_j, \lambda_i, \beta_i$; i.e., $\pi_{ij} = \Pr(y_{ij} = 1 \mid \eta_j, \lambda_i, \beta_i)$ for $i = 1, \ldots, I$ and $j= 1, \ldots, N$. The parameter $\eta_j$ represents the latent ability of the $j$-th individual, while $\beta_i$ and $\lambda_i$ encode the item characteristics for the $i$-th item. In the two-parameter logistic (2PL) model, the probability $\pi_{ij}$ is determined using the logistic function as
\begin{equation}\label{eq:2PL}
  \text{logit} (\pi_{ij}) = \lambda_i(\eta_j - \beta_i),
   \quad i = 1, \ldots, I, \quad j = 1, \ldots, N.
\end{equation}
A further assumption for the latent abilities is that they are independently and identically distributed according to some distribution $G$,
\begin{equation}\label{eq:2PL_latent_abilities}
  \eta_j \stackrel{iid}{\sim} G, \quad j = 1, \ldots, N,
\end{equation}
with $G$ traditionally a standard normal distribution. 
The parameter $\lambda_i > 0$ is often referred to as \emph{discrimination}, since items with a large $\lambda_i$ are better at discriminating between subjects with similar abilities, while $\beta_i$ is called \emph{difficulty} because for any fixed $\eta_j$ the probability of a correct response to item $i$ is decreasing in $\beta_i$.

Often, the log-odds in~\eqref{eq:2PL} are reparameterized as
\begin{equation*}
\lambda_i\eta_j + \gamma_i,
\end{equation*}
with $\gamma_i = -\lambda_i \beta_i$.  The two parameterizations are sometimes referred to as \emph{IRT parameterization} and \emph{slope-intercept (SI) parameterization}, respectively. While the slope-intercept parameterization is often considered for computational convenience, the IRT parameterization is the most traditional in terms of interpretation. In exploring different strategies for Bayesian estimation, we will consider both alternatives and investigate potential differences in terms of computational performance. 

\inlineRevised{Alternative models can be obtained by considering a different number of item parameters. When $\lambda_i = 1$ for all $i = 1, \ldots, I$, the model in~\eqref{eq:2PL} reduces to the one-parameter logistic (1PL) model, also known as Rasch model \citep{rasch1990probabilistic}. In some settings one may wish to account for the probability of answering correctly by chance, by introducing a third set of item parameters, $\upsilon_i, i = 1, \ldots, I$, referred to as \emph{guessing} parameters, so that
\begin{equation*}
    \Pr(y_{ij} = 1 \mid \eta_j, \lambda_i, \beta_i, \upsilon_i) = \upsilon_i + (1 - \upsilon_i) \text{expit}\{\lambda_i(\eta_j - \beta_i)\},
\end{equation*}
where $\text{expit}\{\cdot\}$ denotes the inverse of the logistic function. This model is typically referred to as the three-parameter logistic model (3PL), and is often relevant in educational assessments. } 

\subsection{Semiparametric IRT models}\label{subsec:semiparametric}
The classical formulation of IRT models assumes that the latent abilities in \eqref{eq:2PL_latent_abilities} follow a normal distribution.
This assumption can be relaxed, modeling the distribution of ability as a mixture of normal distributions, where the number of mixture components does not need to be specified in advance but rather is learned from the data. This can be achieved using a Dirichlet process mixture (DPM) model for the distribution of ability. In particular, the distribution of ability $G$ in~\eqref{eq:2PL_latent_abilities} can be specified as a convolution involving a Dirichlet process (DP) prior, i.e.
\begin{align}\label{eq:DP}
    G = \int \mathcal{K}(\eta_j \mid  \theta) F(d \theta),\quad F \sim \mbox{DP}(\alpha, G_0),
\end{align} 
where $\mathcal{K}(\cdot\mid  \theta)$ is a suitable probability kernel indexed by the parameter $\theta$, while $\alpha$ and $G_0$ are, respectively, the concentration parameter and the base distribution of the Dirichlet process. %\inlineRevised{$F$ follows a Dirichlet Process distribution, which results in a set of draws from $G_0$, with potentially many repeats, that represent a distribution for $\theta$.  Hence, $F$ is a distribution (or measure) drawn from a Dirichlet Process distribution. $G$ is a distribution drawn from the Dirichlet Process mixture comprised of the choices $G_0$ and $\mathcal{K}$.}

In the context of binary IRT models, it seems natural to choose a normal kernel for $\mathcal{K}(\cdot\mid  \theta)$, indexed by parameters $\theta = \{\mu, \sigma^2\}$. This \inlineRevised{means the distribution of ability is} a mixture of normal distributions, where the number of mixture components \inlineRevised{and their means and variances are} unknown.  Furthermore, under this choice, taking $\alpha \to 0$ leads to the original parametric model discussed in Section \ref{subsec:2PL}, \inlineRevised{in this case a single normal}. Parameters characterizing each mixture component are drawn from the base distribution, $G_0$. For computational convenience the base distribution is typically the product of conjugate distributions, e.g., a normal distribution for $\mu$ and an inverse-gamma distribution for $\sigma^2$.

%\inlineRevised{The remaining part of the model is the Dirichlet Process, which governs the number of components, including the number of repeats for each one.  The number of repeats is like a weight that governs the clustering of individuals among components.  In theory there is an infinite number of components, but in implementation the MCMC will sample over addition and removal in a finite set of components.} 
We proceed now to discuss the Dirichlet process prior in more detail. There are two main representations of the Dirichlet process, each leading to a different MCMC posterior sampling strategy, namely the stick-breaking representation (SB) \citep{sethuraman1994} and the Chinese Restaurant Process (CRP) \citep{aldous1985exchangeability,pitman1996some,blackwell1973ferguson}. In this work we use the CRP representation. The CRP representation is derived from \eqref{eq:DP} integrating out the random measure $F$. More specifically, let $\theta_1, \ldots, \theta_N$ be an independent sample from $F$, \inlineRevised{with some values possibly repeated}.  Integrating over $F$ one can obtain the joint prior distribution on $(\theta_1, \ldots, \theta_{N})$, which can be written as the product of a sequence of conditional distributions, where
\begin{equation}\label{eq:DP_predictive}
    (\theta_j \mid \theta_{j-1}, \ldots, \theta_1) \sim \frac{\alpha}{\alpha + j -1 } G_0 + \sum_{l = 1}^{j-1} \frac{1}{\alpha + j - 1} \delta_{\theta_l},
\end{equation} 
for $j = 1, \ldots, N$, where $\delta_{a}$ is the Dirac probability measure concentrated at $a$. \inlineRevised{The second term in~\eqref{eq:DP_predictive} represents the probability that a new observation is equal to one of the previous ones, while the first term captures the possibility that we observe a new value, which would be drawn from the base measure $G_0$.} 

The CRP name comes from an analogy often used to describe the process in \eqref{eq:DP_predictive}. Consider a Chinese restaurant with an infinite number of tables, each table serving one dish shared by all customers sitting at that table. In this metaphor, each table represents a possible mixture component, while each dish represents the parameter indexing the distribution associated with the mixture component.  
Customers entering the restaurant can seat themselves at a previously occupied table and share the same dish (with probability proportional to the number of customers already sitting at the table), or go to a new table and order another dish (with probability proportional to $\alpha$). \inlineRevised{The dishes are selected according to the centering distribution $G_0$.}

\inlineRevised{One way to make the Chinese restaurant analogy clearer is by reparameterizing the model. Denote by $\theta_k^*$ the dish served in table $k$ (which is a draw from $G_0$) and let $z_j$ be the variable denoting the table chosen by the $j$th customer. Then}
\begin{align}
\label{eq:crp_allocation}
    p(z_j = k \mid  z_{j -1},\ldots, z_2, z_1,\alpha) = 
    \begin{cases}
        \frac{n_{k}^{j-1}}{\alpha + j -1}, \quad k = 1, \ldots, K^{j-1}, \\
        \frac{\alpha}{\alpha + j - 1}, \quad k = K^{j-1} + 1,
    \end{cases}
\end{align}
where $K^{j-1}$ is the total number of occupied tables by the first $j-1$ customers, and $n^{j-1}_{k}$ is the number of customers at table $k$ among the first $j-1$. \inlineRevised{This new parameterization can be related to the old one by noting that  $\theta_i = \theta_{z_i}^*$.}
The concentration parameter $\alpha$ controls the distribution of the number of tables (components), with larger values favoring more tables. Using the indicators $\mathbf{z} = \{z_j, j = 1, \ldots, N\}$, we can denote by $\mathbf{z}\mid \alpha \sim \mbox{CRP}(\alpha)$ the joint distribution induced by \eqref{eq:crp_allocation}, and rewrite the DPM model for the distribution of ability in \eqref{eq:DP} using
\begin{align*}
%\label{eq:crp_representation}
    \eta_j\mid  z_j, \theta^*_1, \theta^*_2, \ldots  &\stackrel{ind}{\sim} \mathcal{K}(\cdot \mid  \theta^*_{z_j}), \quad j = 1, \ldots, N, \nonumber \\
    \mathbf{z}\mid \alpha &\sim \mbox{CRP}(\alpha), \\
    \theta^*_{k}&\stackrel{iid}{\sim} G_0, \quad k = 1, 2, \ldots.
\end{align*}
\inlineRevised{Together, the probability kernel, base measure, and CRP form a Dirichlet process mixture.} \inlineRevised{Alternatively, the distribution $F$ can be written using the stick-breaking representation:}
\begin{equation*}
    F(\cdot) = \sum_{k=1}^{\infty} w_k \delta_{\tilde{\theta}_k},
\end{equation*} 
where $\tilde{\theta}_1, \tilde{\theta}_2, \ldots$ is a sequence of independent draws from $G_0$ and the weights are constructed by letting $w_k = v_k \prod_{l=1}^{k-1} (1-v_l)$, with $v_1, v_2, \ldots$ being a sequence of independent draws from a $\mbox{Beta}(1,\alpha)$ distribution. This construction makes it clear that, as long as the kernel $\mathcal{K}(\cdot\mid\theta)$ is continuous, the distribution of ability $G$ is also continuous, but $F$ is almost surely discrete, naturally inducing clustering via \inlineRevised{repeats} in the parameter indexing the distribution of ability.

\inlineRevised{Note that an alternative to the formulation described above is to model the distribution of the ability $G$ directly using a DP, e.g., centered around a normal distribution.  Such a model also comprises the standard parametric model as a limiting case (now, when $\alpha \to \infty$) and leads to slightly simpler computational algorithms.  However, we believe that such an approach has some serious drawbacks in the context of most IRT applications.  By definition, realizations from a Dirichlet process are almost surely discrete.  This property has made the Dirichlet process a useful tool in clustering applications. However, in our context, it implies that we believe that two (or more) individuals potentially have exactly the same ability. Not only is this assumption not realistic, but it potentially prevents us from distinguishing individuals based on their abilities, which is one common goal in IRT modeling. The use of a Dirichlet process mixture with a continuous kernel (Gaussian, in this case) sidesteps this issue.}

%%%%%%%%%%%%%%%%%%%%%%%%%%%%

\subsection{Identifiability and constraints}\label{subsec:identifiability}

Without additional constraints, the parameters of \inlineRevised{the models presented in Section~\ref{sec:IRT}} are not identifiable (e.g., see \citealp{geweke1981maximum,bafumi2005practical}, as well as Section A in the Supplementary Materials). For example in the 2PL and 3PL models, increasing all $\eta_j$ and $\beta_i$ values by the same amount yields the same probabilities in (\ref{eq:2PL}) for all $i$ and $j$. More generally, the ability parameters are known up to a linear transformation, and constraints are needed to identify them. To address this problem, traditional work on parametric IRT models assumes that latent abilities in \eqref{eq:2PL_latent_abilities} come from a standard normal distribution, i.e., $G \equiv \mathcal{N}(0, 1)$, and constrains the discrimination parameters $\lambda_i$ for $ i = 1, \ldots,I$ to be positive.

Alternative constraints can also establish identifiability and could yield different computational performance for MCMC sampling. A common alternative considers sum-to-zero constraints for the item parameters \citep{fox2010bayesian}
\begin{equation}\label{eq:sum_to_zero_constraints}
\sum_{i = 1}^I \beta_i = 0, \quad \left(\text{or } \sum_{i = 1}^I \gamma_i = 0 \right), \quad\textbf{} \sum_{i = 1}^I \log(\lambda_i) = 0.     
\end{equation}
Centering the difficulty parameters addresses the invariance to translations, while centering the log of the discrimination parameters (setting their product to one) accounts for the invariance to rescalings of the latent space.  Another potential set of constraints, popular in political science applications, involves fixing the value of the latent traits for two individuals \citep[e.g., see][]{clinton2004statistical}.  Whatever the set of constraints, it is worthwhile to note that they can be either directly incorporated in the model as part of the prior (and, therefore in the structure of the sampling algorithms), or they can be applied as a postprocessing step (after running an unconstrained MCMC). This last approach is typical of parameter-expanded algorithms, which embed the target model in a larger specification. Parameter expansion has been proposed in the literature to accelerate EM \citep{liu1998parameter} and Gibbs sampler \citep{liu1999parameter} convergence, as well as to induce new classes of priors \citep{gelman2004parameterization}. Although targeting the same posterior, constrained priors and parameter expansion can lead to very different results in terms of convergence and mixing of the MCMC algorithms.

Similar arguments apply for the semiparametric extensions using the Dirichlet process mixture. In that setting, one identifiability strategy may be to constrain the base distribution $G_0$, e.g., by letting $G_0 \sim \mathcal{N}(0, 1)$ (for example, see \citealp{duncan2008nonparametric}). However, even if the prior expectation and variance of $G_0$ are zero and one, the corresponding posterior quantities can deviate substantially from these values, leading to biased inference \citep{yang2010bayesian}.
More general centering approaches have been proposed in the literature when a DP distribution is used to model random effects or latent variables in a hierarchical model \citep{yang2010semiparametric,yang2010bayesian,li2011center}.  These approaches rely on parameter expansion by sampling from the unconstrained DP model and then applying a post-processing procedure to the posterior samples. This post-processing procedure requires the analytical evaluation of the posterior mean and variance of the DP random measure, with \cite{li2011center} providing results under the CRP representation and \cite{yang2010semiparametric} under the stick-breaking one. 
Although such strategies are useful for general hierarchical models to avoid identifiability issues, for the semiparametric 1PL, 2PL, and 3PL models it is simpler to use the sum-to-zero constraints on the item parameters in \eqref{eq:sum_to_zero_constraints}, and that is the approach we adopt in this work. As in the parametric case, we can either include these constraints in the prior or use the parameter expansion approach for sampling and then center and rescale the posterior samples as appropriate.

\section{Sampling strategies for logistic IRT models}\label{sec:sampling_strategies}

In this work we explore different \emph{sampling strategies} for Bayesian estimation of the logistic parametric and semiparametric IRT models. We define a \emph{sampling strategy} to include the combination of model parameterization, identifiability constraints and sampling algorithms.
\inlineRevised{
We focus on the case of the 2PL model, as it contains the 1PL as a special case and presents the same identifiability challenges as the 3PL model. The strategies considered are summarized in Table~\ref{tab:model_sampling_strategies}.
}

\begin{table}[!h]
\begin{center}
\begin{adjustbox}{max width=\textwidth}
\begin{tabular}{|Sl|Sc|Sc|Sc|Sc|}
  \hline
  \multirow{3}{*}{\textbf{Identifiability constraints} } 
      & \multicolumn{2}{c|}{\textbf{Parametric}}  &
        \multicolumn{2}{c|}{\textbf{Semi-parametric}} \\
         \cline{2-5}
      &  Slope-intercept & IRT & Slope-intercept & IRT \\
  \hline
      Constrained abilities &   MH/conjugate      &   MH/conjugate     &     &       \\
                            &   Centered     &   HMC (Stan)  &    & \\
  \hline
      Constrained item parameters      &   MH/conjugate      &   MH/conjugate$^*$    &  MH/conjugate &  MH/conjugate$^*$     \\
  \hline
    Unconstrained           &   MH/conjugate      &   MH/conjugate    &  MH/conjugate   &   MH/conjugate    \\
               &   Centered       &       &  Centered  &   \\
  \hline
\end{tabular}
\end{adjustbox}
\end{center}

\caption{Summary of the 14 sampling strategies considered for the parametric and semiparametric 2PL model. Each of the 14 entries is a different strategy with ``MH/conjugate'', ``Centered'', and ``HMC (Stan)'' referring to three different sampling algorithms discussed below. The asterisk symbol denotes the sampling strategies that lead directly to samples parameterized as model  \eqref{eq:baseline_parameterization}. Others need post-processing to correspond to model \eqref{eq:baseline_parameterization}.}\label{tab:model_sampling_strategies}
\end{table}

We explore both parameterizations of the 2PL model mentioned in Section~\ref{subsec:2PL}: the IRT and the slope-intercept parameterization. To compare estimates obtained from different parameterizations on a common scale, we post-process posterior samples (using transformations described in the Supplementary Materials, Section A) to respect the following base parameterization
\begin{align}
\label{eq:baseline_parameterization}
   \text{logit}(\pi_{ij}) &= \lambda_i(\eta_j - \beta_i), \quad i = 1, \ldots, I, \nonumber \\
   \sum_{i = 1}^I \log(\lambda_i) &= 0 \quad \sum_{i = 1}^I \beta_i = 0, \nonumber \\
  \eta_j &\sim G,\quad j = 1, \ldots, N, 
\end{align}
where $G$ denotes a general distribution for the latent abilities, either parametric or nonparametric. The model in \eqref{eq:baseline_parameterization} follows the IRT parameterization with sum-to-zero identifiability constraints, which is typically the target one for inference for interpretability reasons. 

As discussed in Section~\ref{subsec:identifiability}, our target inferential model in \eqref{eq:baseline_parameterization} can be estimated directly, accounting for identifiability constraints. This can be achieved by introducing in the model formulation a set of auxiliary item parameters, $\{\lambda_i^\prime, \beta^\prime_i\}$ for each $i = 1, \ldots, I$ and defining $\{\lambda_i, \beta_i\}$ as
\begin{align}\label{eq:constrained_item}
  \log(\lambda_i) & = \log(\lambda^\prime_i) - \frac{1}{I}\sum_{i = 1}^I \log(\lambda^\prime_i) \quad 
  \beta_i = \beta^\prime_i - \frac{1}{I}\sum_{i = 1}^I \beta^\prime_i,\quad i = 1, \ldots, I.
\end{align}
In Table~\ref{tab:model_sampling_strategies} we label this model as the \emph{constrained item parameters model}. Unconstrained priors are then placed on the auxiliary parameters $\{\lambda_i^*, \beta^*_i\}$. The same formulation applies under the slope-intercept parameterization, where the sum-to-zero constraints are placed on the pairs $\{\log(\lambda_i), \gamma_i\}$ for $i = 1, \ldots, I$. 

Alternatively, we can consider \emph{unconstrained} versions of the 2PL model, treating the unconstrained model as a parameter-expanded version of the target inferential model in \eqref{eq:baseline_parameterization}, where the redundant parameters are the means of the difficulty and log discrimination parameters. Hence we conduct MCMC sampling with known model unidentifiability, and before using the results for inference, we transform samples to follow the model in \eqref{eq:baseline_parameterization}, as described in the Supplementary Materials, Section A. 

Finally, for the parametric case only, we consider the traditional version of the 2PL model that assumes the ability parameters $\eta_j$ for $j = 1, \ldots, N$ follow a standard normal distribution (\emph{constrained abilities model}). 

\subsection{Sampling algorithms}\label{sec:sampling_alg}

\inlineRevised{We focus on general MCMC sampling algorithms available in easy-to-access software tools for Bayesian hierarchical models, such as NIMBLE and Stan, that can flexibly accommodate different choices of prior distributions and link functions.} \inlineRevised{In this section we give an overall description of the algorithms considered for the different sampling strategies. We refer to the Supplementary Materials, Section B, for a detailed summary of the samplers used for each parameter.} 

For both the parametric and semiparametric models we consider NIMBLE's default sampling configuration (\emph{MH/conjugate algorithm}). NIMBLE's MCMC uses an overall one-at-a-time sampling strategy, cycling over individual parameters, or parameter blocks for parameters with a multivariate prior. 
By default, specific sampler types are assigned to the parameters or parameter blocks, \inlineRevised{but the user can choose to change sampler types, control blocking strategies, and modify details of sampling algorithm behavior.} NIMBLE's default MCMC configuration 
assigns a conjugate (sometimes called ``Gibbs'') sampler where possible, sampling from the corresponding full conditional posterior distribution. For non-conjugate continuous-valued parameters, NIMBLE's default sampler assignment is an adaptive random walk Metropolis-Hastings. For the parametric versions of the 2PL model, the strategies using the default NIMBLE assignments (\emph{MH/conjugate algorithm}) correspond to these conjugate and adaptive random walk Metropolis-Hastings samplers, with the latter also used for most parametric components of the semiparametric 2PL.
Specialized samplers are assigned when Bayesian nonparametric priors are considered in the semiparametric 2PL. 

In the case of the slope-intercept parameterization, we take advantage of NIMBLE's flexibility to include user-programmed custom samplers (\emph{centered sampler}). The proposed centered sampler uses an adaptive random walk Metropolis-Hastings sampler with a joint proposal for each pair of item parameters $\{\lambda_i, \gamma_i\}$ for $i = 1, \ldots, I$, thereby accounting for their posterior correlation. The proposal is made under a reparameterization of the model that centers the abilities to have mean zero. Implementation details are provided in the Supplementary Materials, Section B.

Finally, in the parametric setting only, we consider a Hamiltonian Monte-Carlo (HMC) algorithm, as implemented in the Stan software \citep{stan2017jss}. Stan implements an adaptive HMC sampler \citep{betancourt2017geometric} based on the No-U-Turn sampler (NUTS) of \cite{hoffman2014no}. HMC algorithms are known to produce samples that are much less autocorrelated than those of other samplers but at more computational cost given the need to calculate the gradient of the log-posterior. In this work, we limit the comparison to the IRT parameterization with constraints on the abilities distribution, as that is the model provided in the \texttt{edstan} \textsc{R} package \citep{edstan2017}.

\subsection{Aims}\label{sec:aims}

\inlineRevised{In the remainder of the paper, we study the efficiency of the MCMC sampling strategies in Table~\ref{tab:model_sampling_strategies} to fit binary logistic IRT models, and we compare inferential results under parametric and semiparametric specifications. Using both simulated and real-world data, we aim to answer the following questions: }

\begin{enumerate}
    \item[Q.1]  \begin{revised}
    For the parametric binary logistic IRT model, which of the Metropolis-Hastings-based MCMC sampling strategies in Table~\ref{tab:model_sampling_strategies} are most efficient?  Do different strategies work better in different scenarios for the distribution of ability? 
   \end{revised}

	\item[Q.2] \begin{revised}
    For the parametric binary logistic IRT model, how does efficiency of random walk Metropolis-Hastings sampling compare to Hamiltonian Monte Carlo (HMC), as implemented in the popular Stan package?  This question is of interest because HMC is not readily available for semiparametric models using Dirichlet process priors.
   \end{revised}

	\item[Q.3] How does MCMC efficiency of a semiparametric model compare to that of a parametric one?  Does this comparison differ when the parametric model is correctly vs. incorrectly specified?

	\item[Q.4] To what degree does the use of a parametric model when its assumptions are violated yield bad inference?  Does use of a semiparametric model change inference even when a parametric model would be valid?

	\item[Q.5] How much do results differ between the semiparametric and parametric models for the real data examples?
\end{enumerate}

\inlineRevised{In Section~\ref{sec:results_efficiency} we discuss the choice of efficiency metrics to address Q.1-Q.3 and present the results obtained using simulated and real-world data. In Section~\ref{sec:results_inference}, we discuss differences in the inferential results to investigate Q.4-Q.5.}

\section{Choice of prior distributions}\label{sec:prior_distributions}

Past research on Bayesian IRT models has warned about the use of either vague priors or highly informative priors when there is little information about the parameters \citep{sheng2010sensitivity,natesan2016bayesian}.  In particular \cite{natesan2016bayesian} investigated the use of different prior choices in 1PL and 2PL models using MCMC and variational Bayes algorithms and found that the use of vague priors tends to produce biased inference or convergence issues.  Similarly, it is well known that highly informative prior distributions on parameters can strongly affect model comparison procedures.

\inlineRevised{To ensure a fair comparison between results from different strategies, we chose the parameters of the priors in such a way that the induced prior predictive distribution of the data is similar across all the different model parameterizations.}
This ``predictive matching approach'' has been widely used to guide prior elicitation in model comparison settings \citep{berger1996intrinsic,bedrick1996new,ibrahim1997properties}.

\begin{figure}[H]
    \centering
    \includegraphics[width=0.95\textwidth]{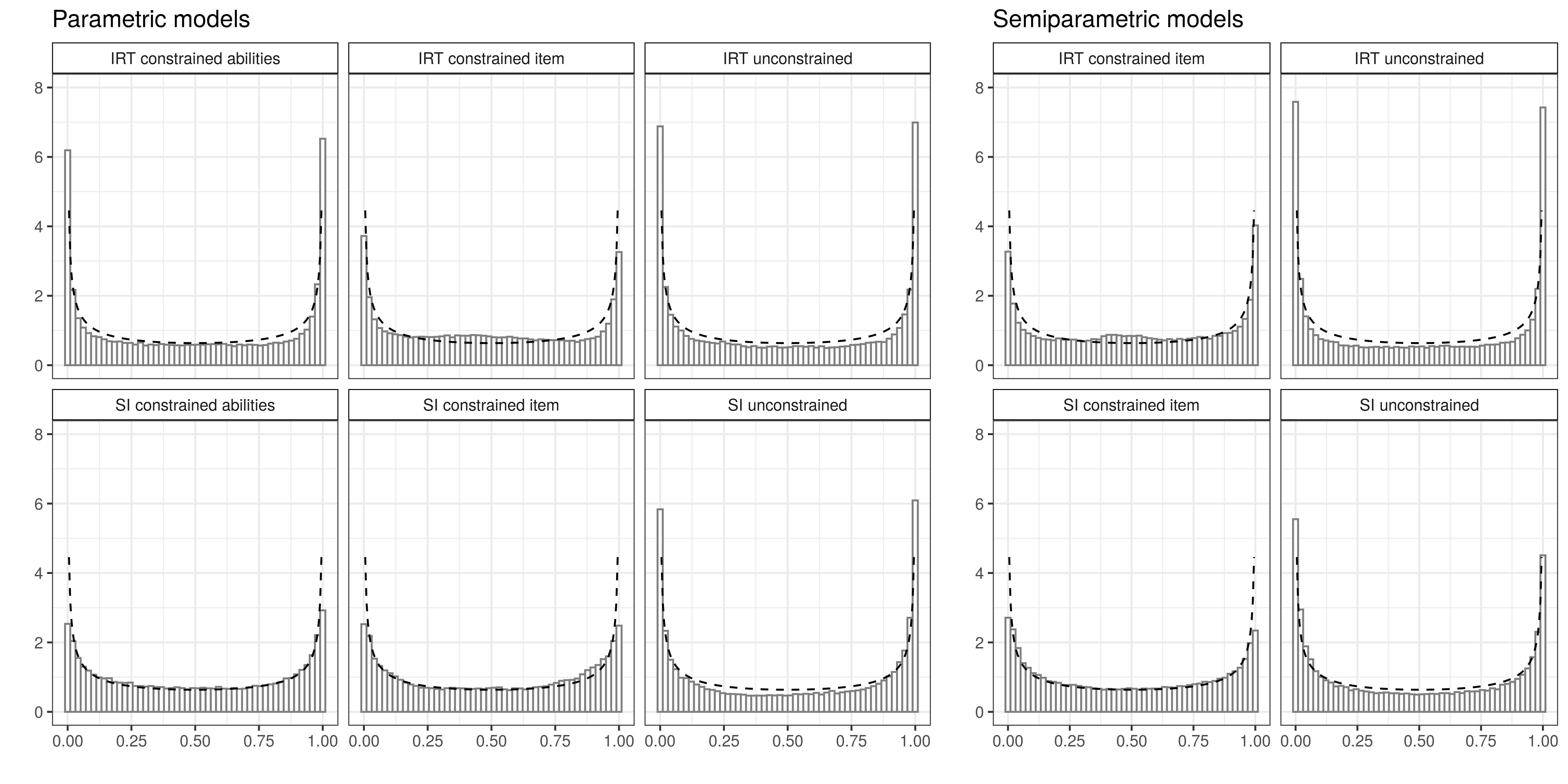}
   \caption{Histogram of samples from the induced prior on $\pi_{ij}$ under each of the considered models. Dashed line indicates the density function of a $\mbox{Beta}(0.5, 0.5)$ distribution. Samples for the semiparametric models use a prior distribution $\mbox{Gamma}(2,4)$ for the DP concentration parameter $\alpha$, but similar results are obtained under the other settings presented Section~\ref{sec:prior_distributions}.}
   \label{fig:priorgraphs}
\end{figure}

In the context of binary logistic IRT models, we aim to match the prior marginal predictive distribution of a response $y_{ij}$, which in turn can be achieved by matching the induced prior distribution on the marginal prior probability of a correct response, $\pi_{ij} = \mbox{expit}\{\lambda_i(\eta_j - \beta_i)\}$. Note that all the priors discussed in this paper are separately exchangeable, which means that this prior marginal will be the same for any values of $i$ and $j$.  In particular, we attempt to match a $\mbox{Beta}(0.5, 0.5)$ distribution, which is both the reference and the Jeffreys prior for the Bernoulli likelihood in the fully exchangeable case \citep{bernardo1979reference,berger2009formal}. A similar approach to prior elicitation in the context of latent space models for networks can be found in \cite{guhaniyogi2020joint} and \cite{sosa2021latent}.
Because there are no analytical expressions available for the prior distribution of $\pi_{ij}$, we use simulations to estimate the shape of the prior distribution and obtain an approximate match. This is facilitated by our implementation in NIMBLE.  Indeed, one of the advantages of the NIMBLE system is that it provides a seamless way to simulate from the model of interest. Histograms of samples from the resulting induced priors can be seen in Figure~\ref{fig:priorgraphs} for a set of parametric and semiparametric models. Further details are presented in the following subsections.

\subsection{Priors for the item parameters}

In Bayesian IRT modeling, normal distributions are typically chosen as priors for the item parameters. This is true under both parameterizations. In addition, the discrimination parameters, $\{\lambda_i\}_{i=1}^{I}$, are typically assumed positive, so we consider a normal distribution on the log-scale. To summarize, priors on the item parameters are:
\begin{equation*}
  \log{\lambda_i} \sim \mathcal{N}(\mu_{\lambda},\sigma^2_{\lambda}), 
  \quad \beta_i \sim \mathcal{N}(0, \sigma^2_{\beta}),
  \quad \gamma_i \sim \mathcal{N}(0, \sigma^2_{\gamma}) \quad i = 1,\ldots, I.
\end{equation*}
By default, we center on the difficulty parameters $\beta_i$ (or the reparameterized version $\gamma_i$) on 0 for $i = 1, \ldots, I$, while we set $\sigma^2_{\beta} = \sigma^2_{\gamma} = 3$. For the discrimination parameters, we set $\mu_{\lambda} = \sigma^2_{\lambda} = 0.5$ such that the prior probability mass on the original scale is mostly in the range $(0.5, 2.5)$. 

\subsection{Priors for the distribution of ability}

In choosing priors for the abilities, we distinguish between the parametric and semiparametric cases. In the parametric case, excluding the strategies in which the distribution is a standard normal, we assume $G \equiv \mathcal{N}(\mu_{\eta}, \sigma^2_{\eta})$. We specify hyperpriors for the unknown mean and variance, using a normal distribution for the mean $\mu_{\eta} \sim \mathcal{N}(0, 3)$, and an inverse-gamma distribution for the variance, $\sigma^2_{\eta} \sim \mbox{InvGamma}(2.01, 1.01)$ as in \cite{paulon2018}, with hyperparameter values implying an a priori marginal expected value of $1$ and an a priori variance equal to $100$. 

In the semiparametric case, we need to specify the base distribution $G_0$ of the DP mixture prior along with the hyperparameters. We choose $G_0 \equiv \mathcal{N}(0, \sigma^2_0) \times \mbox{InvGamma}(\nu_1, \nu_2)$ where $\mbox{InvGamma}(\nu_1, \nu_2)$ denotes an inverse-gamma distribution with shape parameter $\nu_1$ and  mean $\nu_2/(\nu_1 -1)$. In choosing values for the hyperparameters $\{\sigma^2_0, \nu_1 ,\nu_2\}$, we first considered the concentration parameter $\alpha$ as fixed and evaluated the induced prior distribution on $\boldsymbol{\pi}$ for values of $\alpha \in \{0.01, 0.05, 0.5, 1, 1.5, 2\}$. Recall that $\alpha$ controls the prior expectation and variance of the number of clusters induced by the DP, which are both of the order $\alpha \log(N)$. We discuss prior choice for the $\alpha$ in Section~\ref{sec:DP_conc_prior}. As in the parametric case, we center the normal distribution for the mixture component means on $0$ with $\sigma_0^2 = 3$ and set $\nu_1 = 2.01$ and $\nu_2 = 1.01$ for the inverse-gamma distribution. Given these settings, we found that choosing $\alpha \in \{0.01, 0.05, 0.5, 1, 1.5, 2\}$ does not have much effect on the marginal prior distribution of the $\pi_{ij}$s.

\subsection{Prior on the DP concentration parameter}\label{sec:DP_conc_prior}

One may be interested in placing a prior distribution on the concentration parameter $\alpha$ of the Dirichlet process. A typical choice for the DP concentration parameter is a $\mbox{Gamma}(a, b)$, with shape $a >0$ and scale $b >0$, due to its computational convenience \citep{escobar1995bayesian}. As previously stated, the concentration parameter controls the prior distribution of the number of clusters \citep{escobar1995bayesian,liu1996nonparametric}. In choosing values $a$ and $b$, we considered the implied prior mean and variance of the number of clusters.

Let $K_N$ denote the number of clusters for a sample of size $N$. Results from \cite{antoniak1974} and \cite{liu1996nonparametric} show that the expected value and variance of $K_N$ given $\alpha$ is
\begin{equation}\label{eq:number_of_clusters_expectation_variance}
  \mathbb{E}(K_N\mid \alpha) = \sum_{i = 1}^N \frac{\alpha}{\alpha + N - i},   \quad
  \mathbb{V}ar(K_N\mid \alpha) = \sum_{i = 1}^N \frac{\alpha(i - 1)}{(\alpha + N - i)^2}.
\end{equation}
We exploit these results to choose values $a$ and $b$ that lead to reasonable a priori values for the moments of the number of clusters for each of our applications.
For a given N and for different  values of $a$ and $b$, we evaluated the marginal expectation and variance of the quantities in~\eqref{eq:number_of_clusters_expectation_variance} via Monte Carlo approximation. We sample $\alpha_r$ for $r = 1, \ldots, R$ from its prior and compute 
\begin{equation*}
  \widehat{\mathbb{E}}(K_N) = \frac{1}{R} \sum_{r = 1}^R \mathbb{E}[K_N \mid \alpha_r],   \quad
  \widehat{\mathbb{V}ar}(K_N) = \frac{1}{R}\sum_{r = 1}^R \mathbb{V}ar(K_N \mid \alpha_r) +
  \widehat{\mathbb{V}ar} \left(\mathbb{E}[K_N \mid \alpha] \right), 
\end{equation*}
where $\widehat{\mathbb{V}ar} \left(\mathbb{E}[K_N \mid \alpha]\right) = R^{-1}\sum_{r =1}^R \left[\mathbb{E}[K_N \mid \alpha_r] - \widehat{\mathbb{E}}[K_N]\right]^2$. 

We explored a few prior choices and tabulate approximated moments in Table~\ref{tab:prior_values}, for the values of $N$ in our datasets. We consider the popular choice of $a= 2$, $b = 4$ for the hyperparameters as in \cite{escobar1995bayesian} along with values favoring a small number of clusters ($a = 1, b = 3$) and values leading to a more vague prior ($a = 1, b = 1$). For our applications we decided to favor a relatively small number of clusters, choosing $a=2, b = 4$ as hyperparameters for the simulated data, and $a = 1, b = 3$ for the real-world data.

\begin{table}
\resizebox{\textwidth}{!}{%
\begin{tabular}{c|cc|cc|cc|cc}
  $\alpha \sim Gamma(a, b)$ &
  $\mathbb{E}[\alpha]$ &  $\mathbb{V}ar(\alpha)$ &
  $\widehat{\mathbb{E}}(K_{2,000})$  & $\widehat{\mathbb{V}ar}(K_{2,000})$ & 
  $\widehat{\mathbb{E}}(K_{14,525})$ & $\widehat{\mathbb{V}ar}(K_{14,525})$ & 
  $\widehat{\mathbb{E}}(K_{7,377})$  & $\widehat{\mathbb{V}ar}(K_{7,377})$ \\
  \hline
  $a = 2, b = 4$ & $0.5$ & $0.12$ & $4.7$ & $9.3$  & $5.6$ & $14.12$ & $5.3$ & $12.2$ \\
  $a = 1, b = 3$ & $0.3$ & $0.11$ & $3.5$ & $7.6$  & $4.2$ & $11.76$ & $3.9$ & $10.2$ \\
  $a = 1, b = 1$ & $1$   & $1$    & $7.8$ & $43.7$ & $9.8$ & $73.35$ & $9.2$ & $64.5$\\
\end{tabular}
}
\label{tab:prior_values}
\caption{Approximate expectation and variance of the a priori number of clusters, $K_N$,  under different choices of the concentration parameter distribution, for $N = \mbox{2,000, 14,525, 7,377}$ \inlineRevised{as in our data example presented in Section~\ref{sec:data}}.}
\end{table}

\section{Data examples}\label{sec:data}

\subsection{Synthetic data}\label{sec:simulted_data}

We specify three different simulation scenarios for the distribution of ability: \emph{unimodal}, \emph{bimodal} and \emph{multimodal} distributions. For all the scenarios, we simulate responses from $N = 2,000$ individuals on $I = 15$ binary items. Values for the discrimination parameters $\{\lambda_{i}\}_{i = 1}^{15}$ are sampled from a $\mbox{Uniform}(0.5, 1.5)$ distribution, while values for difficulty parameters $\{\beta_{i}\}_{i = 1}^{15}$ are taken to be equally spaced in $(-3, 3)$. We center the log of the discrimination parameters on zero, while the difficulty parameters are already centered based on how they are generated.  
We consider three different underlying distributions for the latent abilities, $\eta_j$ for $j = 1, \ldots, 2,000$. In the \emph{unimodal} scenario, latent abilities are generated from a normal distribution with mean $0$ and variance $(1.25)^2$. In the \emph{bimodal} scenario, we use a equal-weights mixture of two normal distributions with means $\{-2, 2\}$ and common variance $(1.25)^2$. \inlineRevised{Finally, for the \emph{multimodal} scenario, latent abilities are generated from the following mixture
\begin{equation*}
    \frac{1}{5} \mathcal{N}(-2, 1 ) + 
    \frac{2}{5} \mathcal{N}(0, (0.5)^2 ) + 
    \frac{2}{5} \mathcal{SN}(3, 1, -3),
\end{equation*}
where $\mathcal{SN}(\xi, \omega, \zeta)$ indicates a skew-normal distribution \citep{azzalini1985class} with location parameter $\xi$, scale parameter $\omega > 0$ and parameter $\zeta$ that controls the asymmetry of the distribution.}

\inlineRevised{As a sensitivity analysis, we considered other values of $N$ and $I$, simulating data for each scenario following a factorial design with $I \in \{10, 30\}$ and $N \in \{1,000, 5,000\}$. We discuss efficiency results of the different sampling strategies in Section~\ref{sec:results_efficiency} and report results in the Supplementary Materials, Section D.}

\subsection{Real world data}

The first example is a subset of data from the 1996 Health Survey for England \citep{healthdata1996}, a survey conducted yearly to collect information concerning health and behavior of households in England. In particular, we have data for 10 items measuring Physical Functioning (PF-10), which is a sub-scale of the SF-36 Health Survey \citep{ware2003sf36} administered to people aged $16$ and above. In this case the latent trait quantifies the physical status of a given individual \citep{mchorney1997evaluation,hays2000item}.

\begin{figure}[H]
    \centering
     \includegraphics[width=\textwidth]{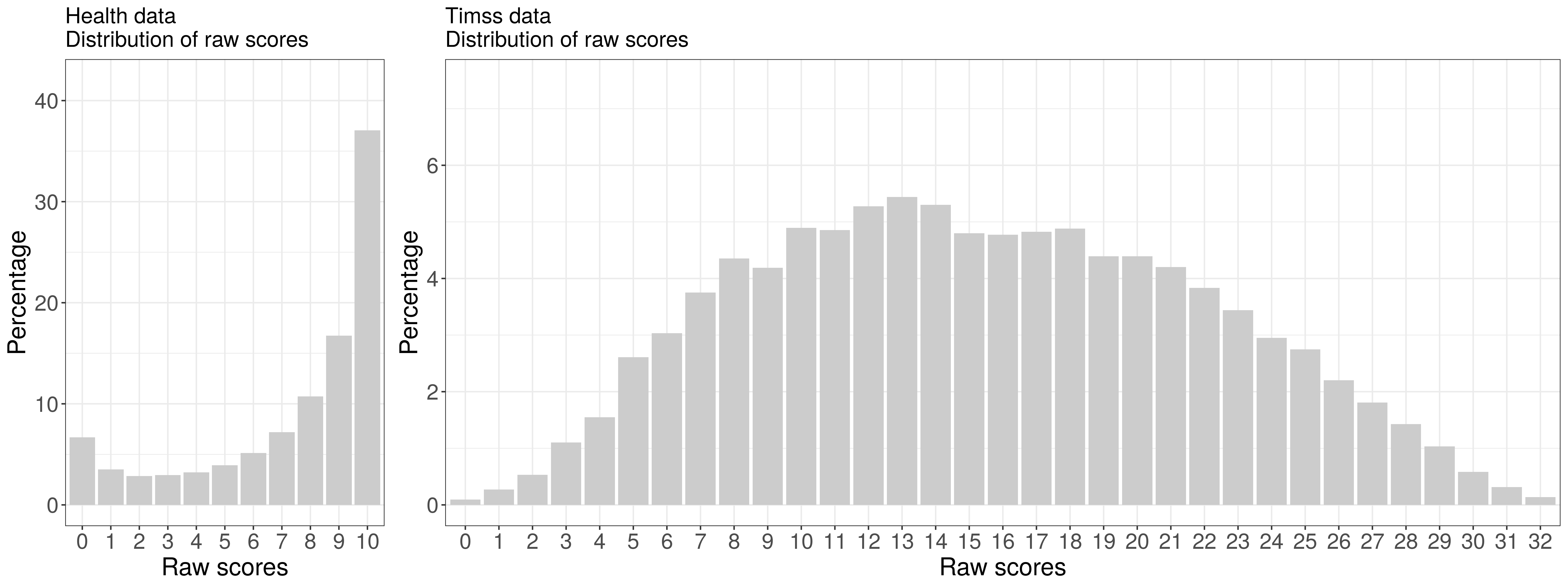}
    \caption{Distribution of the raw scores (total of correct answers) for the real data examples: health data (left panel) and TIMSS data (right panel).}\label{fig:real_data_raw_scores}
\end{figure}

Participants in the survey were asked whether they perceived limitations in a variety of physical activities (e.g., running, walking, lifting heavy objects) and if so the degree of limitation. We list the original questions in the the Supplementary Materials, Section C. Answers to items comprised three possible responses (``yes, a lot'', ``yes, limited a little'', ``no, not limited at all''); however, in our analysis we consider the dichotomous indicator for not being limited at all. The left panel of Figure~\ref{fig:real_data_raw_scores} shows the distribution of raw scores, i.e. the total of correct answers. 
\inlineRevised{We consider the 2PL model for this data, as it reasonable to assume that some of the questions are more informative in defining individuals with high physical impairment (see Supplementary Materials, Section C).}
For simplicity, we analyzed complete case data from $14,525$ individuals out of $15,592$ respondents, although the model can easily accommodate missing data. 

The second example uses data from the TIMSS (Trends in International Mathematics and Science Study) survey, which is an international comparative educational survey dedicated to improving teaching and learning in mathematics and science for students around the world (\url{http://timssandpirls.bc.edu/TIMSS2007/about.html}). We used data from the 2007 eighth-grade mathematics assessment for the United States (N = $7,377$), publicly available at \url{https://timssandpirls.bc.edu/TIMSS2007/idb_ug.html}. The dataset comprises $214$ items, with $192$ of them dichotomous, while the remaining $22$ have three category responses (``incorrect'', ``partially correct'', ``correct''). We dichotomized these latter questions, considering partially correct answers as incorrect ones. Like other large-scale assessments, participants in TIMSS only received a subset of the items according to a booklet design, resulting in $28$-$32$ item responses per student. Distribution of the raw scores for the data are shown in the right panel of Figure~\ref{fig:real_data_raw_scores}. \inlineRevised{As for the previous example, it is reasonable to assume that some items discriminate differently between students with high and low ability. However, in the context of educational testing, the 3PL model is often considered because it accounts for the probability of answering correctly by chance. Hence we consider both the 2PL and the 3PL models when evaluating the different strategies.}

Finally, we note that participants in the survey were sampled according to a complex two-stage clustered sampling design that we did not consider in our application. In other contexts the design is typically taken into account using sampling weights for model estimation, as discussed for example by \cite{rutkowski2010international}.

\section{Comparing results in terms of efficiency}\label{sec:results_efficiency}

MCMC performance is often evaluated in terms of mixing, often by calculating the  effective sample size (ESS), which is the equivalent number of independent samples that would contain the same statistical information as the actual non-independent samples. 
However, comparison between different MCMC algorithms based solely on mixing can be misleading, as different samplers can vary greatly in terms of computational cost \citep{nguyen2020}.
\inlineRevised{Hence, it is appropriate to consider ESS per computation time (in seconds), the rate at which effectively independent samples are generated.  A second issue is how to combine ESS results for multiple parameters.  For this purpose, we used the \emph{multivariate ESS} (mESS) recently introduced by \cite{vats2019multivariate}, which accounts for cross-correlations among parameters.}

Computation time is typically measured for the actual MCMC run, not accounting for steps to prepare for a run, thereby focusing on the algorithms of interest \inlineRevised{rather than unrelated aspects of the software}. Comparison between HMC and MCMC algorithms raises the question of how to fairly account for computation times, given that these two classes of algorithms use different tuning phases. \inlineRevised{Since there is not an established way to compare these two algorithms in the literature}, we decided to consider different timings when using the two algorithms: (i) \emph{sampling time}, which accounts only for the time to draw the posterior samples, hence discarding the time needed for the burn-in and warm-up phases of the two algorithms and (ii) \emph{total time} comprising also the burn-in and warm-up phases. 
\inlineRevised{Although one can use alternative metrics for the comparison, this choice can provide interesting and useful insights.} When computing efficiency based on the sampling time, we can assess pure efficiency of sampling from the posterior. Using total time accounts for potentially different times needed for warm-up/burn-in by the different algorithms but introduces the difficulty of determining the optimal burn-in/warm-up time, which we avoided here in favor of using basic defaults.

\inlineRevised{Estimate of the mESS is based on the multivariate batch means estimator as described in \cite{vats2019multivariate} and implemented in the \texttt{mcmcse} \citep{mcmcse} package. Since we used different specifications for the distribution of ability across the different sampling strategies, we calculate the mESS considering  only the common parameters (i.e., the item parameters and sampled abilities) after transforming samples to the parameterization of our target inferential model~(\ref{eq:baseline_parameterization}). The mESS provides a single scalar measure of joint mixing for all the parameters of interest in a model, but it does not necessarily reflect univariate ESS values of each parameter.  For example, mESS can be larger than all the univariate ESS values (see Supplementary Materials, Section D, Figures~\ref{fig:ESSunimodal}-\ref{fig:ESSmultimodal} for some insights). Given this, it can be useful as a simple overall performance metric but does not replace ESS for specific parameters of interest.}

\inlineRevised{Given the large number of examples and sampling strategies, we performed a preliminary experiment to choose the number of iterations and number of burn-in or warm-up samples. In particular, for a portion of the simulations we used multiple runs to determine the number of iterations and samples needed to obtain reliable estimates of the mESS.
For all MCMCs using NIMBLE, we decided to use a total of $50,000$ iterations, with a $10\%$ burn-in of $5,000$ for all examples. When running the HMC algorithm via Stan, we used a total of $15,000$ iterations, with the first $5,000$ iterations as warm-up steps.
However, we observed highly variable values of the mESS estimates when using the HMC algorithm. We also found that values of mESS are correlated with those of the HMC tuning parameters (see Supplementary Materials, Section D). Given this variability in mixing performance, we decided to limit comparisons with HMC to the simulation scenarios, reporting results from the run with the median ESS across multiple replications.}
\inlineRevised{All the models were estimated using a Linux cluster with 4 nodes having 24 cores and 128 GB RAM per node (Intel(R) Xeon(R) CPU E5-2643 v2 @ 3.50GHz). Across simulations, the running times ranged between 15-94 minutes for parametric models and between 37-124 minutes for semiparametric ones. For the data applications, running times ranged between 85-384 minutes for the parametric models, and 410-610 minutes for semiparametric ones.}

\subsection{Efficiency results for simulated data}

\begin{figure}[H]
	\centering
	 \includegraphics[width=\textwidth]{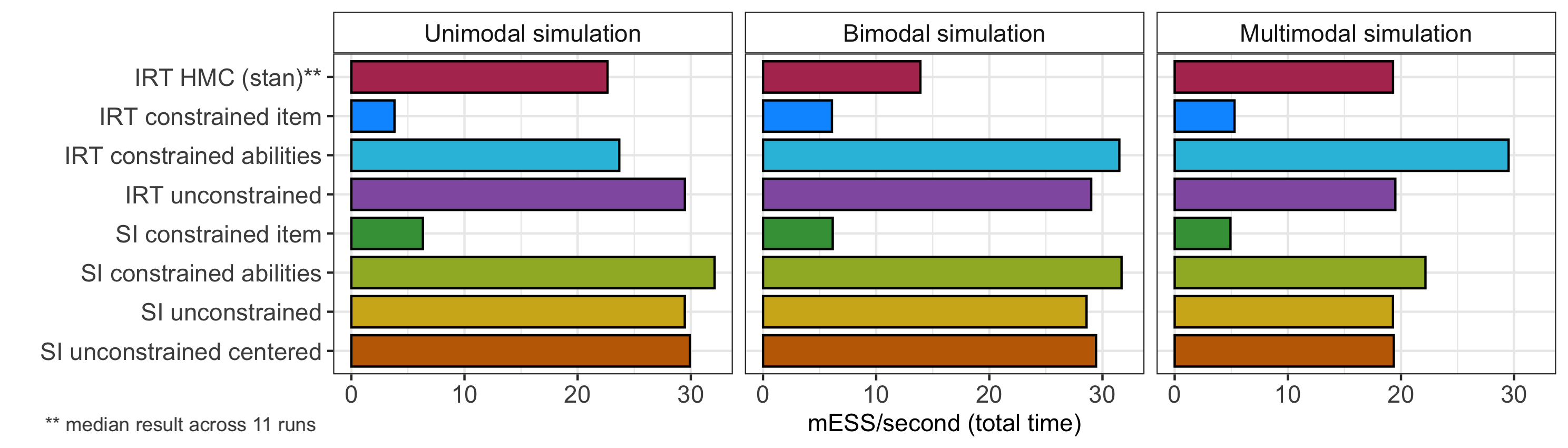}
	 \includegraphics[width=\textwidth]{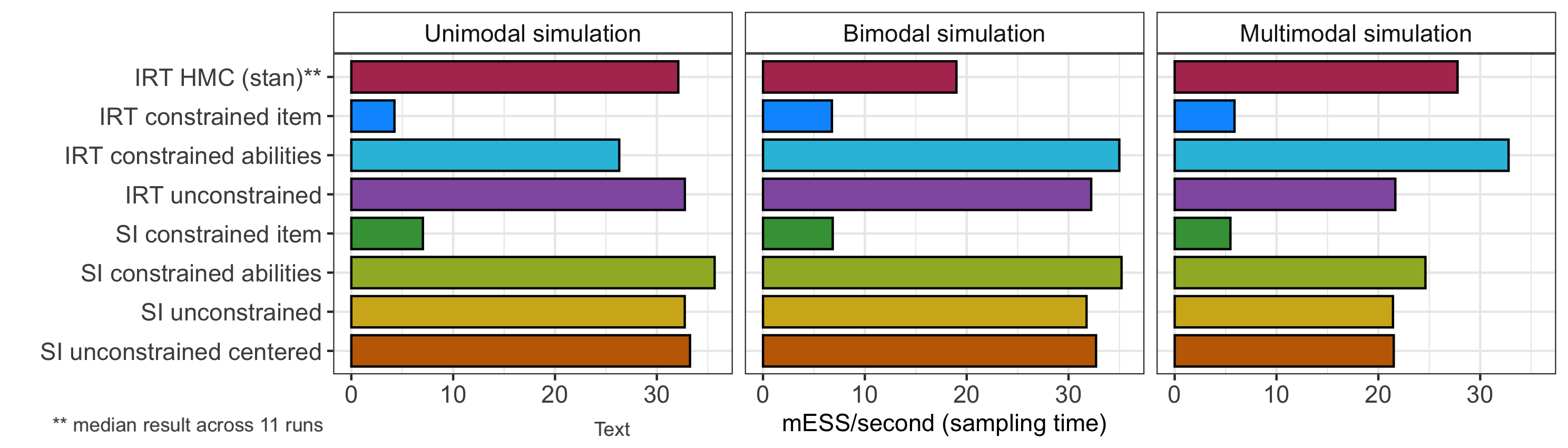}
	\caption{Multivariate ESS per second for various sampling strategies used to estimate the 2PL parametric model for the unimodal (left column), bimodal (middle column), and multimodal (right column) scenarios. Results are computed using total time (top row) and sampling time (bottom row). Note that SI stands for ``slope-intercept''.}\label{fig:efficiency_parametric}
\end{figure}

For the three simulation scenarios we estimated the 2PL parametric model using the different sampling strategies summarized in Table~\ref{tab:model_sampling_strategies}. Figure~\ref{fig:efficiency_parametric} compares efficiency for all these strategies using the multivariate ESS per second, computed with respect to both the total and sampling  time. 
Using different time baselines when computing efficiency changes the ranking of the MCMC algorithms for the unimodal and multimodal simulation, highlighting the trade-off for the HMC algorithm between sampling efficiency and computational cost. While the HMC is highly efficient in producing samples with low correlation, warm-up steps are computationally expensive. \inlineRevised{Recall that efficiency values presented for the HMC strategy are relative to a median performance across multiple runs.}
Amongst the non-HMC strategies, unconstrained scenarios generally mix well, as do scenarios with constraints on the abilities. However, imposing constraints on the item parameters directly in the sampling performs poorly because obtaining each sample is time-consuming. This is because the constraints in~\eqref{eq:constrained_item} require calculation of all the likelihood terms for each parameter update, whereas for other strategies only the likelihood terms for individuals' responses on the item under consideration need to be calculated. While incorporating constraints on the item parameters is time-consuming, such a strategy could be useful in more complicated hierarchical models, in particular when it is unclear how to rescale posterior samples.  The centered strategy for the slope-intercept parameterization seems to have little impact across the scenarios.

Moving to the semiparametric models, recall that we did not consider identifiability constraints on the abilities. We either included identifiability constraints on the item parameters in the sampling or sampled from the unconstrained model and rescaled the posterior samples.

\begin{figure}[t]
	\centering
	 \includegraphics[width=\textwidth]{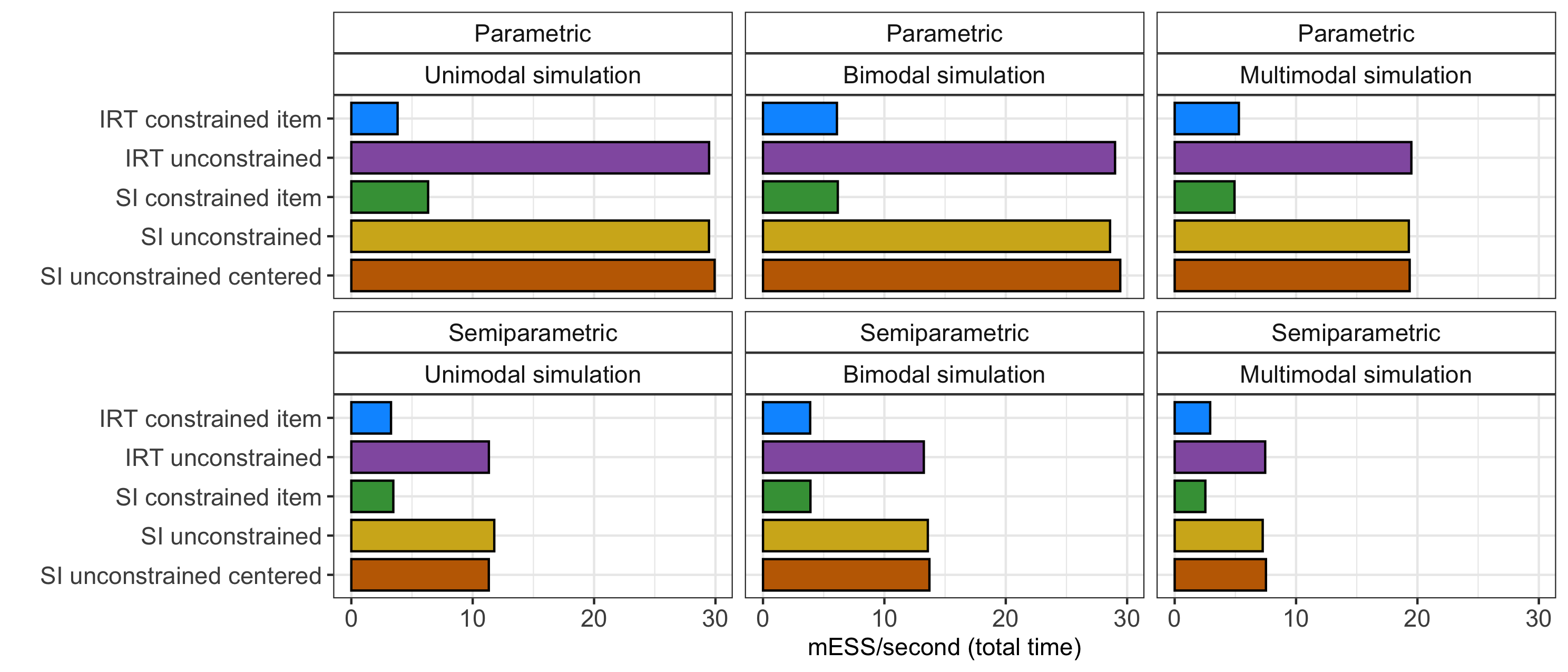}
	\caption{Multivariate ESS per second (computed using the total time)  for semiparametric 2PL models (bottom row) in comparison with their parametric version (top row) under the two simulation scenarios. Note that SI stands for ``slope-intercept''.}\label{fig:efficiency_parametric_bnp}
\end{figure}

As expected when using a more complicated model, we observed some reduction in efficiency in the semiparametric model compared to the parametric model, but not a drastic one (Figure~\ref{fig:efficiency_parametric_bnp}). Results for the semiparametric case are similar in relative terms, but not in absolute magnitudes, when comparing the different parameterizations and constraints.

\inlineRevised{For the non-HMC strategies, we also looked at how different combinations of numbers of items and individuals affect efficiency of the different sampling strategies (as discussed in Section 5.1), with results shown in  the Supplementary Materials, Section D. We found that the ranking across strategies is generally stable across the different scenarios; strategies using constraints on items are the worst overall, and the benefit of using other strategies is most evident when the number of individuals $N$ is low.}

\subsection{Efficiency results for real-world data}

\begin{figure}[hb]
    \centering
        \includegraphics[width=\textwidth]{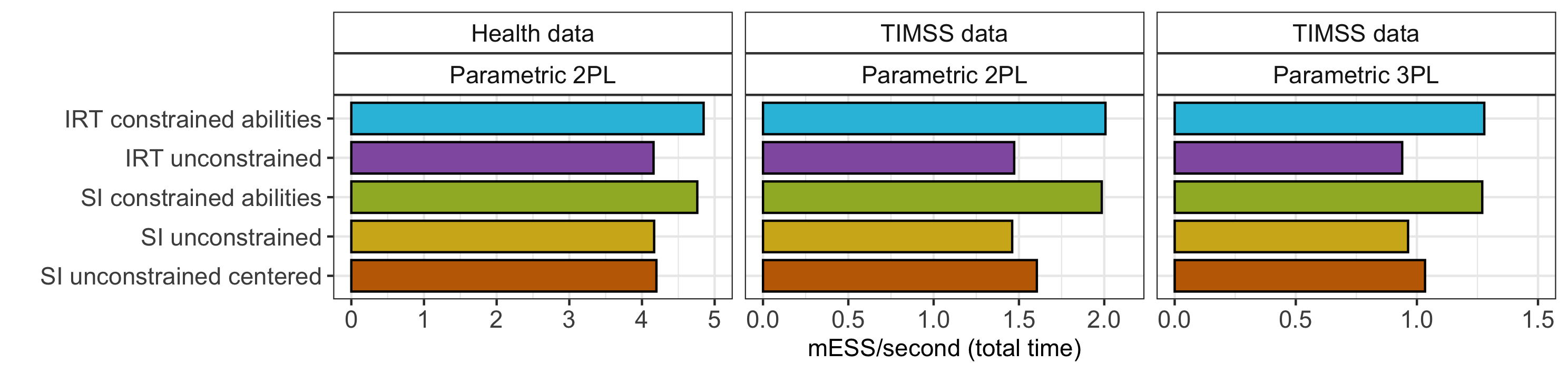}
        \includegraphics[width=\textwidth]{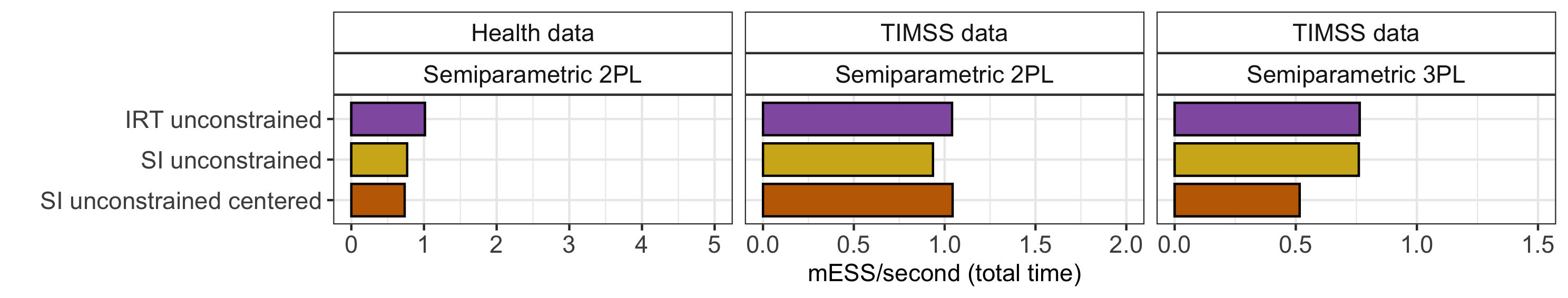}
    \caption{Multivariate ESS per second (computed using the total time) for the health data (left column) parametric and semiparametric 2PL models, and TIMSS data for the 2PL (middle column) and 3PL (right column) models. Note the scales are different for the different rows and that SI stands for ``slope-intercept''.}\label{fig:efficiency_data}
\end{figure}

We did similar comparisons using the real-data examples (Figure~\ref{fig:efficiency_data}), noting that we excluded the constrained items strategy, given its poor performance on the simulated datasets, \inlineRevised{and the HMC strategy, because of the high variability in mixing performance.}

The efficiency is lower than for the simulated datasets because the real data have many more individuals or items, and therefore more parameters. \inlineRevised{The same consideration applies when comparing efficiency of the 2PL and 3PL models for the TIMSS data. We also noted that for the health data,} when considering the posteriors for the abilities in the semiparametric model, we saw evidence for multimodality and some difficulty moving between modes for individuals with high raw scores. The multimodality is likely related to it being difficult for the semiparametric model to identify the exact magnitude of the ability for such individuals. The use of more informative priors, with careful elicitation of the prior distribution, may be important in such cases.

\section{Comparing results in terms of statistical inference}\label{sec:results_inference}

In this section we compare results for the parametric and semiparametric models in terms of statistical inference, regardless of the sampling strategy used to obtain posterior samples. All posterior samples follow the parameterization in \eqref{eq:baseline_parameterization}, and we use samples from the most efficient sampling strategy for each dataset. We use posterior means as point estimates for the item and ability parameters. For the simulated datasets, we measure how well the models recover the (known) true value of the parameters using absolute error, e.g., $|\hat{\beta_i} - \beta_i|$, and squared error, e.g., $(\hat{\beta_i} - \beta_i)^2$. 

A crucial point of this paper is to make inference on the distribution of latent abilities. An estimate of this distribution is sometimes based on the posterior means of the abilities \citep{duncan2008nonparametric,bambirra2018bayesian}, and histograms or kernel density plots are reported. Such an estimate ignores uncertainty in the estimates of individual abilities. Instead, one should directly obtain the point estimate of the posterior distribution of the latent abilities $p(\eta\mid  \mathbf{Y})$ (for any value of $\eta$) using the posterior samples. In the parametric case, this reduces to:
\begin{equation}\label{eq:parametric_estimate_latent_distribution}
    \widehat{p(\eta\mid  \mathbf{Y})} = \frac{1}{T} \sum_{t= 1}^T \mathcal{N}(\eta; \mu^{(t)}, \sigma^{2(t)}),
\end{equation}
with $\mathcal{N}(\cdot ; \mu, \sigma^{2})$ indicating the probability density function of a normal distribution with mean $\mu$ and variance $\sigma^{2}$, and $t = 1, \ldots, T$ denoting an MCMC iteration.
In the semiparametric case, a point estimate of $p(\eta\mid \mathbf{Y})$ is the posterior mean of the mixing measure $G$ of the Dirichlet process. This can be obtained using posterior samples, averaging over the DP conditional distribution in~\eqref{eq:DP_predictive} computed for each iteration $t = 1, \ldots, T$, 
\begin{equation}\label{eq:DP_estimate_latent_distribution}
	p\widehat{(\eta\mid  \mathbf{Y})} = \frac{1}{T} \sum_{t= 1}^T \left\{ \left[ \sum_{k = 1}^{K^{(t)}} 
	\frac{n_k^{(t)}}{\alpha^{(t)} + N} \mathcal{N}(\eta ; \mu_k^{(t)}, \sigma_k^{2^{(t)}}) \right]+ 
	\frac{\alpha^{(t)}}{\alpha^{(t)} + N} \mathcal{N}(\eta ; \mu_{K^{(t)} + 1}, \sigma_{K^{(t)} + 1}^{2})  \right\},
\end{equation}
with $n_k^{(t)}$ the number of observations in cluster $k$ at iteration $t$, $K^{(t)}$ the total number of clusters at iteration $t$, and $\mu_{K^{(t)} + 1}$ and $\sigma_{K^{(t)} + 1}^{2}$ sampled from $G_0$ (conditional on the data). We graphically compare estimates for the distribution of ability resulting from \eqref{eq:parametric_estimate_latent_distribution}--\eqref{eq:DP_estimate_latent_distribution} with the estimates obtained using the posterior means. 

It is possible to make full inference on $p(\eta\mid \mathbf{Y})$ in the semiparametric setting; this requires sampling from the posterior of the mixing distribution $F$. A computational approach to obtain the entire posterior distribution has been presented in \cite{gelfand2002}, a version of whose algorithm is implemented in NIMBLE in the function \texttt{getSamplesDPMeasure}. This function provides samples of a truncated version of the infinite mixture to a level $L$. The value of $L$ varies at each iteration of the MCMC's output when $\alpha$ is random, while it is the same at each iteration when $\alpha$ is fixed. In our case, for every MCMC iteration, we can obtain samples of the vector of mixture weights $\{w_1^{(t)}, \ldots, w_{L^{(t)}}^{(t)}\}$ and parameters of the mixture components.
We can use these samples to make inference on functionals of the distribution, such as the percentile for an individual, $100 \times p_j$, where $p_j = \int_{-\infty}^{\eta_j} p(\eta \mid  \mathbf{Y})d\eta$, typically paired with test scores when giving results for educational assessments. For an individual $j$ for $j = 1, \ldots, N$ we estimate $p_j$ at each MCMC iteration as
\begin{equation}\label{eq:percentile_estimator_bnp}
	p_j^{(t)} =  \sum_{l= 1}^{L^{(t)}} w_l^{(t)} F_{\mathcal{N}}(\eta_j^{(t)}; \mu_{l}^{(t)}, \sigma_{l}^{2(t)}),
\end{equation}
where $F_{\mathcal{N}}$ denotes the distribution function of the normal distribution. For comparison, we define the parametric counterpart as $p_j^{(t)} = F_{\mathcal{N}}(\eta_j^{(t)}; \mu^{(t)}, \sigma^{2(t)})$.

%%%%%%%%%%%%%%%%%%%%%%%%%%%%%%%%%%%%%%%%%%%%%%%%%%%%%%%%%%%%%%%%%%%%%%%%%%%%%
\begin{table}[t]
\resizebox{\textwidth}{!}{%
\begin{tabular}{|l|cc|cc|cc|cc|cc|cc|}
  \hhline{~------------}
  % \multirow{3}{*}{ }      & 
       \multicolumn{1}{c|}{}& 
       \multicolumn{4}{c|}{\textbf{Unimodal Simulation}}  &
       \multicolumn{4}{c|}{\textbf{Bimodal Simulation}} & 
       \multicolumn{4}{c|}{\textbf{Multimodal Simulation}}  \\
  \hhline{~------------}
  % \multirow{3}{*}{ } & 
      \multicolumn{1}{c|}{}&
      \multicolumn{2}{c|}{\textbf{Parametric}}        &
        \multicolumn{2}{c|}{\textbf{Semi-parametric}} &
        \multicolumn{2}{c|}{\textbf{Parametric}}        &
        \multicolumn{2}{c|}{\textbf{Semi-parametric}} &
        \multicolumn{2}{c|}{\textbf{Parametric}} &
        \multicolumn{2}{c|}{\textbf{Semi-parametric}} \\
  \hhline{~------------}
      \multicolumn{1}{c|}{}&
      MAE    & MSE    & MAE    & MSE    & MAE    & MSE    & MAE    & MSE  & MAE    & MSE & MAE    & MSE    \\
  \hline
\thead{Difficulty\\parameters}     & 0.0996 & 0.0185 & 0.0988 & 0.0183 & 0.0895 & 0.0137 & 0.0673 & 0.0083 & 0.0647 & 0.0080 & 0.0737  & 0.0103\\
\thead{Discrimination\\parameters} & 0.0734 & 0.0069 & 0.0731 & 0.0070 & 0.0832 & 0.0105 & 0.0397 & 0.0020 & 0.0721 & 0.0082 & 0.0677  & 0.0062\\
\thead{Ability\\parameters}        & 0.4836 & 0.3719 & 0.4836 & 0.3720 & 0.5944 & 0.5571 & 0.5501 & 0.4775 & 0.5477 & 0.4753 & 0.5212  & 0.4444\\
   \hline
\end{tabular}
}
\caption{MAE and MSE for the item and ability parameters estimates, under the three simulation scenarios, using samples from most efficient MCMC-based sampling strategies.}\label{tab:errors_point_estimates}
\end{table}

%%%%%%%%%%%%%%%%%%%%%%%%%%%%%%%%%%%%%%%%%%%%%%%%%%%%%%%%%%%%%%%%%%%%%%%%%%%%%

\subsection{Inferential results for the simulated datasets}

We report results using posterior samples from the unconstrained sampling strategy under the IRT parameterization. Using the absolute error and the squared error for each parameter, we report in Table~\ref{tab:errors_point_estimates} the mean absolute error (MAE) and the mean squared error (MSE) across item and ability parameters. 

%%%%%%%%%%%%%%%%%%%%%%%%%%%%%%%%%%%%%%%%%%%%%%%%%%%%%%%%%%%%%%%%%%%%%%%%%%%%%
\begin{figure}[H]
	\centering
	 \includegraphics[width=\textwidth]{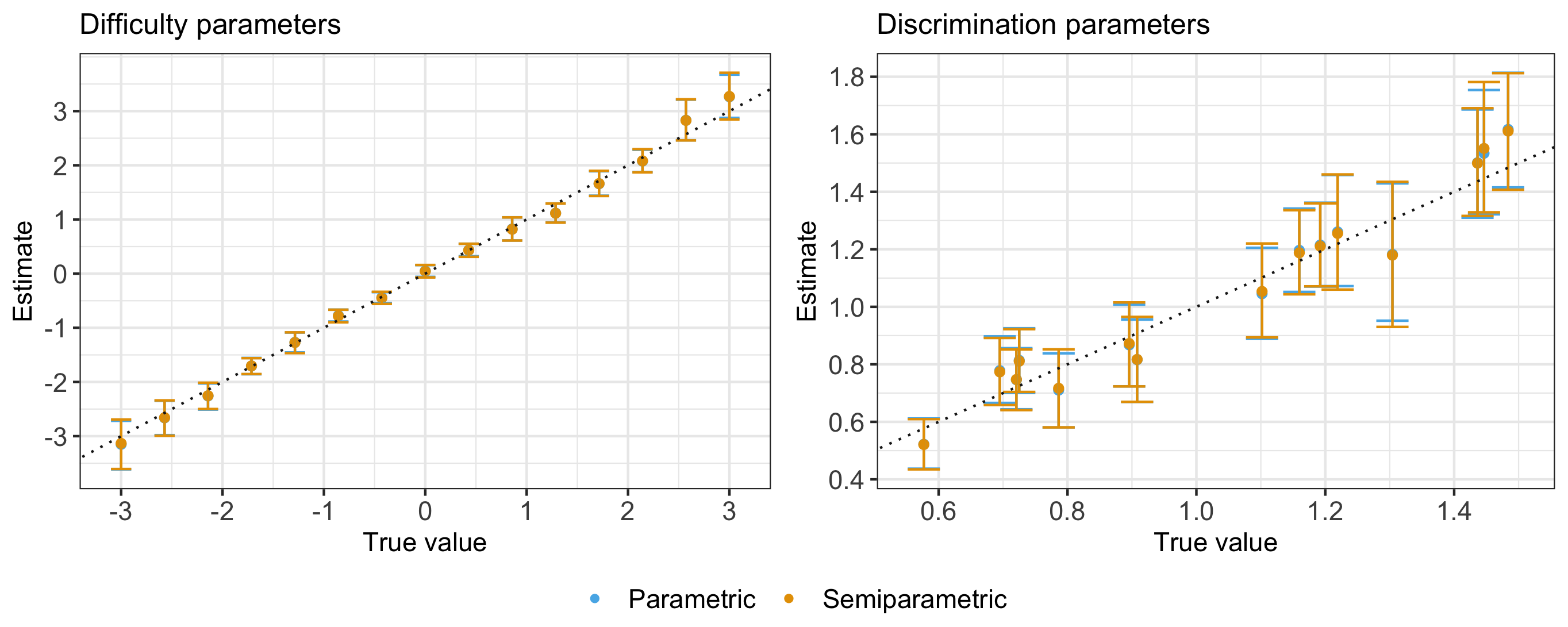}
	 \caption{Unimodal simulated data. Comparison of the posterior mean estimates (with $95\%$ credible interval) of item parameters (difficulties $\boldsymbol{\beta}$ and discriminations $\boldsymbol{\lambda}$) for parametric and semiparametric 2PL models and true simulated values. Note that some estimates from the parametric model overlap almost exactly with semiparametric ones.}
	 \label{fig:item_parameters_unimodal}
\end{figure}
%%%%%%%%%%%%%%%%%%%%%%%%%%%%%%%%%%%%%%%%%%%%%%%%%%%%%%%%%%%%%%%%%%%%%%%%%%%%%

\begin{figure}[H]
	\centering
	 \includegraphics[width=\textwidth]{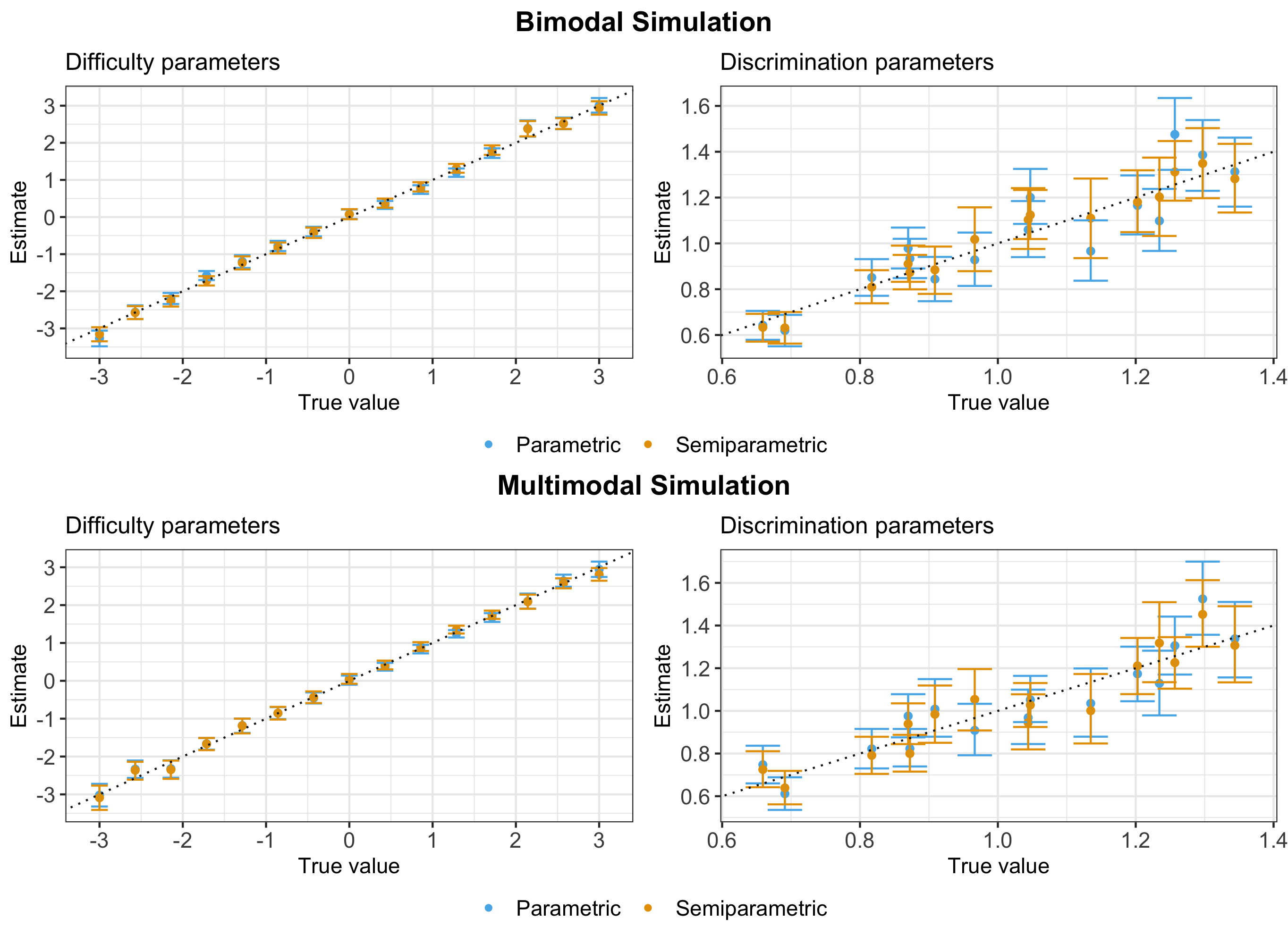}     
	 \caption{Bimodal Simulation (top row) and multimodal simulation (bottom row). Comparison of the posterior mean estimates (with $95\%$ credible interval) of item parameters (difficulties $\boldsymbol{\beta}$ and discriminations $\boldsymbol{\lambda}$) for parametric and semiparametric 2PL models and true simulated values.}
	 \label{fig:item_parameters_bimodal}
\end{figure}

In the unimodal scenario, we observe similar performance for the parametric and semiparametric 2PL in estimating item parameters (Figure~\ref{fig:item_parameters_unimodal}). As expected, we observe some differences when considering the bimodal and multimodal scenario (Figure~\ref{fig:item_parameters_bimodal}). The use of a model with a more flexible distribution improves recovery of both item and ability parameters in the bimodal scenarios (Table~\ref{tab:errors_point_estimates}). This is especially evident when comparing estimates of the discrimination parameters, in particular for larger values. Instead, in the multimodal case the inference for the item parameters seems relatively insensitive to the specification. 

%%%%%%%%%%%%%%%%%%%%%%%%%%%%%%%%%%%%%%%%%%%%%%%%%%%%%%%%%%%%%%%%%%%%%%%%%%%%%
%%%%%%%%%%%%%%%%%%%%%%%%%%%%%%%%%%%%%%%%%%%%%%%%%%%%%%%%%%%%%%%%%%%%%%%%%%%%%
       
\begin{figure}[H]
	\centering
	 \includegraphics[width=\textwidth]{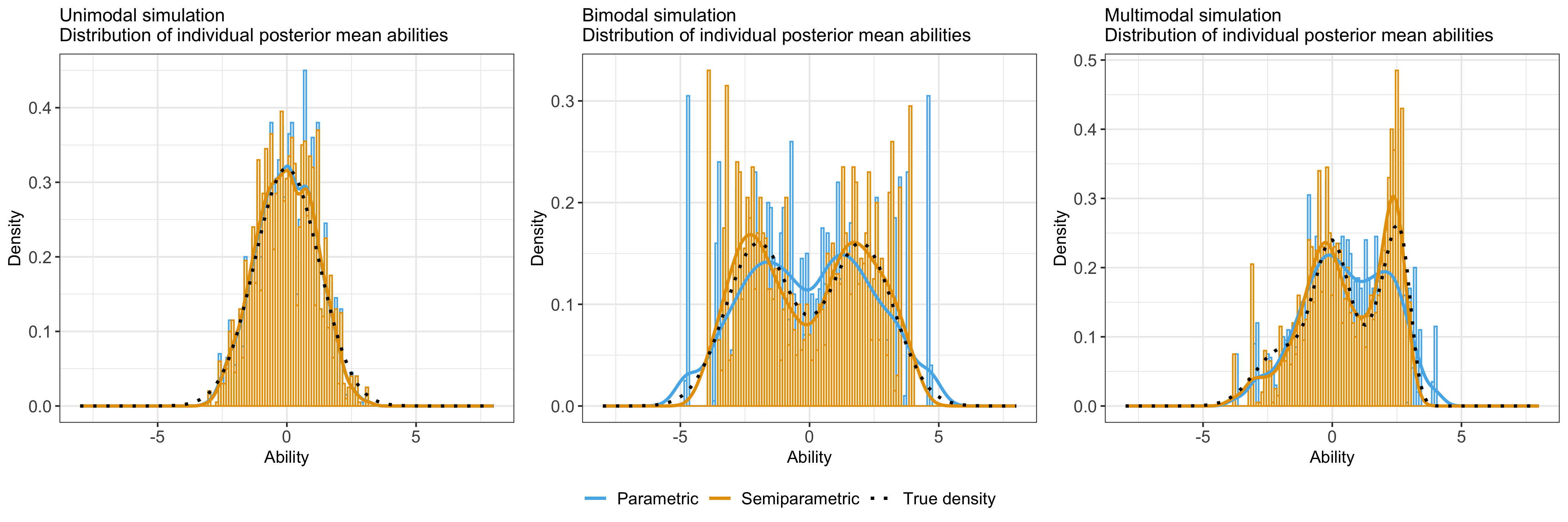}
	\caption{Histogram and density estimate of individual posterior mean abilities, under unimodal (left column), bimodal (middle column), and multimodal (right column) scenarios compared with the true density (dotted line).}
	\label{fig:abilities_hist_simu}
\end{figure}

\begin{figure}[H]
	\centering
	 \includegraphics[width=\textwidth]{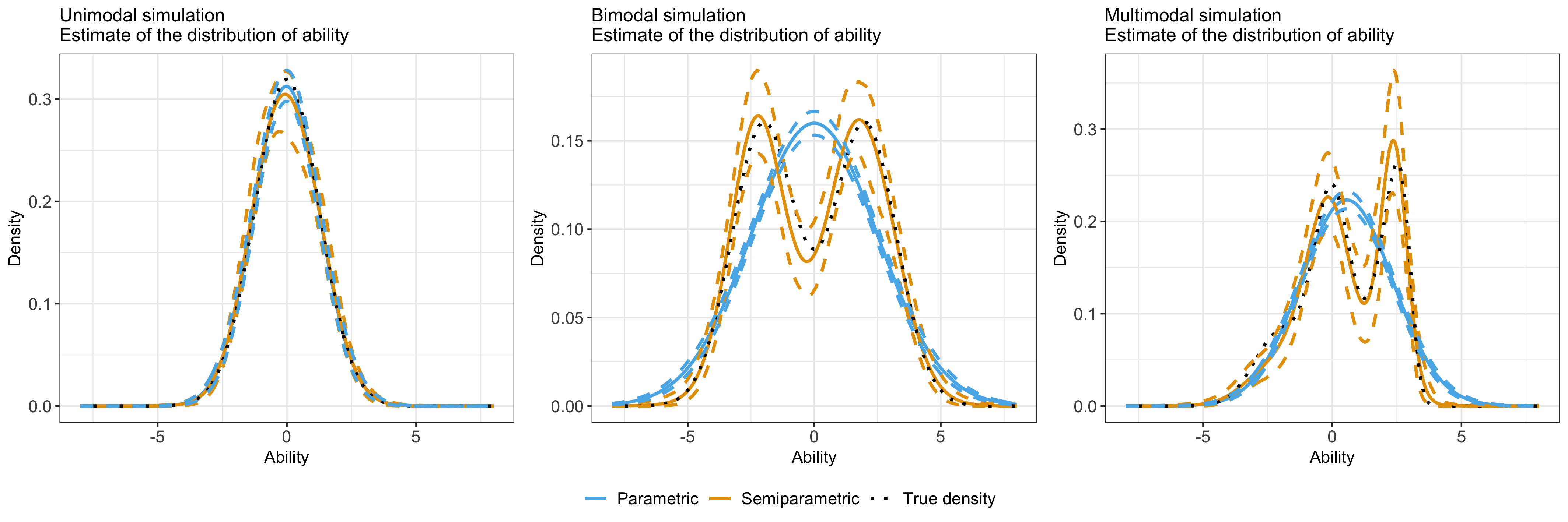}
	\caption{Distribution of ability estimated under the unimodal (left column), bimodal (middle column), and multimodal (right column) scenarios, compared with the true density (dotted line). Dashed lines indicate $95\%$ credible intervals for the estimated distributions.}
	\label{fig:abilities_posterior predictive_simu}
\end{figure}

Results for  the ability parameters are similar when estimating abilities using the posterior means of the individual abilities (Figure~\ref{fig:abilities_hist_simu}). However, results are very different when one looks at estimates of the distribution of ability (Figure~\ref{fig:abilities_posterior predictive_simu}). 
%%%%%%%%%%%%%%%%%%%%%%%%%%%%%%%%%%%%%%%%%%%%%%%%%%%%%%%%%%%%%%%%%%%%%%%%%%%%%

%%%%%%%%%%%%%%%%%%%%%%%%%%%%%%%%%%%%%%%%%%%%%%%%%%%%%%%%%%%%%%%%%%%%%%%%%%%%%
The normality assumption of the parametric model leads to unimodal density estimates, inconsistent with the true distribution, whereas the semiparametric model can recover it. The posterior means of individual abilities in Figure~\ref{fig:abilities_hist_simu} are a compromise between the inferred distribution of ability and the information in the data, so with sufficient observations, one can obtain estimates of the distribution that are reasonable even with severe model mis-specification. In other words, for mis-specified parametric models, the in-sample predictions for observed individuals can be reasonable, while the out-of-sample predictions based on \eqref{eq:parametric_estimate_latent_distribution} for new individuals are poor. Note that when using the parametric model, inspection of the posterior means of individual abilities can be used to assess model mis-specification relative to the assumed parametric distribution.

\begin{figure}[H]
	\centering
	 \includegraphics[width=\textwidth]{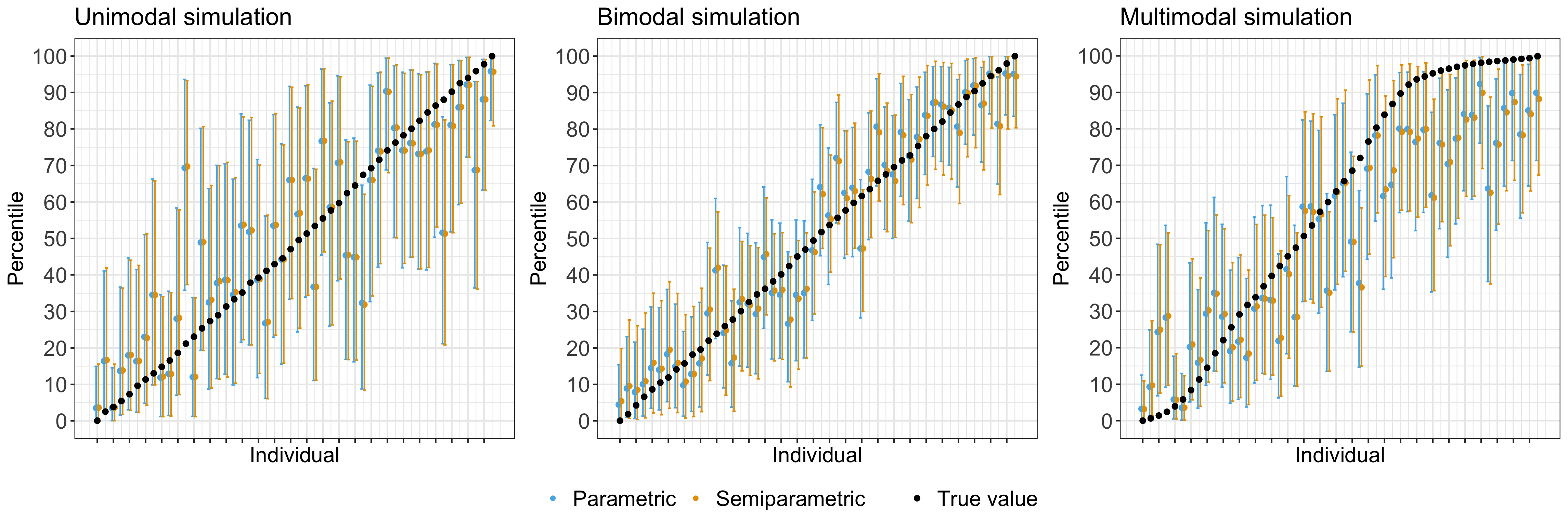}
	 \caption{Estimates of individual percentiles (with $95\%$ credible interval) for a subset of $50$ individuals with varying (true) ability levels under the unimodal (left column), bimodal (middle column), and multimodal (right column). Black dots correspond to true percentiles.}\label{fig:simulationPercentiles}
\end{figure}

Mis-specification of the distribution of ability has limited effect when estimating individual percentiles. In Figure~\ref{fig:simulationPercentiles} we compare the posterior mean estimates of individual percentiles with the percentiles calculated using the true distribution assumed for the simulation, for a subset of $50$ individuals. \inlineRevised{Overall, the parametric and semiparametric estimates produce similar results even when the estimated density is largely different. This is because estimation of percentile is basically ranking the individuals; it makes sense that ranking is relatively insensitive to the estimation of the distribution of ability.}

%%%%%%%%%%%%%%%%%%%%%%%%%%%%%%%%%%%%%%%%%%%%%%%%%%%%%%%%%%%%%%%%%%%%%%%%%%%%%

%%%%%%%%%%%%%%%%%%%%%%%%%%%%%%%%%%%%%%%%%%%%%%%%%%%%%%%%%%%%%%%%%%%%%%%%%%%%%

\subsection{Inferential results for real-world data}

For the real data examples we graphically inspect results from the parametric and semiparametric models. To compare the overall model fit in the parametric and semiparametric cases, we computed the Widely Applicable Information Criterion (WAIC) \citep{watanabe2010asymptotic}. We refer to the NIMBLE user manual for details about WAIC calculation \citep[see Section 7.7]{nimble-manual:2021}.

We found that for the health data the semiparametric model performs better than the parametric one (WAIC of 68,527 versus 70,414), while for the TIMSS data the best model is the semiparametric 3PL (WAIC of 228,871; parametric 3PL: 229,123; semiparametric 2PL: 229,529; parametric 2PL: 229,575). \inlineRevised{Hence in this Section, we show inferential results using samples from the IRT unconstrained model for both the health and TIMSS data, using the 3PL model for TIMMS. We also report inferential results for the 2PL model in the Supplementary Materials (Section E)}.

\begin{figure}[H]
	\centering
		\includegraphics[width = \textwidth]{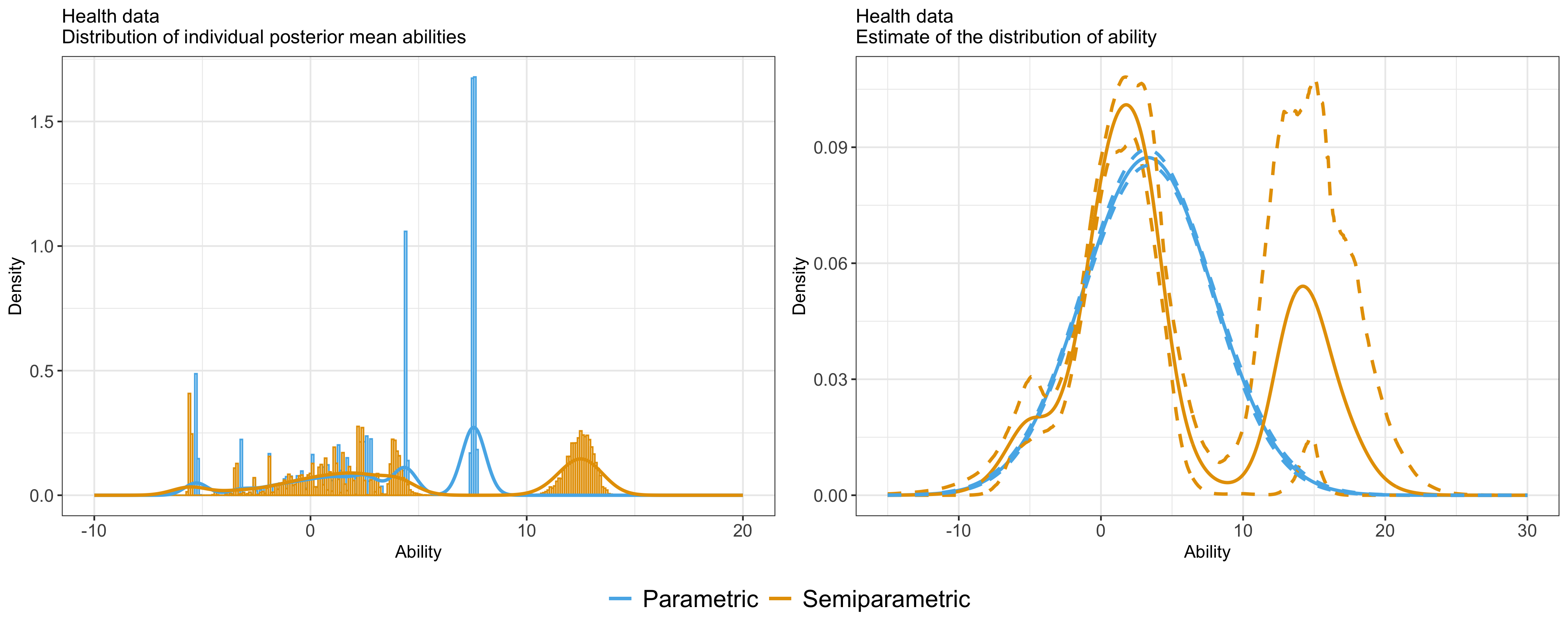}
	 \caption{Health Data. Histogram and density estimate of the posterior means of the latent abilities (left panel) and estimate of the posterior distribution for the latent abilities (right panel). Dashed lines indicate $95\%$ credible intervals for the estimated distributions.}
	\label{fig:abilities_health_data}
\end{figure}

In Figure~\ref{fig:item_parameters_health} we compare item parameter estimates from the two models for the health data application, while Figure~\ref{fig:abilities_health_data} shows estimates for the distribution of abilities. Recall that, in this case, we interpret the latent ability as physical ability, with high values characterizing healthy individuals. As with the bimodal simulation, estimates from the parametric model of the distribution of physical ability are quite different than the distribution of individual posterior mean abilities. It is clear that the parametric model is badly mis-specified and would produce bad out-of-sample predictions. In contrast, the semiparametric model seems to nicely characterize multi-modality in the latent distribution. We observe in Figure~\ref{fig:abilities_health_data} large credible intervals for high values of this distribution that can be explained by the presence of many individuals with high raw scores, (i.e., 9 or 10 out of 10, Figure~\ref{fig:real_data_raw_scores}) for whom the model can clearly determine that their physical abilities are high, but with the exact magnitudes being difficult to identify. 

The two modeling assumptions yield different estimates of the item parameters (Figure~\ref{fig:item_parameters_health}), with this difference being higher for extreme values. However, the relative ranking of the items is roughly the same in both cases, with for example item 1 (\emph{Vigorous activities}) being the most difficult item and the one with lowest value of the discrimination parameter.
According to the parametric model, discrimination parameters for item $3$ (\emph{Lift/carry}) and item $10$ (\emph{Bathing/dressing}) should have similar values, while the semiparametric model separates them. 

%%%%%%%%%%%%%%%%%%%%%%%%%%%%%%%%%%%%%%%%%%%%%%%%%%%%%%%%%%%%%%%%%%%%%%%%%%%%%
\begin{figure}[H]
	\centering
	 \includegraphics[width = \textwidth]{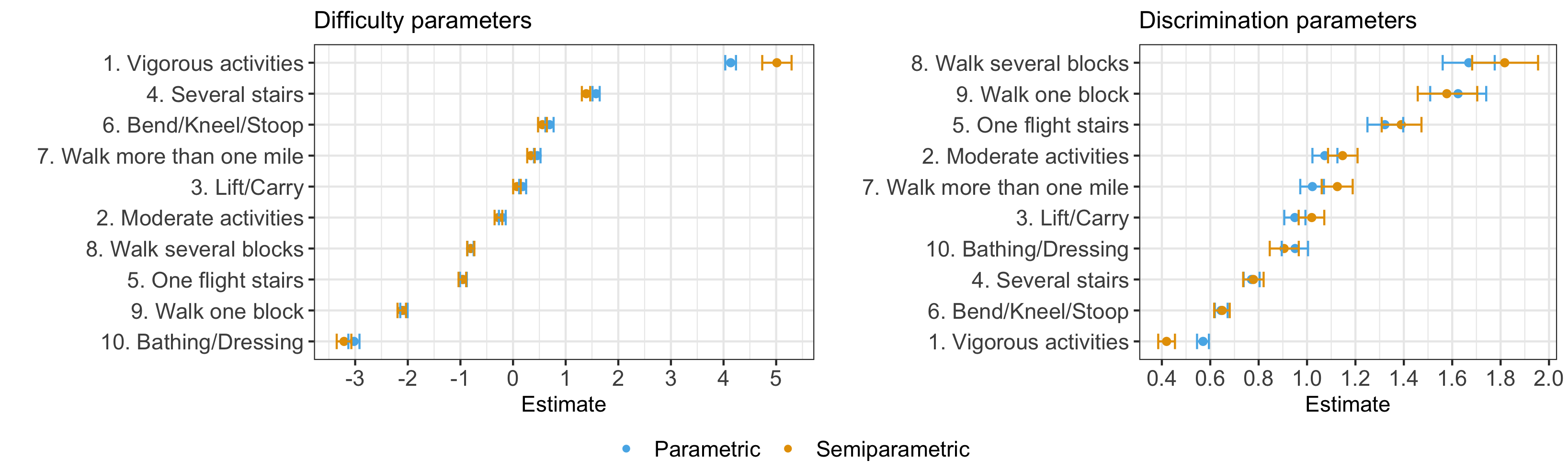}
	  \caption{Health data. Comparison of item parameter estimates from the parametric and semiparametric models. In each panel items are ordered by increasing values of the parameter estimate under the semiparametric model.}
	 \label{fig:item_parameters_health}
\end{figure}
%%%%%%%%%%%%%%%%%%%%%%%%%%%%%%%%%%%%%%%%%%%%%%%%%%%%%%%%%%%%%%%%%%%
%%%%%%%%%%%%%%%%%%%%%%%%%%%%%%%%%%%%%%%%%%%%%%%%%%%%%%%%%%%%%%%%%%%

\begin{figure}[H]
	\centering
	 \includegraphics[width=\textwidth]{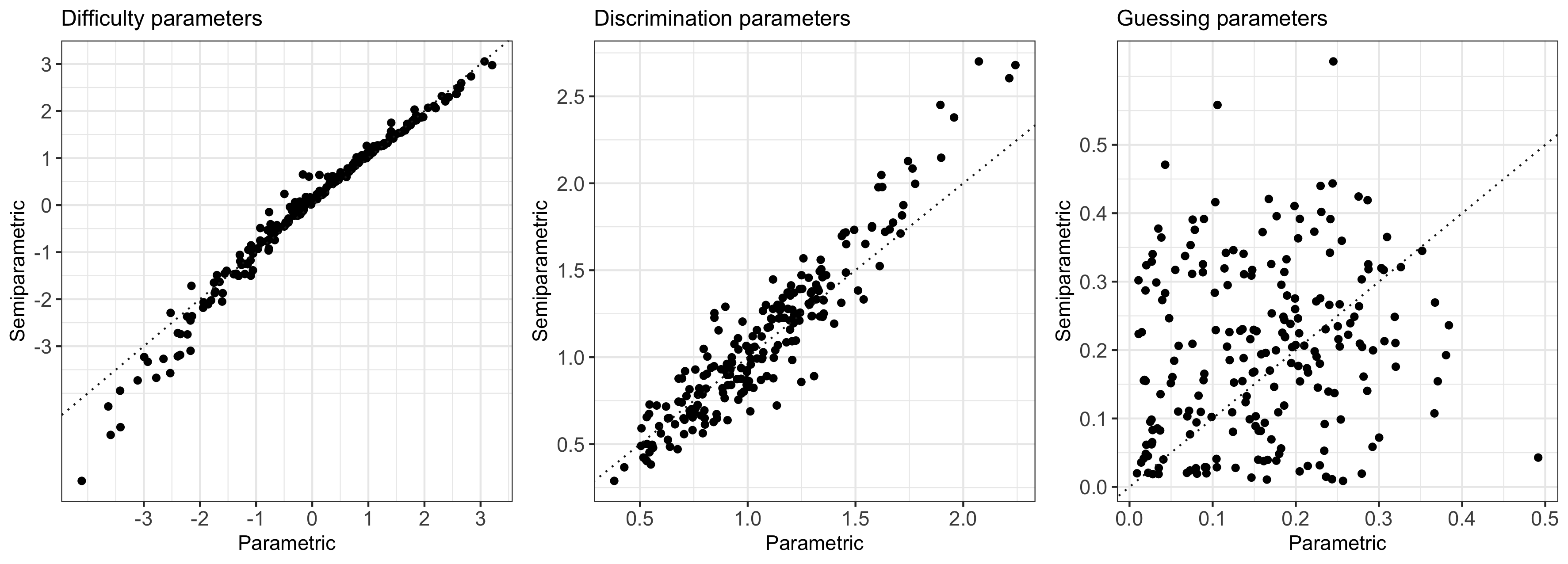}
	  \caption{TIMSS data. Comparison of posterior estimates of the item parameters between the parametric and semiparametric 3PL model both using the SI unconstrained centered sampling strategy.}
	 \label{fig:item_parameters_timss}
\end{figure}

\begin{figure}[H]
	\centering
	 \includegraphics[width=\textwidth]{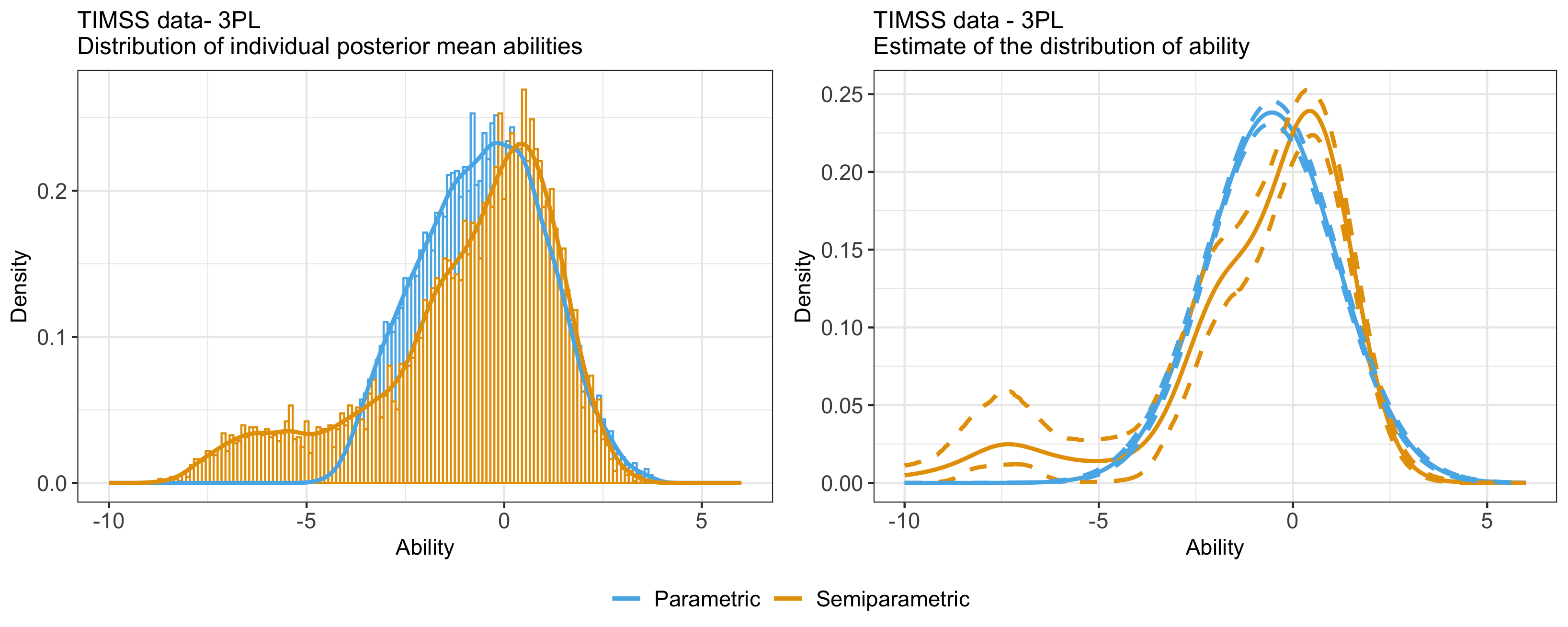}
	\caption{TIMSS Data. Histogram and density estimate of the posterior means of the latent abilities (left panel), and estimate of the posterior distribution for the latent abilities (right panel). Dashed lines indicate $95\%$ credible intervals for the estimated distributions.}
	\label{fig:abilities_timss_data}
\end{figure}

For the TIMSS data we only compare point estimates of the item parameters (Figure~\ref{fig:item_parameters_timss}), due to the large number of parameters. \inlineRevised{Estimates for the difficulty parameters and discrimination parameters have the largest difference for low/high values of the parameters, while estimates of the guessing parameters are quite different. Figure~\ref{fig:abilities_timss_data} shows estimates of the distribution of ability. In this case the semiparametric model estimate shows departure from the normal parametric assumption, with multimodality in the estimated distribution. Differences in the distribution of ability between the semiparametric and parametric model may explain the large discrepancies between the guessing parameter estimates; however, further investigation of this matter is outside the scope of the paper. We found also that the estimate of the ability distribution is quite different under the semiparametric 3PL and semiparametric 2PL model with the distribution under the 2PL model being unimodal and right-skewed instead of multimodal (see Supplementary Materials, Section F).}
 
%%%%%%%%%%%%%%%%%%%%%%%%%%%%%%%%%%%%%%%%%%%%%%%%%%%%%%%%%%%%%%%%%%%%
%%%%%%%%%%%%%%%%%%%%%%%%%%%%%%%%%%%%%%%%%%%%%%%%%%%%%%%%%%%%%%%%%%%%
%%%%%%%%%%%%%%%%%%%%%%%%%%%%%%%%%%%%%%%%%%%%%%%%%%%%%%%%%%%%%%%%%%%%%%%%%%%%%

\begin{figure}[H]
	\centering
 	\includegraphics[width=\textwidth]{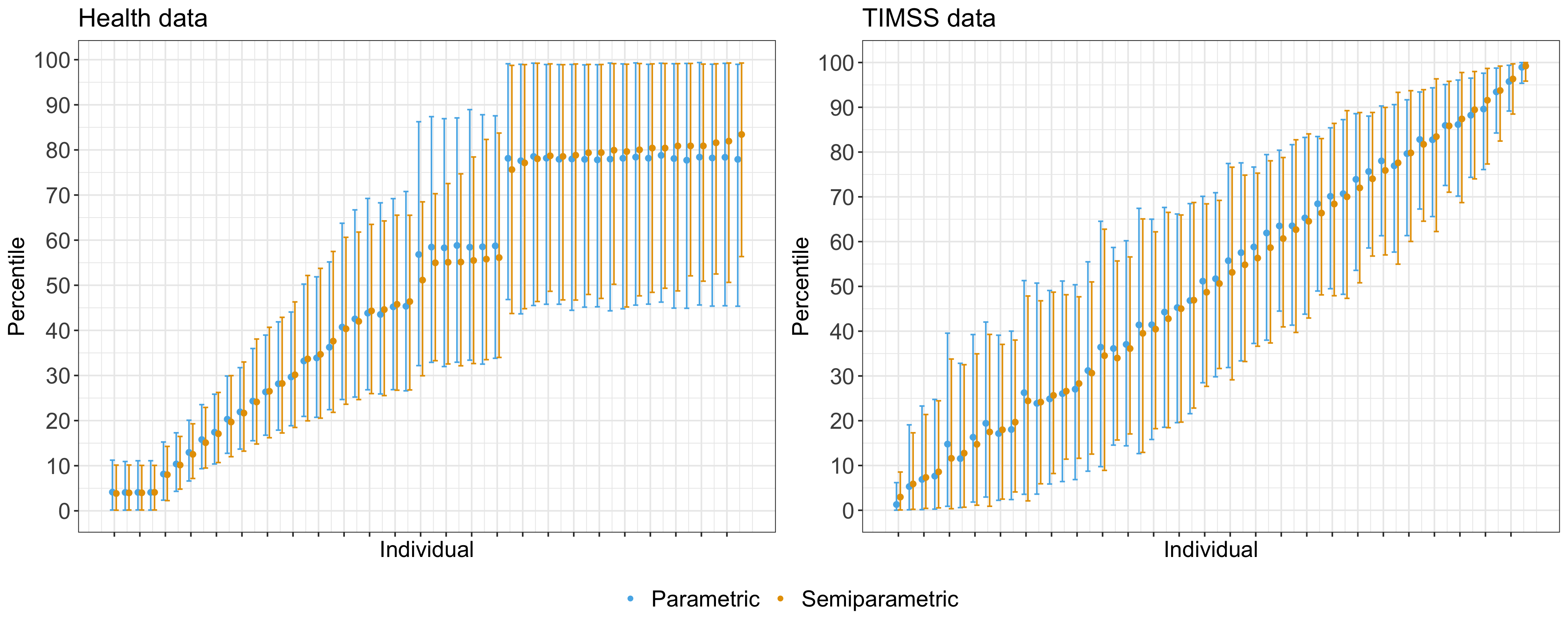}
	\caption{Estimates of individual percentiles (with $95\%$ credible interval) for a subset of $50$ individuals, for the health data (left panel) and TIMSS data (right panel).}
	\label{fig:data_percentiles}
\end{figure}

Figure~\ref{fig:data_percentiles} compares estimates of the percentiles for both the health and TIMSS data for a sample of $50$ individuals sorted according to the point estimates of the abilities from the semiparametric model. There are moderate differences in percentile values and individual ordering between the parametric and semiparametric models; in particular some estimates are associated with larger intervals than in the semiparametric case.

\section{Discussion}\label{sec:discussion}

In this paper, we consider a semiparametric extension for \inlineRevised{binary logistic IRT models}, using Dirichlet process mixtures as a nonparametric prior to flexibly characterize the distribution of ability. We provide an overview of these models and study how different sets of constraints can address identifiability issue and lead to different MCMC estimation strategies.

Focusing on the 2PL and 3PL models, we compare efficiency and inferential results under different sampling strategies based on model parametrization, constraints and sampling algorithms. We find that MCMC performance across strategies can vary in relation to underlying shape of the latent distribution \inlineRevised{and the total number of parameters.}

When moving to semiparametric modeling, the computational cost can be high for large datasets, given that sampling from the Dirichlet process requires iteration through all individuals. However we find computational costs to be reasonable in our applications in light of the better inferential results.

In particular under model mis-specification, inference for item parameters worsens noticeably in the parametric model compared to the semiparametric model. With sufficient data, inference for the abilities of observed individuals can be decent even under mis-specification of the distribution of ability, but inference for the unknown latent distribution (i.e., the predictive distribution for new individuals) as a whole can be quite bad. \inlineRevised{Although parametric IRT models can work well in applications in educational assessment, having access to semiparametric models can be broadly useful as it allows inference on the entire underlying distribution of ability and its functionals. This is particularly relevant in contexts where the distribution of the individual latent trait is more complicated, for example, when measuring health \citep{smits2020study} or psychological outcomes \citep{reise2016item}.}

\inlineRevised{Results of this work potentially can be generalized to versions of binary IRT models using different prior distributions or link functions (e.g. probit), since we considered general MCMC sampling algorithms as opposed to algorithms tailored to specific choices of such model components. As a general recommendation for IRT models, we found that sampling strategies using parameter expansion are more efficient than those embedding the constraints.}

In this work we extensively use the NIMBLE software for hierarchical modeling, with code  reproducing results in the paper available at \url{https://github.com/salleuska/IRT_nimble_code}. 
Although there are other software solutions enabling Bayesian nonparametric modeling, these are often limited in the type of algorithms or in the class of models available. NIMBLE offers a high degree of flexibility in that the models considered in this paper could be easily embedded in more complicated ones. Sampler assignment can be highly customized by the user, including user-defined sampling algorithms. This customizability makes NIMBLE a powerful platform for comparing different sampling strategies. At the same time, NIMBLE allows easy sharing of the most successful strategies as block-box implementations for end users.

% \subsection*{Acknowledgements}
% \textcolor{red}{This work was supported by funding from
% grants ACI-1550488, DMS-1622444, DMS-2114727 and DMS-2023495 from the National Science Foundation.}

\newpage

\vspace{\fill}\pagebreak

%% ITEM 9 [See the "howto.tex" file.]
\appendix
\renewcommand{\theequation}{A\arabic{equation}}
\setcounter{equation}{0}
\renewcommand{\thesection}{\Alph{subsection}}
\setcounter{section}{0}
\section*{Supplementary materials}

\section*{A. Identifiability}

The 2PL model is not identifiable based on the likelihood. Here we demonstrate the non-identifiability for the two parameterizations we consider, showing how different linear transformations lead to the same probabilities.
Note that these transformations are defined for every parameter associated with each item $i = 1, \ldots, I$ and individual $j = 1, \ldots, N$.

Under the IRT parameterization:
\begin{enumerate}
	\item $\eta_j^\prime = \eta_j/s$ and $\lambda_i^\prime = s\lambda_i$ 	
			\begin{equation*}
				\lambda_i^\prime(\eta_j^\prime - \beta_i) = 
				s\lambda_i(\eta_j/s- \beta_i) = 
				\lambda_i\eta_j  - \lambda_i\beta_i = \lambda_i(\eta_j - \beta_i),
			\end{equation*}
	\item $\eta_j^\prime = \eta_j + c$ and $\beta_i^\prime = \beta_i + c$, 	
			\begin{equation*}
				\lambda_i(\eta_j^\prime - \beta_i^\prime) = 
				\lambda_i(\eta_j + c - (\beta_i + c)) = 
				\lambda_i(\eta_j  - \beta_i).
			\end{equation*}
\end{enumerate}
Under the slope-intercept parameterization:
\begin{enumerate}
	\item  $\eta_j^\prime = \eta_j/s$ and $\lambda_i^\prime = s\lambda_i$, 	
			\begin{equation*}
				\lambda_i^\prime \eta_j^\prime + \gamma_i = 
				s\lambda_i \eta_j/s + \gamma_i = 
				\lambda_i\eta_j  + \gamma_i, 
			\end{equation*}
	\item $(\lambda_i\eta_j)^\prime = \lambda_i\eta_j + c$ and $\gamma_i^\prime = \gamma_i - c$, 
		  or $\eta_j^\prime = \eta_j + c$ and $\gamma_i^\prime = \gamma_i -\lambda_ic$ 	
			\begin{equation*}
				\lambda_i\eta_j^\prime + \gamma_i^\prime = 
				\lambda_i(\eta_j + c) + \gamma_i -\lambda_i c  = 
				\lambda_i\eta_j  +  \gamma_i.
			\end{equation*}
\end{enumerate}

\subsection*{Post-processing to satisfy identifiability constraints}
 
This section reports the transformations we apply to item and ability parameters in order to satisfy the identifiability constraints in our base parameterization~\eqref{eq:baseline_parameterization}. These transformations are applied to each posterior sample. 

Under the IRT parameterization, the set of transformations for each posterior sample of $\{\lambda_i, \beta_i, \eta_j\}$ for $i = 1, \ldots, I, j = 1, \ldots, N$ takes these forms:
\begin{align*}
  &\lambda^*_i = s\lambda_i,               \quad
  \beta^*_i = \frac{\beta_i -b}{s}, \quad
  \eta^*_j = \frac{\eta_j - b}{s},
\end{align*}
subject to $\prod_{i = 1}^I \lambda^*_i = 1$, $\sum_{i = 1}^I \beta^*_i = 0$. By solving the system of equations given by the transformations and the set of identifiability constraints, we obtain 
\begin{align}
 s = \exp\left\{\sum_{i = 1}^I \log(\lambda_i)/I \right\}, \quad 
 b = \frac{\sum_{i = 1}^I \beta_i}{I}.
\end{align}

Under the slope-intercept parameterization, the set of transformations for each posterior sample of $\{\lambda_i, \gamma_i, \eta_j\}$ for $i = 1, \ldots, I, j = 1, \ldots, N$ takes these forms:
\begin{align*}
	&\tilde{\lambda}_i = s\lambda_i,              \quad
	\tilde{\gamma}_i = \gamma_i -\lambda_i c,       \quad
	\tilde{\eta}_j = \frac{\eta_j + c}{s}, 
\end{align*}
subject to $\prod_{i = 1}^I \tilde{\lambda}_i = 1$, $\sum_{i = 1}^I \tilde{\gamma}_i = 0$. Similarly, by solving the system of equations given by the transformations and the set of identifiability constraints, we obtain
\begin{align}
 s = \exp\left\{\sum_{i = 1}^I \log(\lambda_i)/I \right\}, \quad c = \frac{\sum_{i = 1}^I \gamma_i}{\sum_{i = 1}^I \lambda_i}.
\end{align}

Finally, to get from the slope-intercept parameterization to the IRT parameterization, we define $\tilde{\beta}_i\vcentcolon= -\tilde{\gamma}_i/\tilde{\lambda}_i$ and then calculate $\beta_i^* = \tilde{\beta}_i - \sum_i \tilde{\beta}_i/I$.

\subsection*{Rescaling the DP density}

We can obtained posterior samples from the mixing distribution $F$ via NIMBLE's \texttt{getSamplesDPmeasure} function, allowing us to estimate the density for the latent ability distribution. However, when comparing these estimated densities between models, for some of the sampling strategies, we need to transform the estimated density to account for the transformations of the abilities from the scale on which sampling is done to the scale in (\ref{eq:baseline_parameterization}). 

As an example, consider the IRT parameterization without constraints. From the MCMC output we can obtain $p(\tilde{\eta})$ evaluated for different values of $\tilde{\eta}$, but we want $p(\tilde{\eta}^*)$ with $\tilde{\eta}^* = (\tilde{\eta} - b)/s$. To do so we need the Jacobian of the transformation, which is simply $s$. Then, we obtain $p(\tilde{\eta}^*)$

\begin{equation}
 p(\tilde{\eta}^*) = p_{\tilde{\eta}}(s\tilde{\eta}^*+b) \left|\frac{\partial (s\tilde{\eta}^* + b) }{\partial \tilde{\eta}^*}\right | = p_{\tilde{\eta}}(s\tilde{\eta}^*+b) s.
\end{equation}

\section*{B. Details about the sampling algorithms}

Tables~\ref{tab:parametric2PLdetails}-\ref{tab:semiparametric2PLdetails} summarize the sampling algorithms used for each sampling strategy in Table~\ref{tab:model_sampling_strategies}. 

\begin{table}[!h]
\caption{Summary of the sampling algorithms used for each parameter under the sampling strategies considered for the parametric 2PL model.}\label{tab:parametric2PLdetails}
\begin{center}
\begin{adjustbox}{max width=\textwidth}
\begin{tabular}{|Sl|Sc|Sc|Sc|Sc|Sc|Sc|}
  \hline
  \multirow{3}{*}{\textbf{Model constraints} } 
      & \multicolumn{2}{c|}{\textbf{IRT parameterization}}  &
        \multicolumn{2}{c|}{\textbf{SI parameterization}} \\
         \cline{2-5}
      &  MH/conjugate  & HMC (Stan) & MH/conjugate & Centered \\
  \hline
      Constrained abilities  & 
      \thead{Adaptive MH $\{\log(\lambda_i), \beta_i, \eta_j\}$} 
       & \thead{HMC $\left\{ \{\log(\lambda_i)\},  \{\beta_i\},  \{\eta_j\}\right\} $} & 
        \thead{Adaptive MH $\{\log(\lambda_i), \gamma_i, \eta_j\}$} &
       \thead{Centered sampler for pairs $\{\log(\lambda_i), \gamma_i \}$ \\ Adaptive MH $\{\eta_j\}$}   
       \\
  \hline
      Constrained item  & 
      \thead{Adaptive MH $\{\log(\lambda^*_i), \beta^*_i, \eta_j\}$} 
      & 
      -  
      & 
      \thead{Adaptive MH $\{\log(\lambda^*_i), \gamma^*_i, \eta_j\}$} 
      & - 
      \\
  \hline
   Unconstrained  & 
      \thead{Adaptive MH $\{\log(\lambda_i), \beta_i, \eta_j\}$ \\ Conjugate $\{\mu, \sigma^2 \}$} 
       &  - & 
       \thead{Adaptive MH $\{\log(\lambda_i), \gamma_i, \eta_j\}$\\ Conjugate $\{\mu, \sigma^2 \}$}
      &
      \thead{Centered sampler for pairs $\{\log(\lambda_i), \gamma_i \}$ \\ Adaptive MH $\{\eta_j\}$ \\ Conjugate $\{\mu, \sigma^2 \}$}  \\
  \hline
\end{tabular}
\end{adjustbox}
\end{center}
\end{table}

\begin{table}[H]
\caption{Summary of the sampling algorithms used for each parameter under the sampling strategies considered for the semiparametric 2PL model.}\label{tab:semiparametric2PLdetails}
\begin{center}
\begin{adjustbox}{max width=\textwidth}
\begin{tabular}{|Sl|Sc|Sc|Sc|Sc|Sc|}
  \hline
  \multirow{3}{*}{\textbf{Model constraints} } 
      & \multicolumn{1}{c|}{\textbf{IRT parameterization}}  &
        \multicolumn{2}{c|}{\textbf{SI parameterization}} \\
         \cline{2-4}
      &  MH/conjugate  & MH/conjugate & Centered \\
  \hline
      Constrained item  & 
      \thead{Adaptive MH $\{\log(\lambda^*_i), \beta^*_i, \eta_j\}$ \\ 
      CRP sampler  $\{\alpha, \{z_j\}\}$ \\ 
      Conjugate $\{ \{\mu^*_{k}\}, \{\sigma^{2*}_{k}\}\}$} 
      & 
      \thead{Adaptive MH $\{\log(\lambda^*_i), \gamma^*_i, \eta_j\}$\\ 
      CRP sampler  $\{\alpha, \{z_j\}\}$ \\ 
      Conjugate $\{ \{\mu^*_{k}\}, \{\sigma^{2*}_{k}\}\}$} & -
      \\
  \hline
   Unconstrained  & 
      \thead{Adaptive MH $\{\log(\lambda_i), \beta_i, \eta_j\}$ \\ 
      CRP sampler  $\{\alpha, \{z_j\}\}$ \\ 
      Conjugate $\{ \{\mu^*_{k}\}, \{\sigma^{2*}_{k}\}\}$} 
       &  
       \thead{Adaptive MH $\{\log(\lambda_i), \gamma_i, \eta_j\}$ \\
      CRP sampler  $\{\alpha, \{z_j\}\}$ \\ 
      Conjugate $\{ \{\mu^*_{k}\}, \{\sigma^{2*}_{k}\}\}$}
      &
       \thead{Centered sampler for pairs $\{\log(\lambda_i), \gamma_i \}$ \\ 
       Adaptive MH $\{\eta_j\}$ \\
      CRP sampler  $\{\alpha, \{z_j\}\}$ \\ 
      Conjugate $\{ \{\mu^*_{k}\}, \{\sigma^{2*}_{k}\}\}$}  \\
  \hline
\end{tabular}
\end{adjustbox}
\end{center}
\end{table}

\paragraph{Description of the sampling algorithms}
\begin{itemize}
    \item \textbf{Conjugate sampler}: for the models used in the paper, we exploit conjugancy results for the Normal distribution with Normal Inverse-gamma priors for the mean and variance. 
    \item \textbf{Adaptive MH} (Metropolis-Hastings): uses a normal proposal distribution, with initial proposal variance equal to $1$ and and adaptation interval of $200$ iterations. The adaptation routine is implemented as given in \cite{shaby2010}.
    \item \textbf{Centered sampler}: this is a custom defined sampler implemented by the authors. Details are given in the below.
    \item \textbf{CRP sampler}: under the CRP specification, the random measure $G$ is integrated out from the model and NIMBLE assigns a collapsed sampler. 
    \begin{itemize}
        \item clustering indicators $\mathbf{z}$ are updated as in described in \cite{neal2000markov};
        \item the DP concentration parameter $\alpha$ is sampled as described in \cite{escobar1995bayesian} (Section 6) when a Gamma prior is used, as in the models considered in the paper. If another prior is considered, NIMBLE uses a random walk Metropolis-Hastings. 
    \end{itemize}
\end{itemize}

\subsection*{Centered sampler}
We consider a custom sampler for the 2PL model under the slope-intercept parameterization. Intuition for this sampling strategy comes from the resemblance to a linear model. In order to sample $\{\lambda_i, \gamma_i\}$ efficiently, we propose centering the implied covariate, $\eta_j$, to have mean zero. This is analogous to centering covariates in a linear model, but in this case the "covariate" values are not fixed, so the centering needs to be done in each iteration. For a given item $i$ for $i = 1, \ldots, I$ we can rewrite
\begin{align*}
\lambda_i \eta_j + \gamma_i &= \lambda_i (\eta_j-\bar{\eta}) + \lambda_i\bar{\eta} + \gamma_i, \\
&= \lambda_i \eta_j^{c} + \gamma_i^{c},
\end{align*}
such that the quantity $\eta_j^{c} = \eta_j-\bar{\eta}$ is centered. The idea is to propose a new value $\lambda_i^*$ in this new parameterization at each MCMC iteration, using a random walk on the log scale. Translating to the original parameterization, we have:
\begin{align*}
\lambda_i^* \eta_j^{c} + \gamma_i^{c} &= \lambda_i^* (\eta_j-\bar{\eta})  + \lambda_i\bar{\eta} + \gamma_i, \\ 
&= \lambda_i^* \eta_j - \lambda_i^*\bar{\eta} + \lambda_i\bar{\eta} + \gamma_i.
\end{align*}
This means that we are proposing $\gamma_i^* = \gamma_i + \bar{\eta}(\lambda_i - \lambda_i^*)$. Thus we have a joint proposal $(\lambda_i^*, \gamma_i^*)$ that accounts for the usual correlation in a regression between intercept and slope. Apart from accounting for sampling $\lambda_i$ on the log scale, the proposal is symmetric, so no Hastings correction is needed. The original sampler for $\gamma_i$ can stay the same. This is because in the reparameterization with $\gamma_i^{c}$ above, shifting $\gamma_i$ by a certain amount is equivalent to shifting $\gamma_i^{c}$.

\section*{C. Health data questions}

The following items are about activities you might do during a typical day. Does your
health now limit you in these activities? If so, how much?
\begin{enumerate}
\setlength\itemsep{0.2cm}
\item Vigorous activities: Vigorous activities, such as running, lifting heavy objects, participating in strenuous sports.
\item Moderate activities: Moderate activities, such as moving a table, pushing a vacuum cleaner, bowling or playing golf.
\item Lift/Carry: Lifting or carrying groceries.
\item Several stairs: Climbing several flights of stairs.
\item One flight stairs: Climbing one flight of stairs.
\item Bend/Kneel/Stoop: Bending, kneeling, or stooping.
\item Walk more mile: Walking more than a mile.
\item Walk several blocks: Walking several blocks.
\item Walk one block: Walking one block.
\item Bathing/Dressing: Bathing or dressing yourself.
\end{enumerate}

\section*{D. A note on efficiency comparisons}

\subsection*{Comments on the multivariate ESS}

\begin{figure}[H]
    \includegraphics[width=\textwidth]{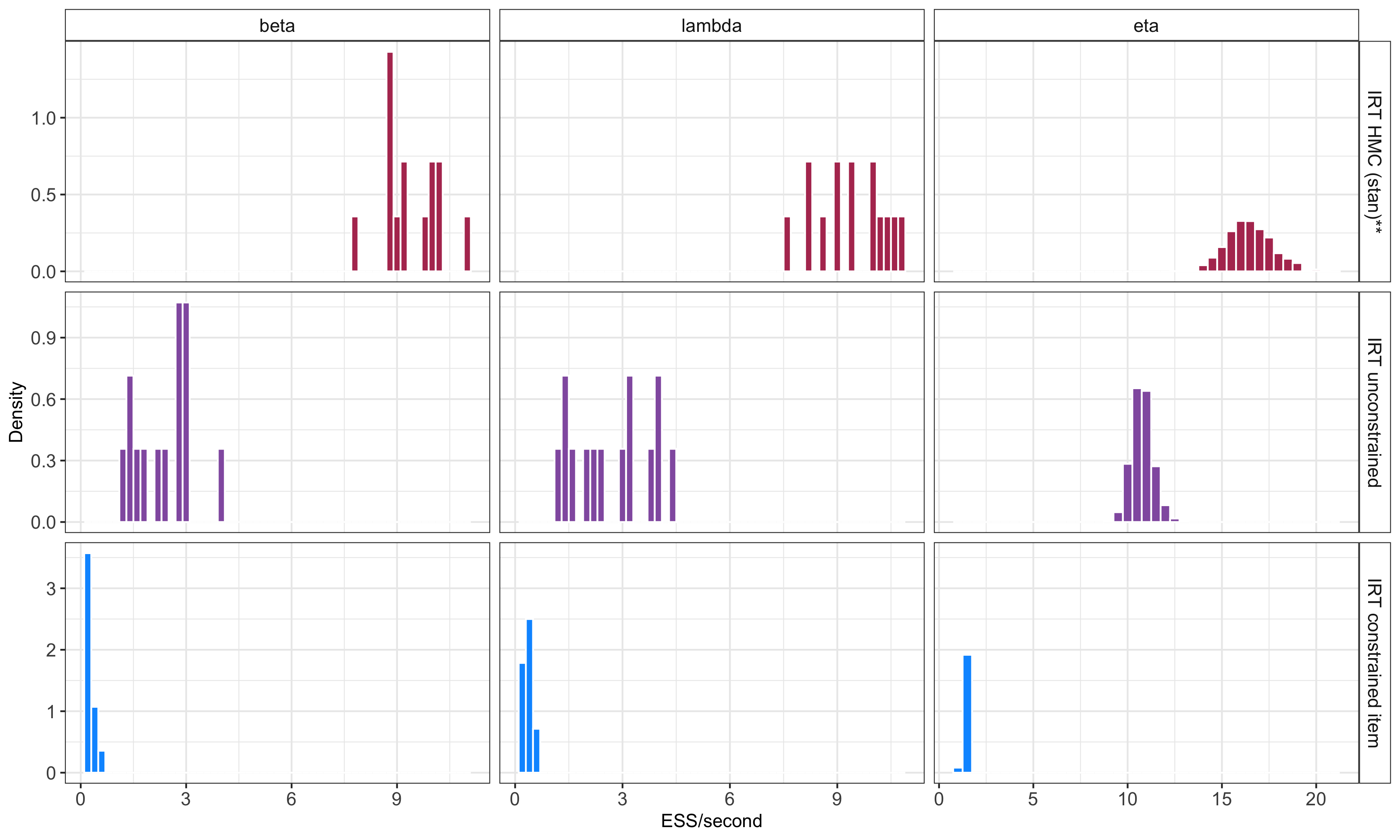}
    \begin{center}         
     \resizebox{0.8\textwidth}{!}{%
     \begin{tabular}{|c|c|c|c|c|}
        \hline
        Strategy & mESS & total time (second) & efficiency (mESS/second) \\
        \hline
        IRT HMC (Stan)**  	  & 31802  & 1405 &   23 \\
        IRT unconstrained     & 27540  &  934 &   29 \\
        IRT constrained item  & 27904  & 7318 &   4 \\
        \hline
      \end{tabular}
      }
    \end{center}
  \caption{Unimodal simulation ($N = 2000, I = 15)$. Distribution of univariate efficiencies (univariate ESS/seconds) for each group of parameters used to compute the multivariate efficiency (mESS/second) using the total time. The symbol ** denotes median results across 11 runs.}\label{fig:ESSunimodal}
\end{figure}

\begin{figure}[H]
    \includegraphics[width=\textwidth]{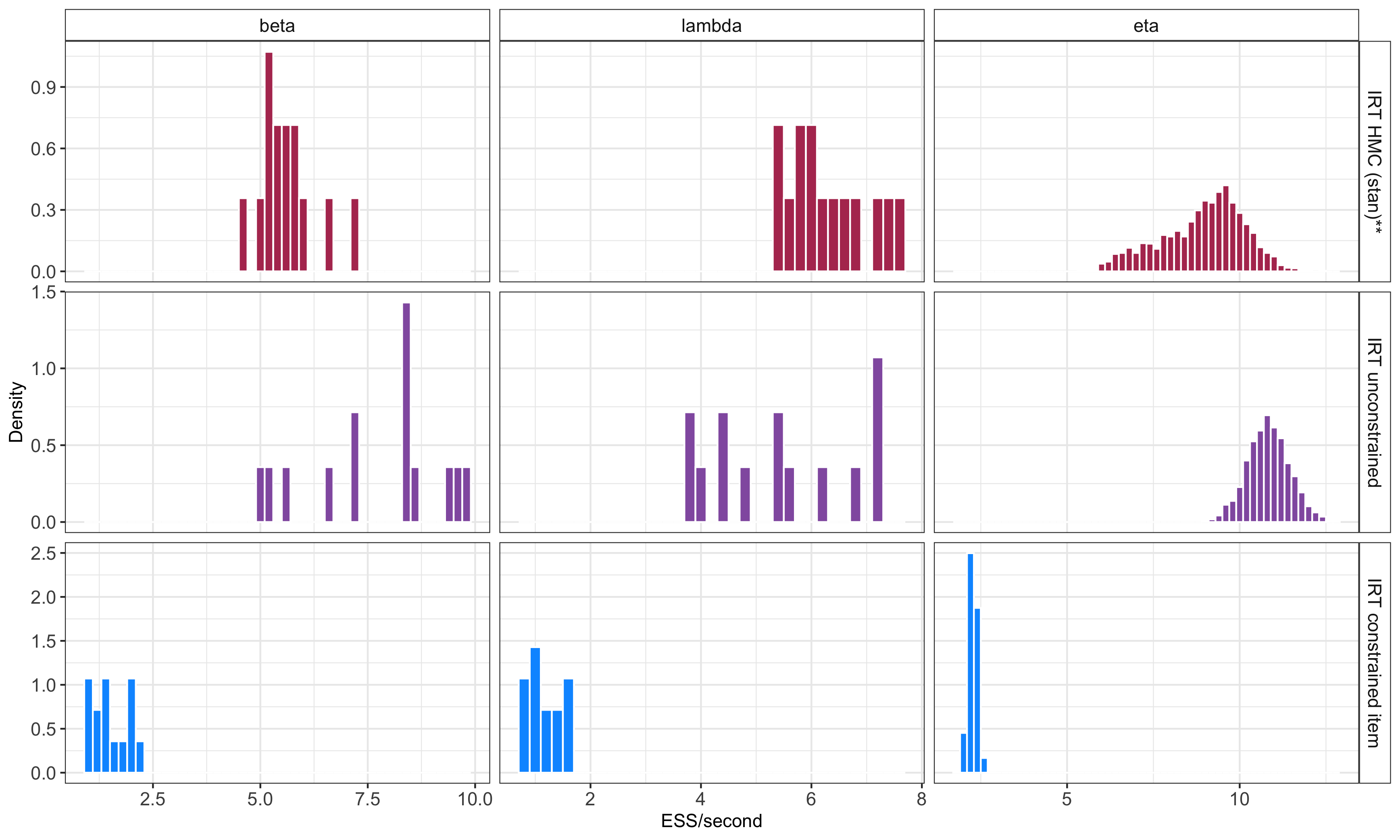}
        \begin{center}         
        \resizebox{0.8\textwidth}{!}{%
        \begin{tabular}{|c|c|c|c|c|}
        \hline
        Strategy & mESS & total time (second) & efficiency (mESS/second) \\
        \hline
        IRT HMC (Stan)**      &  30473 & 2191 & 14  \\ 
        IRT unconstrained     &  27167 &  937 & 29  \\
        IRT constrained item  &  27463 & 4508 & 6 \\
        \hline
      \end{tabular}
      }
    \end{center}

    \caption{Bimodal simulation ($N = 2000, I = 15)$. Distribution of univariate efficiencies (univariate ESS/seconds) for each group of parameters used to compute the multivariate efficiency (mESS/second) using the  total time. The symbol ** denotes median results across 11 runs.}\label{fig:ESSbimodal}
\end{figure}

\begin{figure}[H]
    \includegraphics[width=\textwidth]{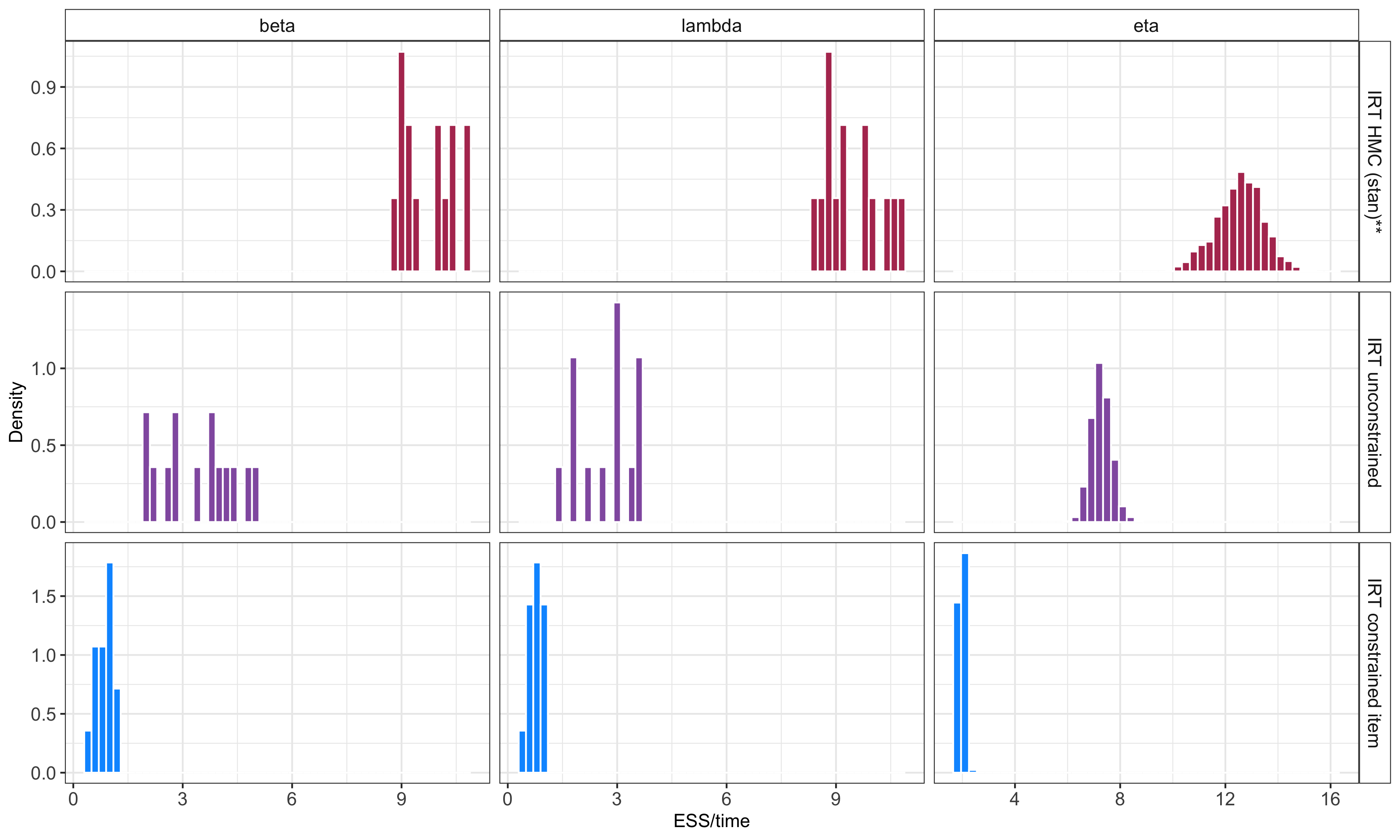}
     \begin{center}         
        \resizebox{0.8\textwidth}{!}{%
        \begin{tabular}{|c|c|c|c|c|}
        \hline
        Strategy & mESS & total time (second) & efficiency (mESS/second) \\
        \hline
        IRT HMC (Stan)**     & 22698 & 1176  & 19  \\
        IRT unconstrained    & 27442 & 1407  & 19  \\
        IRT constrained item & 27756 & 5248  &  5 \\
        \hline
        \end{tabular}
      }
    \end{center}
    \caption{Multimodal simulation ($N = 2000, I = 15)$. Distribution of univariate efficiencies (univariate ESS/seconds) for each group of parameters used to compute the multivariate efficiency (mESS/second)  using the  total time. The symbol ** denotes median results across 11 runs.
    }\label{fig:ESSmultimodal}
\end{figure}

In this section we compare univariate and multivariate efficiency metrics. In particular, we compare efficiency values based on the mESS (multivariate efficiency), with the distribution of efficiencies calculated for each parameter (univariate efficiency) using the total time. We report these metrics for each simulation scenario, selecting three representative strategies under the IRT parameterization for the parametric 2PL model. Figures~\ref{fig:ESSunimodal}-\ref{fig:ESSmultimodal} show the distribution of univariate efficiency for difficulty, discrimination and ability parameters, along with  a table reporting information for the multivariate efficiency. There are some differences between the multivariate and univariate efficiency results. This is expected because the mESS provides a single scalar measure of mixing performance that accounts for cross-correlation among the parameters and does not necessarily reflect the distribution of univariate ESSs. In fact, values of the mESS can be larger than all the univariate ESSs. For example, the IRT HMC strategy has larger univariate ESS values compared to other sampling strategies, but lower mESS.


\subsection*{Variability of mESS when using HMC from the Stan software}
\begin{figure}[H]
    \includegraphics[width=\textwidth]{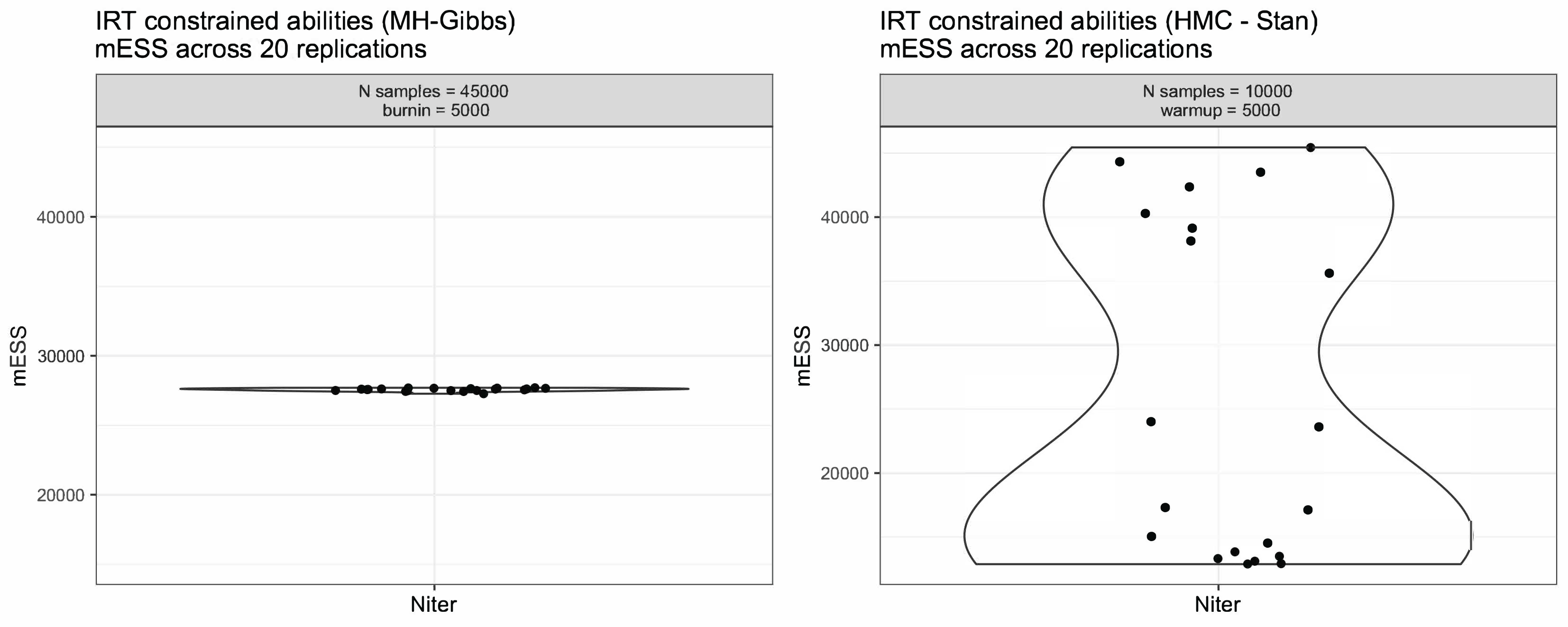}
    \caption{Comparison of mESS estimates across multiple runs for the IRT constrained abilities approach using the unimodal simulation scenario.}\label{fig:mESSvariability}
\end{figure}

When deciding on the number of posterior , burn-in and warm-up samples, we tried to obtain a reliable estimate of the multivariate ESS. To ensure the chains were long enough, we used multiple runs for some of the experiments. We found that mESS estimates based on the chosen settings of the MCMC algorithm (i.e., number of iterations,  number of burn-in or warm-up samples) have negligible variability across multiple runs for all strategies with the exception of using HMC as implemented in Stan. As an example, Figure \ref{fig:mESSvariability} shows the distribution of multivariate ESS for the IRT constrained abilities approach. 

\begin{figure}[H]
    \includegraphics[width=0.49\textwidth]{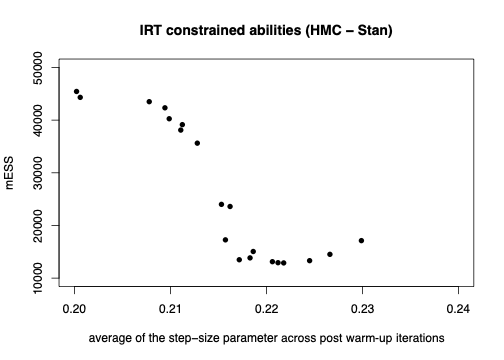}
    \includegraphics[width=0.49\textwidth]{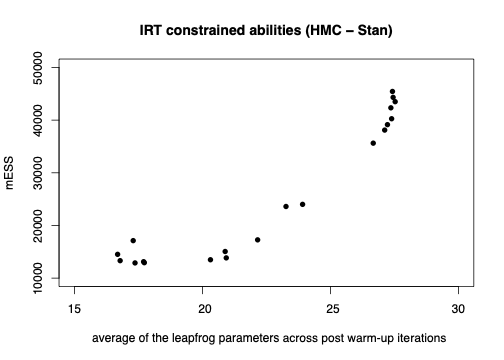}
    \caption{Estimates of the mESS versus average values of the HMC tuning parameters (step-size and leapfrog) across post warm-up iterations.}\label{fig:mESSandStanParams}
\end{figure}

We  also found that lower and higher values of ESS for the strategies using the HMC are highly correlated with values of the tuning parameters of the HMC (i.e., leapfrog and step-size parameters), which are typically estimated during the warm-up phase (see Figure~\ref{fig:mESSandStanParams}). 

\section*{E. Results from additional simulations}

We investigated how different combinations of numbers of items and individuals affect efficiency of the different sampling strategies. In particular, we simulated data under the three scenarios presented in Section~\ref{sec:simulted_data} following a factorial design with $I \in \{10, 30\}$ and individuals $N \in \{1,000, 5,000\}$. We omit results for the strategy using HMC due to the high variability in estimating the mESS.

\begin{figure}[H]
    \includegraphics[width=\textwidth]{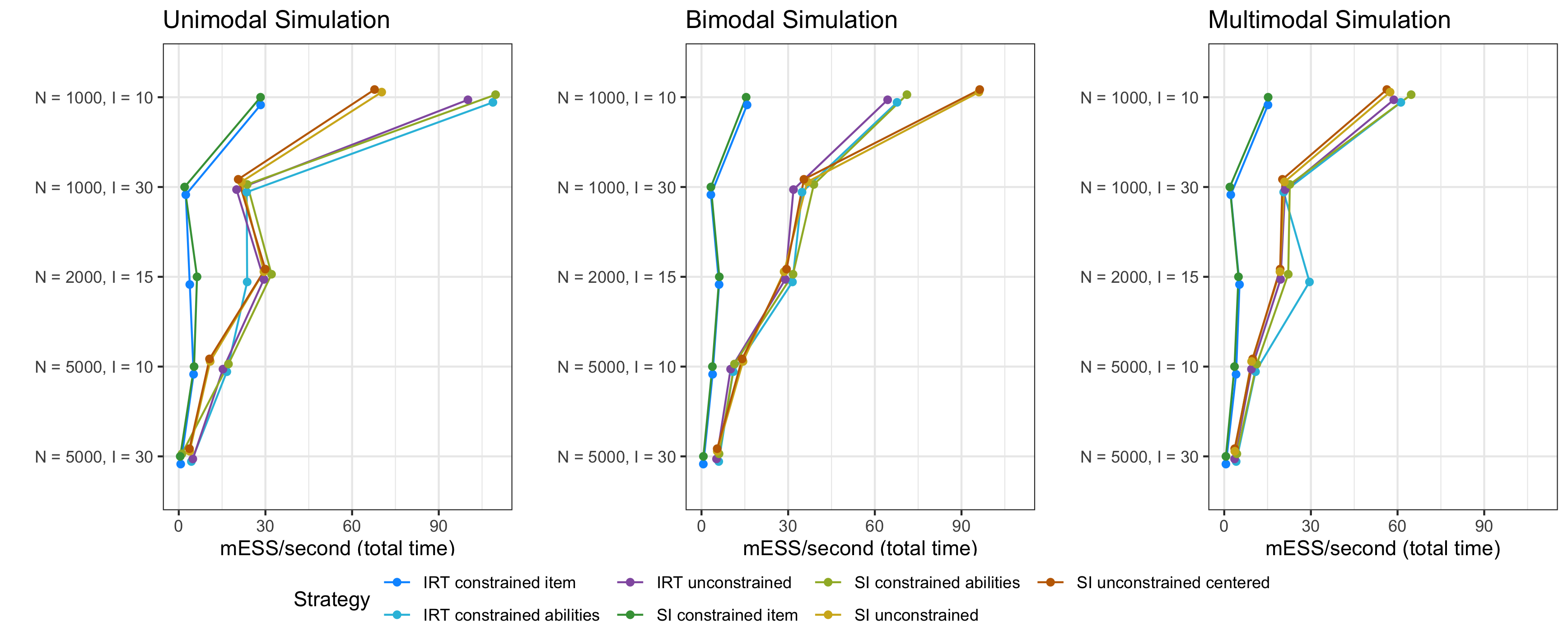}
  \caption{Multivariate ESS per second of the parametric sampling strategies (excluding HMC) across different combinations of numbers of items and individuals.}\label{figSM:parametric}
\end{figure}

\begin{figure}[H]
    \includegraphics[width=\textwidth]{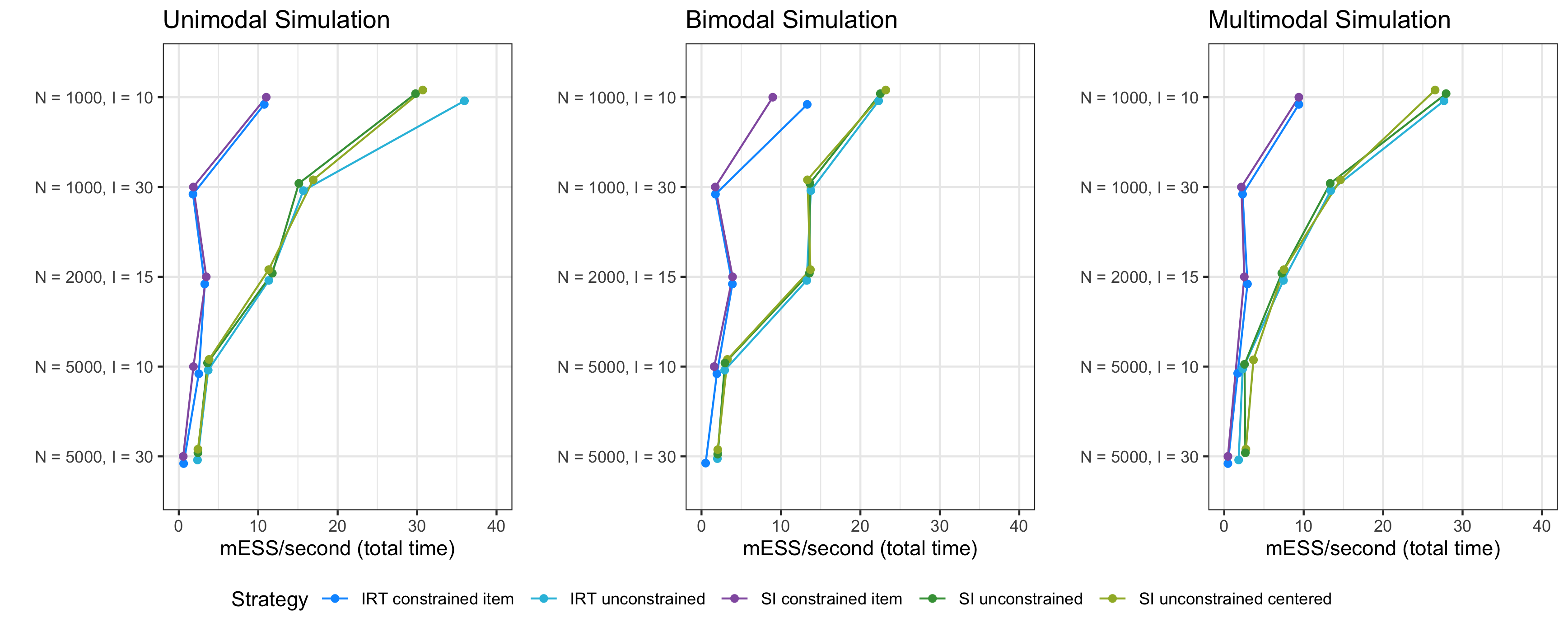}
  \caption{Multivariate ESS per second of the semiparametric sampling strategies (excluding HMC) across different combinations of numbers of items and individuals.}\label{figSM:semiparametric}
\end{figure}

\section*{F. Inferential results for the TIMSS data using the 2PL model}

%%%%%%%%%%%%%%%%%%%%%%%%%%%%%%%%%%%%%%%%%%%%%%%%%%%%%%%%%%%%%%%%%%%%
\begin{figure}[H]
    \centering
     \includegraphics[width=\textwidth]{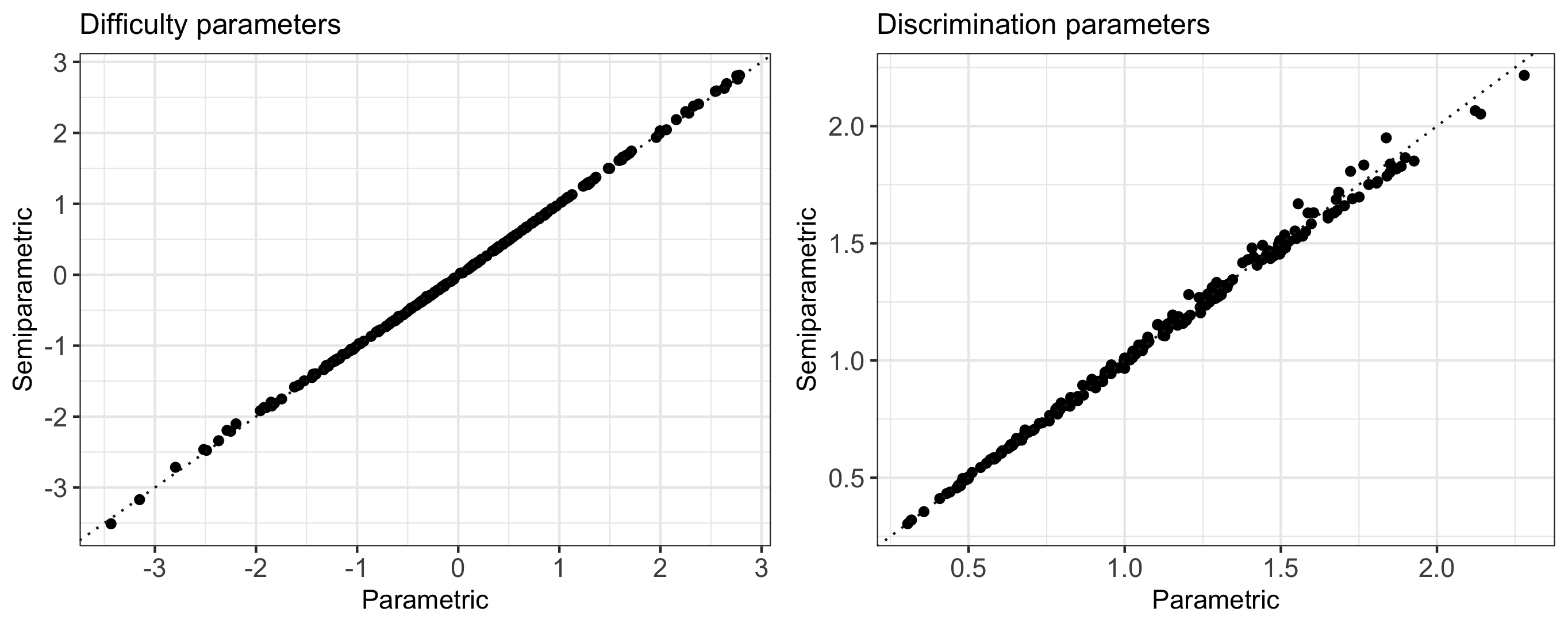}
      \caption{TIMSS data, 2PL model. Comparison of posterior estimates of the item parameters between the parametric and semiparametric model both using the SI unconstrained centered sampling strategy.}
     \label{fig:item_parameters_timss_2PL}
\end{figure}
%%%%%%%%%%%%%%%%%%%%%%%%%%%%%%%%%%%%%%%%%%%%%%%%%%%%%%%%%%%%%%%%%%%%
\begin{figure}[H]
    \centering
     \includegraphics[width=\textwidth]{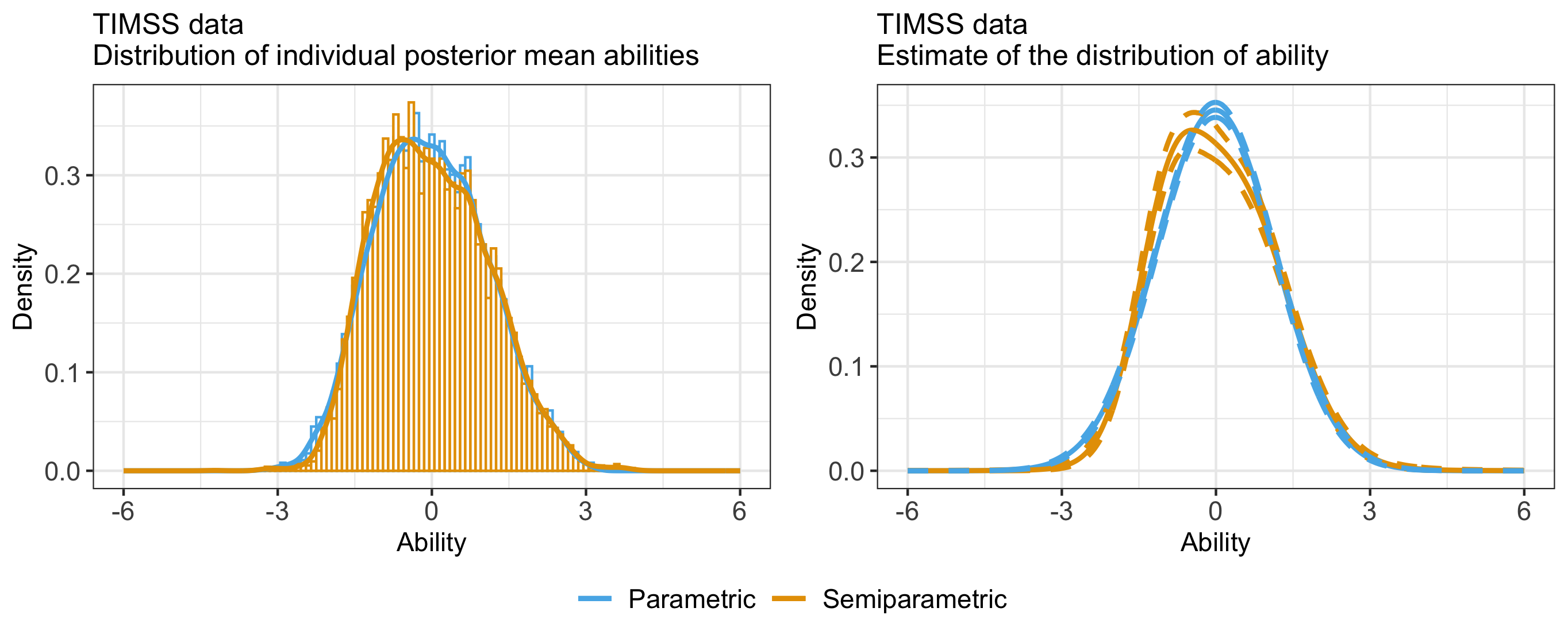}
    \caption{TIMSS Data, 2PL model. Histogram and density estimate of the posterior means of the latent abilities (left panel), and estimate of the posterior distribution for the latent abilities (right panel). Dashed lines indicate $95\%$ credible intervals for the estimated distributions.}
    \label{fig:abilities_timss_data_2PL}
\end{figure}
%%%%%%%%%%%%%%%%%%%%%%%%%%%%%%%%%%%%%%%%%%%%%%%%%%%%%%%%%%%%%%%%%%%%%%%%%%%%%

\begin{figure}[H]
    \centering
    \includegraphics[width=\textwidth]{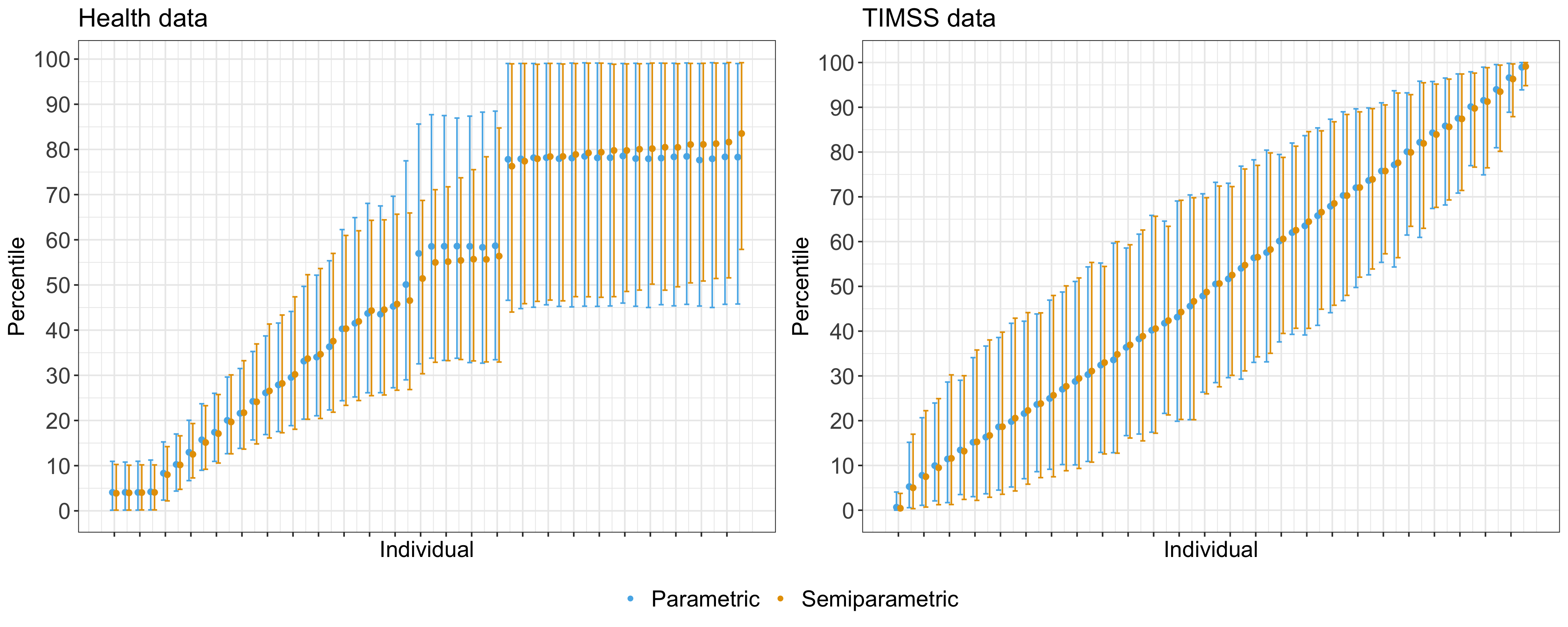}
    \caption{Estimates of individual percentiles (with $95\%$ credible interval) for a subset of $50$ individuals, for the health data (left panel) and TIMSS data (right panel) under the 2PL model.}
    \label{fig:data_percentiles_2PL}
\end{figure}
%\vspace{\fill}\pagebreak

% %% ITEM 10 [See the "howto.tex" file.]
\bibliography{bibliography}

\begin{thebibliography}{}

\bibitem [\protect \citeauthoryear {%
Aldous%
}{%
Aldous%
}{%
{\protect \APACyear {1985}}%
}]{%
aldous1985exchangeability}
\APACinsertmetastar {%
aldous1985exchangeability}%
\begin{APACrefauthors}%
Aldous, D\BPBI J.%
\end{APACrefauthors}%
\unskip\
\newblock
\APACrefYearMonthDay{1985}{}{}.
\newblock
{\BBOQ}\APACrefatitle {Exchangeability and related topics} {Exchangeability and
  related topics}.{\BBCQ}
\newblock
\BIn{} \APACrefbtitle {{\'E}cole d'{\'E}t{\'e} de Probabilit{\'e}s de
  Saint-Flour XIII—1983} {{\'E}cole d'{\'e}t{\'e} de probabilit{\'e}s de
  saint-flour xiii—1983}\ (\BPGS\ 1--198).
\newblock
\APACaddressPublisher{}{Springer}.
\PrintBackRefs{\CurrentBib}

\bibitem [\protect \citeauthoryear {%
Antoniak%
}{%
Antoniak%
}{%
{\protect \APACyear {1974}}%
}]{%
antoniak1974}
\APACinsertmetastar {%
antoniak1974}%
\begin{APACrefauthors}%
Antoniak, C\BPBI E.%
\end{APACrefauthors}%
\unskip\
\newblock
\APACrefYearMonthDay{1974}{11}{}.
\newblock
{\BBOQ}\APACrefatitle {Mixtures of {D}irichlet {P}rocesses with Applications to
  {B}ayesian Nonparametric Problems} {Mixtures of {D}irichlet {P}rocesses with
  applications to {B}ayesian nonparametric problems}.{\BBCQ}
\newblock
\APACjournalVolNumPages{Annals of Statistics}{2}{6}{1152--1174}.
\newblock
\begin{APACrefURL} \url{https://doi.org/10.1214/aos/1176342871}
  \end{APACrefURL}
\newblock
\begin{APACrefDOI} \doi{10.1214/aos/1176342871} \end{APACrefDOI}
\PrintBackRefs{\CurrentBib}

\bibitem [\protect \citeauthoryear {%
Azevedo%
, Bolfarine%
\BCBL {}\ \BBA {} Andrade%
}{%
Azevedo%
\ \protect \BOthers {.}}{%
{\protect \APACyear {2011}}%
}]{%
azevedo2011bayesian}
\APACinsertmetastar {%
azevedo2011bayesian}%
\begin{APACrefauthors}%
Azevedo, C\BPBI L.%
, Bolfarine, H.%
\BCBL {}\ \BBA {} Andrade, D\BPBI F.%
\end{APACrefauthors}%
\unskip\
\newblock
\APACrefYearMonthDay{2011}{}{}.
\newblock
{\BBOQ}\APACrefatitle {{B}ayesian inference for a skew-normal {IRT} model under
  the centred parameterization} {{B}ayesian inference for a skew-normal {IRT}
  model under the centred parameterization}.{\BBCQ}
\newblock
\APACjournalVolNumPages{Computational Statistics \& Data
  Analysis}{55}{1}{353--365}.
\PrintBackRefs{\CurrentBib}

\bibitem [\protect \citeauthoryear {%
Azzalini%
}{%
Azzalini%
}{%
{\protect \APACyear {1985}}%
}]{%
azzalini1985class}
\APACinsertmetastar {%
azzalini1985class}%
\begin{APACrefauthors}%
Azzalini, A.%
\end{APACrefauthors}%
\unskip\
\newblock
\APACrefYearMonthDay{1985}{}{}.
\newblock
{\BBOQ}\APACrefatitle {A class of distributions which includes the normal ones}
  {A class of distributions which includes the normal ones}.{\BBCQ}
\newblock
\APACjournalVolNumPages{Scandinavian Journal of Statistics}{}{}{171--178}.
\PrintBackRefs{\CurrentBib}

\bibitem [\protect \citeauthoryear {%
Bafumi%
, Gelman%
, Park%
\BCBL {}\ \BBA {} Kaplan%
}{%
Bafumi%
\ \protect \BOthers {.}}{%
{\protect \APACyear {2005}}%
}]{%
bafumi2005practical}
\APACinsertmetastar {%
bafumi2005practical}%
\begin{APACrefauthors}%
Bafumi, J.%
, Gelman, A.%
, Park, D\BPBI K.%
\BCBL {}\ \BBA {} Kaplan, N.%
\end{APACrefauthors}%
\unskip\
\newblock
\APACrefYearMonthDay{2005}{}{}.
\newblock
{\BBOQ}\APACrefatitle {Practical issues in implementing and understanding
  {B}ayesian ideal point estimation} {Practical issues in implementing and
  understanding {B}ayesian ideal point estimation}.{\BBCQ}
\newblock
\APACjournalVolNumPages{Political Analysis}{13}{2}{171--187}.
\PrintBackRefs{\CurrentBib}

\bibitem [\protect \citeauthoryear {%
Bambirra~Gon{\c{c}}alves%
, da Costa Campos~Dias%
\BCBL {}\ \BBA {} Machado~Soares%
}{%
Bambirra~Gon{\c{c}}alves%
\ \protect \BOthers {.}}{%
{\protect \APACyear {2018}}%
}]{%
bambirra2018bayesian}
\APACinsertmetastar {%
bambirra2018bayesian}%
\begin{APACrefauthors}%
Bambirra~Gon{\c{c}}alves, F.%
, da Costa Campos~Dias, B.%
\BCBL {}\ \BBA {} Machado~Soares, T.%
\end{APACrefauthors}%
\unskip\
\newblock
\APACrefYearMonthDay{2018}{}{}.
\newblock
{\BBOQ}\APACrefatitle {{B}ayesian item response model: a generalized approach
  for the abilities' distribution using mixtures} {{B}ayesian item response
  model: a generalized approach for the abilities' distribution using
  mixtures}.{\BBCQ}
\newblock
\APACjournalVolNumPages{Journal of Statistical Computation and
  Simulation}{88}{5}{967--981}.
\PrintBackRefs{\CurrentBib}

\bibitem [\protect \citeauthoryear {%
Bedrick%
, Christensen%
\BCBL {}\ \BBA {} Johnson%
}{%
Bedrick%
\ \protect \BOthers {.}}{%
{\protect \APACyear {1996}}%
}]{%
bedrick1996new}
\APACinsertmetastar {%
bedrick1996new}%
\begin{APACrefauthors}%
Bedrick, E\BPBI J.%
, Christensen, R.%
\BCBL {}\ \BBA {} Johnson, W.%
\end{APACrefauthors}%
\unskip\
\newblock
\APACrefYearMonthDay{1996}{}{}.
\newblock
{\BBOQ}\APACrefatitle {A new perspective on priors for generalized linear
  models} {A new perspective on priors for generalized linear models}.{\BBCQ}
\newblock
\APACjournalVolNumPages{Journal of the American Statistical
  Association}{91}{436}{1450--1460}.
\PrintBackRefs{\CurrentBib}

\bibitem [\protect \citeauthoryear {%
Berger%
, Bernardo%
\BCBL {}\ \BBA {} Sun%
}{%
Berger%
\ \protect \BOthers {.}}{%
{\protect \APACyear {2009}}%
}]{%
berger2009formal}
\APACinsertmetastar {%
berger2009formal}%
\begin{APACrefauthors}%
Berger, J\BPBI O.%
, Bernardo, J\BPBI M.%
\BCBL {}\ \BBA {} Sun, D.%
\end{APACrefauthors}%
\unskip\
\newblock
\APACrefYearMonthDay{2009}{}{}.
\newblock
{\BBOQ}\APACrefatitle {The Formal Definition of Reference Priors} {The formal
  definition of reference priors}.{\BBCQ}
\newblock
\APACjournalVolNumPages{The Annals of Statistics}{37}{2}{905--938}.
\newblock
\begin{APACrefURL} \url{http://www.jstor.org/stable/30243652} \end{APACrefURL}
\PrintBackRefs{\CurrentBib}

\bibitem [\protect \citeauthoryear {%
Berger%
\ \BBA {} Pericchi%
}{%
Berger%
\ \BBA {} Pericchi%
}{%
{\protect \APACyear {1996}}%
}]{%
berger1996intrinsic}
\APACinsertmetastar {%
berger1996intrinsic}%
\begin{APACrefauthors}%
Berger, J\BPBI O.%
\BCBT {}\ \BBA {} Pericchi, L\BPBI R.%
\end{APACrefauthors}%
\unskip\
\newblock
\APACrefYearMonthDay{1996}{}{}.
\newblock
{\BBOQ}\APACrefatitle {The intrinsic {B}ayes factor for model selection and
  prediction} {The intrinsic {B}ayes factor for model selection and
  prediction}.{\BBCQ}
\newblock
\APACjournalVolNumPages{Journal of the American Statistical
  Association}{91}{433}{109--122}.
\PrintBackRefs{\CurrentBib}

\bibitem [\protect \citeauthoryear {%
Bernardo%
}{%
Bernardo%
}{%
{\protect \APACyear {1979}}%
}]{%
bernardo1979reference}
\APACinsertmetastar {%
bernardo1979reference}%
\begin{APACrefauthors}%
Bernardo, J\BPBI M.%
\end{APACrefauthors}%
\unskip\
\newblock
\APACrefYearMonthDay{1979}{}{}.
\newblock
{\BBOQ}\APACrefatitle {Reference posterior distributions for {B}ayesian
  inference} {Reference posterior distributions for {B}ayesian
  inference}.{\BBCQ}
\newblock
\APACjournalVolNumPages{Journal of the Royal Statistical Society: Series B
  (Methodological)}{41}{2}{113--128}.
\PrintBackRefs{\CurrentBib}

\bibitem [\protect \citeauthoryear {%
Betancourt%
, Byrne%
, Livingstone%
\BCBL {}\ \BBA {} Girolami%
}{%
Betancourt%
\ \protect \BOthers {.}}{%
{\protect \APACyear {2017}}%
}]{%
betancourt2017geometric}
\APACinsertmetastar {%
betancourt2017geometric}%
\begin{APACrefauthors}%
Betancourt, M.%
, Byrne, S.%
, Livingstone, S.%
\BCBL {}\ \BBA {} Girolami, M.%
\end{APACrefauthors}%
\unskip\
\newblock
\APACrefYearMonthDay{2017}{}{}.
\newblock
{\BBOQ}\APACrefatitle {The geometric foundations of {H}amiltonian {M}onte
  {C}arlo} {The geometric foundations of {H}amiltonian {M}onte {C}arlo}.{\BBCQ}
\newblock
\APACjournalVolNumPages{Bernoulli}{23}{4A}{2257--2298}.
\PrintBackRefs{\CurrentBib}

\bibitem [\protect \citeauthoryear {%
Blackwell%
\ \BBA {} MacQueen%
}{%
Blackwell%
\ \BBA {} MacQueen%
}{%
{\protect \APACyear {1973}}%
}]{%
blackwell1973ferguson}
\APACinsertmetastar {%
blackwell1973ferguson}%
\begin{APACrefauthors}%
Blackwell, D.%
\BCBT {}\ \BBA {} MacQueen, J\BPBI B.%
\end{APACrefauthors}%
\unskip\
\newblock
\APACrefYearMonthDay{1973}{}{}.
\newblock
{\BBOQ}\APACrefatitle {Ferguson distributions via {P\'o}lya urn schemes}
  {Ferguson distributions via {P\'o}lya urn schemes}.{\BBCQ}
\newblock
\APACjournalVolNumPages{The Annals of Statistics}{1}{2}{353--355}.
\PrintBackRefs{\CurrentBib}

\bibitem [\protect \citeauthoryear {%
Bolt%
, Cohen%
\BCBL {}\ \BBA {} Wollack%
}{%
Bolt%
\ \protect \BOthers {.}}{%
{\protect \APACyear {2001}}%
}]{%
bolt2001mixture}
\APACinsertmetastar {%
bolt2001mixture}%
\begin{APACrefauthors}%
Bolt, D\BPBI M.%
, Cohen, A\BPBI S.%
\BCBL {}\ \BBA {} Wollack, J\BPBI A.%
\end{APACrefauthors}%
\unskip\
\newblock
\APACrefYearMonthDay{2001}{}{}.
\newblock
{\BBOQ}\APACrefatitle {A Mixture Item Response Model for Multiple-Choice Data}
  {A mixture item response model for multiple-choice data}.{\BBCQ}
\newblock
\APACjournalVolNumPages{Journal of Educational and Behavioral
  Statistics}{26}{4}{381--409}.
\newblock
\begin{APACrefURL} \url{http://www.jstor.org/stable/3648167} \end{APACrefURL}
\PrintBackRefs{\CurrentBib}

\bibitem [\protect \citeauthoryear {%
Bürkner%
}{%
Bürkner%
}{%
{\protect \APACyear {2021}}%
}]{%
burker2021bayesian}
\APACinsertmetastar {%
burker2021bayesian}%
\begin{APACrefauthors}%
Bürkner, P\BHBI C.%
\end{APACrefauthors}%
\unskip\
\newblock
\APACrefYearMonthDay{2021}{}{}.
\newblock
{\BBOQ}\APACrefatitle {Bayesian Item Response Modeling in R with brms and Stan}
  {Bayesian item response modeling in r with brms and stan}.{\BBCQ}
\newblock
\APACjournalVolNumPages{Journal of Statistical Software}{100}{5}{1–54}.
\newblock
\begin{APACrefURL}
  \url{https://www.jstatsoft.org/index.php/jss/article/view/v100i05}
  \end{APACrefURL}
\newblock
\begin{APACrefDOI} \doi{10.18637/jss.v100.i05} \end{APACrefDOI}
\PrintBackRefs{\CurrentBib}

\bibitem [\protect \citeauthoryear {%
Carpenter%
\ \protect \BOthers {.}}{%
Carpenter%
\ \protect \BOthers {.}}{%
{\protect \APACyear {2017}}%
}]{%
stan2017jss}
\APACinsertmetastar {%
stan2017jss}%
\begin{APACrefauthors}%
Carpenter, B.%
, Gelman, A.%
, Hoffman, M.%
, Lee, D.%
, Goodrich, B.%
, Betancourt, M.%
\BDBL {}Riddell, A.%
\end{APACrefauthors}%
\unskip\
\newblock
\APACrefYearMonthDay{2017}{}{}.
\newblock
{\BBOQ}\APACrefatitle {Stan: A Probabilistic Programming Language} {Stan: A
  probabilistic programming language}.{\BBCQ}
\newblock
\APACjournalVolNumPages{Journal of Statistical Software,
  Articles}{76}{1}{1--32}.
\newblock
\begin{APACrefURL} \url{https://www.jstatsoft.org/v076/i01} \end{APACrefURL}
\newblock
\begin{APACrefDOI} \doi{10.18637/jss.v076.i01} \end{APACrefDOI}
\PrintBackRefs{\CurrentBib}

\bibitem [\protect \citeauthoryear {%
Clinton%
, Jackman%
\BCBL {}\ \BBA {} Rivers%
}{%
Clinton%
\ \protect \BOthers {.}}{%
{\protect \APACyear {2004}}%
}]{%
clinton2004statistical}
\APACinsertmetastar {%
clinton2004statistical}%
\begin{APACrefauthors}%
Clinton, J.%
, Jackman, S.%
\BCBL {}\ \BBA {} Rivers, D.%
\end{APACrefauthors}%
\unskip\
\newblock
\APACrefYearMonthDay{2004}{}{}.
\newblock
{\BBOQ}\APACrefatitle {The statistical analysis of roll call data} {The
  statistical analysis of roll call data}.{\BBCQ}
\newblock
\APACjournalVolNumPages{American Political Science Review}{}{}{355--370}.
\PrintBackRefs{\CurrentBib}

\bibitem [\protect \citeauthoryear {%
{de Valpine}%
\ \protect \BOthers {.}}{%
{de Valpine}%
\ \protect \BOthers {.}}{%
{\protect \APACyear {2022}}%
}]{%
nimble-manual:2021}
\APACinsertmetastar {%
nimble-manual:2021}%
\begin{APACrefauthors}%
{de Valpine}, P.%
, Paciorek, C.%
, Turek, D.%
, Michaud, N.%
, Anderson-Bergman, C.%
, Obermeyer, F.%
\BDBL {}Hug, J.%
\end{APACrefauthors}%
\unskip\
\newblock
\APACrefYearMonthDay{2022}{}{}.
\newblock
{\BBOQ}\APACrefatitle {NIMBLE User Manual} {Nimble user manual}{\BBCQ}\
  [\bibcomputersoftwaremanual].
\newblock
\begin{APACrefURL} \url{https://r-nimble.org} \end{APACrefURL}
\newblock
\begin{APACrefDOI} \doi{10.5281/zenodo.1211190} \end{APACrefDOI}
\PrintBackRefs{\CurrentBib}

\bibitem [\protect \citeauthoryear {%
{de Valpine}%
\ \protect \BOthers {.}}{%
{de Valpine}%
\ \protect \BOthers {.}}{%
{\protect \APACyear {2020}}%
}]{%
nimble-software2020}
\APACinsertmetastar {%
nimble-software2020}%
\begin{APACrefauthors}%
{de Valpine}, P.%
, Paciorek, C\BPBI J.%
, Turek, D.%
, Michaud, N.%
, Anderson-Bergman, C.%
, Obermeyer, F.%
\BDBL {}Paganin, S.%
\end{APACrefauthors}%
\unskip\
\newblock
\APACrefYearMonthDay{2020}{}{}.
\newblock
\APACrefbtitle {{NIMBLE}: {MCMC}, Particle Filtering, and Programmable
  Hierarchical Modeling.} {{NIMBLE}: {MCMC}, particle filtering, and
  programmable hierarchical modeling.}
\newblock
\begin{APACrefURL} \url{https://cran.r-project.org/package=nimble}
  \end{APACrefURL}
\newblock
\APACrefnote{{R} package version 0.9.1}
\newblock
\begin{APACrefDOI} \doi{10.5281/zenodo.1211190} \end{APACrefDOI}
\PrintBackRefs{\CurrentBib}

\bibitem [\protect \citeauthoryear {%
de Valpine%
\ \protect \BOthers {.}}{%
de Valpine%
\ \protect \BOthers {.}}{%
{\protect \APACyear {2017}}%
}]{%
devalpine2017nimble}
\APACinsertmetastar {%
devalpine2017nimble}%
\begin{APACrefauthors}%
de Valpine, P.%
, Turek, D.%
, Paciorek, C\BPBI J.%
, Anderson-Bergman, C.%
, Lang, D\BPBI T.%
\BCBL {}\ \BBA {} Bodik, R.%
\end{APACrefauthors}%
\unskip\
\newblock
\APACrefYearMonthDay{2017}{}{}.
\newblock
{\BBOQ}\APACrefatitle {Programming With Models: Writing Statistical Algorithms
  for General Model Structures With NIMBLE} {Programming with models: Writing
  statistical algorithms for general model structures with nimble}.{\BBCQ}
\newblock
\APACjournalVolNumPages{Journal of Computational and Graphical
  Statistics}{26}{2}{403-413}.
\newblock
\begin{APACrefURL} \url{https://doi.org/10.1080/10618600.2016.1172487}
  \end{APACrefURL}
\newblock
\begin{APACrefDOI} \doi{10.1080/10618600.2016.1172487} \end{APACrefDOI}
\PrintBackRefs{\CurrentBib}

\bibitem [\protect \citeauthoryear {%
Duncan%
\ \BBA {} MacEachern%
}{%
Duncan%
\ \BBA {} MacEachern%
}{%
{\protect \APACyear {2008}}%
}]{%
duncan2008nonparametric}
\APACinsertmetastar {%
duncan2008nonparametric}%
\begin{APACrefauthors}%
Duncan, K\BPBI A.%
\BCBT {}\ \BBA {} MacEachern, S\BPBI N.%
\end{APACrefauthors}%
\unskip\
\newblock
\APACrefYearMonthDay{2008}{}{}.
\newblock
{\BBOQ}\APACrefatitle {Nonparametric {B}ayesian modelling for item response}
  {Nonparametric {B}ayesian modelling for item response}.{\BBCQ}
\newblock
\APACjournalVolNumPages{Statistical Modelling}{8}{1}{41--66}.
\PrintBackRefs{\CurrentBib}

\bibitem [\protect \citeauthoryear {%
Escobar%
\ \BBA {} West%
}{%
Escobar%
\ \BBA {} West%
}{%
{\protect \APACyear {1995}}%
}]{%
escobar1995bayesian}
\APACinsertmetastar {%
escobar1995bayesian}%
\begin{APACrefauthors}%
Escobar, M\BPBI D.%
\BCBT {}\ \BBA {} West, M.%
\end{APACrefauthors}%
\unskip\
\newblock
\APACrefYearMonthDay{1995}{}{}.
\newblock
{\BBOQ}\APACrefatitle {{B}ayesian density estimation and inference using
  mixtures} {{B}ayesian density estimation and inference using
  mixtures}.{\BBCQ}
\newblock
\APACjournalVolNumPages{Journal of the American Statistical
  Association}{90}{430}{577--588}.
\PrintBackRefs{\CurrentBib}

\bibitem [\protect \citeauthoryear {%
Ferguson%
}{%
Ferguson%
}{%
{\protect \APACyear {1973}}%
}]{%
ferguson1973}
\APACinsertmetastar {%
ferguson1973}%
\begin{APACrefauthors}%
Ferguson, T\BPBI S.%
\end{APACrefauthors}%
\unskip\
\newblock
\APACrefYearMonthDay{1973}{03}{}.
\newblock
{\BBOQ}\APACrefatitle {A {B}ayesian Analysis of Some Nonparametric Problems} {A
  {B}ayesian analysis of some nonparametric problems}.{\BBCQ}
\newblock
\APACjournalVolNumPages{The Annals of Statistics}{1}{2}{209--230}.
\newblock
\begin{APACrefDOI} \doi{10.1214/aos/1176342360} \end{APACrefDOI}
\PrintBackRefs{\CurrentBib}

\bibitem [\protect \citeauthoryear {%
Finch%
\ \BBA {} Edwards%
}{%
Finch%
\ \BBA {} Edwards%
}{%
{\protect \APACyear {2016}}%
}]{%
finch2016rasch}
\APACinsertmetastar {%
finch2016rasch}%
\begin{APACrefauthors}%
Finch, H.%
\BCBT {}\ \BBA {} Edwards, J\BPBI M.%
\end{APACrefauthors}%
\unskip\
\newblock
\APACrefYearMonthDay{2016}{}{}.
\newblock
{\BBOQ}\APACrefatitle {Rasch Model Parameter Estimation in the Presence of a
  Nonnormal Latent Trait Using a Nonparametric {B}ayesian Approach} {Rasch
  model parameter estimation in the presence of a nonnormal latent trait using
  a nonparametric {B}ayesian approach}.{\BBCQ}
\newblock
\APACjournalVolNumPages{Educational and Psychological
  Measurement}{76}{4}{662-684}.
\newblock
\begin{APACrefURL} \url{https://doi.org/10.1177/0013164415608418}
  \end{APACrefURL}
\newblock
\begin{APACrefDOI} \doi{10.1177/0013164415608418} \end{APACrefDOI}
\PrintBackRefs{\CurrentBib}

\bibitem [\protect \citeauthoryear {%
Flegal%
\ \protect \BOthers {.}}{%
Flegal%
\ \protect \BOthers {.}}{%
{\protect \APACyear {2021}}%
}]{%
mcmcse}
\APACinsertmetastar {%
mcmcse}%
\begin{APACrefauthors}%
Flegal, J\BPBI M.%
, Hughes, J.%
, Vats, D.%
, Dai, N.%
, Gupta, K.%
\BCBL {}\ \BBA {} Maji, U.%
\end{APACrefauthors}%
\unskip\
\newblock
\APACrefYearMonthDay{2021}{}{}.
\newblock
{\BBOQ}\APACrefatitle {mcmcse: {M}onte {C}arlo Standard Errors for {MCMC}}
  {mcmcse: {M}onte {C}arlo standard errors for {MCMC}}{\BBCQ}\
  [\bibcomputersoftwaremanual].
\newblock
\APACaddressPublisher{Riverside, CA, and Kanpur, India}{}.
\newblock
\APACrefnote{R package version 1.5-0}
\PrintBackRefs{\CurrentBib}

\bibitem [\protect \citeauthoryear {%
Fox%
}{%
Fox%
}{%
{\protect \APACyear {2010}}%
}]{%
fox2010bayesian}
\APACinsertmetastar {%
fox2010bayesian}%
\begin{APACrefauthors}%
Fox, J\BHBI P.%
\end{APACrefauthors}%
\unskip\
\newblock
\APACrefYear{2010}.
\newblock
\APACrefbtitle {{B}ayesian Item Response modeling: Theory and Applications}
  {{B}ayesian item response modeling: Theory and applications}.
\newblock
\APACaddressPublisher{}{Springer Science \& Business Media}.
\PrintBackRefs{\CurrentBib}

\bibitem [\protect \citeauthoryear {%
Furr%
}{%
Furr%
}{%
{\protect \APACyear {2017}}%
}]{%
edstan2017}
\APACinsertmetastar {%
edstan2017}%
\begin{APACrefauthors}%
Furr, D\BPBI C.%
\end{APACrefauthors}%
\unskip\
\newblock
\APACrefYearMonthDay{2017}{}{}.
\newblock
{\BBOQ}\APACrefatitle {edstan: Stan models for {I}tem {R}esponse {T}heory}
  {edstan: Stan models for {I}tem {R}esponse {T}heory}{\BBCQ}\
  [\bibcomputersoftwaremanual].
\newblock
\begin{APACrefURL} \url{https://CRAN.R-project.org/package=edstan}
  \end{APACrefURL}
\newblock
\APACrefnote{R package version 1.0.6}
\PrintBackRefs{\CurrentBib}

\bibitem [\protect \citeauthoryear {%
Gelfand%
\ \BBA {} Kottas%
}{%
Gelfand%
\ \BBA {} Kottas%
}{%
{\protect \APACyear {2002}}%
}]{%
gelfand2002}
\APACinsertmetastar {%
gelfand2002}%
\begin{APACrefauthors}%
Gelfand, A\BPBI E.%
\BCBT {}\ \BBA {} Kottas, A.%
\end{APACrefauthors}%
\unskip\
\newblock
\APACrefYearMonthDay{2002}{}{}.
\newblock
{\BBOQ}\APACrefatitle {A Computational Approach for Full Nonparametric
  {B}ayesian Inference Under {D}irichlet {P}rocess Mixture Models} {A
  computational approach for full nonparametric {B}ayesian inference under
  {D}irichlet {P}rocess mixture models}.{\BBCQ}
\newblock
\APACjournalVolNumPages{Journal of Computational and Graphical
  Statistics}{11}{2}{289-305}.
\newblock
\begin{APACrefURL} \url{https://doi.org/10.1198/106186002760180518}
  \end{APACrefURL}
\newblock
\begin{APACrefDOI} \doi{10.1198/106186002760180518} \end{APACrefDOI}
\PrintBackRefs{\CurrentBib}

\bibitem [\protect \citeauthoryear {%
Gelman%
}{%
Gelman%
}{%
{\protect \APACyear {2004}}%
}]{%
gelman2004parameterization}
\APACinsertmetastar {%
gelman2004parameterization}%
\begin{APACrefauthors}%
Gelman, A.%
\end{APACrefauthors}%
\unskip\
\newblock
\APACrefYearMonthDay{2004}{}{}.
\newblock
{\BBOQ}\APACrefatitle {Parameterization and {B}ayesian modeling}
  {Parameterization and {B}ayesian modeling}.{\BBCQ}
\newblock
\APACjournalVolNumPages{Journal of the American Statistical
  Association}{99}{466}{537--545}.
\PrintBackRefs{\CurrentBib}

\bibitem [\protect \citeauthoryear {%
Geweke%
\ \BBA {} Singleton%
}{%
Geweke%
\ \BBA {} Singleton%
}{%
{\protect \APACyear {1981}}%
}]{%
geweke1981maximum}
\APACinsertmetastar {%
geweke1981maximum}%
\begin{APACrefauthors}%
Geweke, J\BPBI F.%
\BCBT {}\ \BBA {} Singleton, K\BPBI J.%
\end{APACrefauthors}%
\unskip\
\newblock
\APACrefYearMonthDay{1981}{}{}.
\newblock
{\BBOQ}\APACrefatitle {Maximum likelihood ``confirmatory'' factor analysis of
  economic time series} {Maximum likelihood ``confirmatory'' factor analysis of
  economic time series}.{\BBCQ}
\newblock
\APACjournalVolNumPages{International Economic Review}{}{}{37--54}.
\PrintBackRefs{\CurrentBib}

\bibitem [\protect \citeauthoryear {%
Guhaniyogi%
\ \BBA {} Rodriguez%
}{%
Guhaniyogi%
\ \BBA {} Rodriguez%
}{%
{\protect \APACyear {2020}}%
}]{%
guhaniyogi2020joint}
\APACinsertmetastar {%
guhaniyogi2020joint}%
\begin{APACrefauthors}%
Guhaniyogi, R.%
\BCBT {}\ \BBA {} Rodriguez, A.%
\end{APACrefauthors}%
\unskip\
\newblock
\APACrefYearMonthDay{2020}{}{}.
\newblock
{\BBOQ}\APACrefatitle {Joint modeling of longitudinal relational data and
  exogenous variables} {Joint modeling of longitudinal relational data and
  exogenous variables}.{\BBCQ}
\newblock
\APACjournalVolNumPages{Bayesian Analysis}{15}{2}{477--503}.
\PrintBackRefs{\CurrentBib}

\bibitem [\protect \citeauthoryear {%
Hays%
, Morales%
\BCBL {}\ \BBA {} Reise%
}{%
Hays%
\ \protect \BOthers {.}}{%
{\protect \APACyear {2000}}%
}]{%
hays2000item}
\APACinsertmetastar {%
hays2000item}%
\begin{APACrefauthors}%
Hays, R\BPBI D.%
, Morales, L\BPBI S.%
\BCBL {}\ \BBA {} Reise, S\BPBI P.%
\end{APACrefauthors}%
\unskip\
\newblock
\APACrefYearMonthDay{2000}{}{}.
\newblock
{\BBOQ}\APACrefatitle {Item response theory and health outcomes measurement in
  the 21st century} {Item response theory and health outcomes measurement in
  the 21st century}.{\BBCQ}
\newblock
\APACjournalVolNumPages{Medical Care}{38}{9 Suppl}{II28}.
\PrintBackRefs{\CurrentBib}

\bibitem [\protect \citeauthoryear {%
Hoffman%
\ \BBA {} Gelman%
}{%
Hoffman%
\ \BBA {} Gelman%
}{%
{\protect \APACyear {2014}}%
}]{%
hoffman2014no}
\APACinsertmetastar {%
hoffman2014no}%
\begin{APACrefauthors}%
Hoffman, M\BPBI D.%
\BCBT {}\ \BBA {} Gelman, A.%
\end{APACrefauthors}%
\unskip\
\newblock
\APACrefYearMonthDay{2014}{}{}.
\newblock
{\BBOQ}\APACrefatitle {The {No-U-Turn} sampler: adaptively setting path lengths
  in {H}amiltonian {M}onte {C}arlo.} {The {No-U-Turn} sampler: adaptively
  setting path lengths in {H}amiltonian {M}onte {C}arlo.}{\BBCQ}
\newblock
\APACjournalVolNumPages{Journal of Machine Learning
  Research}{15}{1}{1593--1623}.
\PrintBackRefs{\CurrentBib}

\bibitem [\protect \citeauthoryear {%
Ibrahim%
}{%
Ibrahim%
}{%
{\protect \APACyear {1997}}%
}]{%
ibrahim1997properties}
\APACinsertmetastar {%
ibrahim1997properties}%
\begin{APACrefauthors}%
Ibrahim, J\BPBI G.%
\end{APACrefauthors}%
\unskip\
\newblock
\APACrefYearMonthDay{1997}{}{}.
\newblock
{\BBOQ}\APACrefatitle {On Properties of Predictive Priors in Linear Models} {On
  properties of predictive priors in linear models}.{\BBCQ}
\newblock
\APACjournalVolNumPages{The American Statistician}{51}{4}{333-337}.
\newblock
\begin{APACrefURL}
  \url{https://www.tandfonline.com/doi/abs/10.1080/00031305.1997.10474408}
  \end{APACrefURL}
\newblock
\begin{APACrefDOI} \doi{10.1080/00031305.1997.10474408} \end{APACrefDOI}
\PrintBackRefs{\CurrentBib}

\bibitem [\protect \citeauthoryear {%
Jara%
, Hanson%
, Quintana%
, M{\"u}ller%
\BCBL {}\ \BBA {} Rosner%
}{%
Jara%
\ \protect \BOthers {.}}{%
{\protect \APACyear {2011}}%
}]{%
jara2011dppackage}
\APACinsertmetastar {%
jara2011dppackage}%
\begin{APACrefauthors}%
Jara, A.%
, Hanson, T\BPBI E.%
, Quintana, F\BPBI A.%
, M{\"u}ller, P.%
\BCBL {}\ \BBA {} Rosner, G\BPBI L.%
\end{APACrefauthors}%
\unskip\
\newblock
\APACrefYearMonthDay{2011}{}{}.
\newblock
{\BBOQ}\APACrefatitle {{DP}package: {B}ayesian semi-and nonparametric modeling
  in {R}} {{DP}package: {B}ayesian semi-and nonparametric modeling in
  {R}}.{\BBCQ}
\newblock
\APACjournalVolNumPages{Journal of Statistical Software}{40}{5}{1}.
\PrintBackRefs{\CurrentBib}

\bibitem [\protect \citeauthoryear {%
Johnson%
}{%
Johnson%
}{%
{\protect \APACyear {2007}}%
}]{%
johnson2007modeling}
\APACinsertmetastar {%
johnson2007modeling}%
\begin{APACrefauthors}%
Johnson, M\BPBI S.%
\end{APACrefauthors}%
\unskip\
\newblock
\APACrefYearMonthDay{2007}{}{}.
\newblock
{\BBOQ}\APACrefatitle {Modeling dichotomous item responses with free-knot
  splines} {Modeling dichotomous item responses with free-knot splines}.{\BBCQ}
\newblock
\APACjournalVolNumPages{Computational Statistics \& Data
  Analysis}{51}{9}{4178--4192}.
\PrintBackRefs{\CurrentBib}

\bibitem [\protect \citeauthoryear {%
{Joint Health Surveys Unit of Social and Community Planning Research and
  University College London}%
}{%
{Joint Health Surveys Unit of Social and Community Planning Research and
  University College London}%
}{%
{\protect \APACyear {2017}}%
}]{%
healthdata1996}
\APACinsertmetastar {%
healthdata1996}%
\begin{APACrefauthors}%
{Joint Health Surveys Unit of Social and Community Planning Research and
  University College London}.%
\end{APACrefauthors}%
\unskip\
\newblock
\APACrefYearMonthDay{2017}{}{}.
\newblock
\APACrefbtitle {Health Survey for England, 1996. [data collection].} {Health
  survey for england, 1996. [data collection].}
\newblock
\APACaddressPublisher{}{5th Edition. UK Data Service. SN: 3886}.
\newblock
\begin{APACrefURL} \url{http://doi.org/10.5255/UKDA-SN-3886-2} \end{APACrefURL}
\PrintBackRefs{\CurrentBib}

\bibitem [\protect \citeauthoryear {%
Karabatsos%
}{%
Karabatsos%
}{%
{\protect \APACyear {2017}}%
}]{%
karabastos2017}
\APACinsertmetastar {%
karabastos2017}%
\begin{APACrefauthors}%
Karabatsos, G.%
\end{APACrefauthors}%
\unskip\
\newblock
\APACrefYearMonthDay{2017}{}{}.
\newblock
{\BBOQ}\APACrefatitle {{B}ayesian Nonparametric Response Models} {{B}ayesian
  nonparametric response models}.{\BBCQ}
\newblock
\BIn{} W\BPBI J.~Van~der Linden\ (\BED), \APACrefbtitle {Handbook of item
  response theory, Volume One: Models} {Handbook of item response theory,
  volume one: Models}\ (\BPG~323-336).
\newblock
\APACaddressPublisher{}{CRC Press}.
\newblock
\begin{APACrefURL}
  \url{https://www.taylorfrancis.com/books/e/9781466514423/chapters/10.1201/9781315374512-32}
  \end{APACrefURL}
\PrintBackRefs{\CurrentBib}

\bibitem [\protect \citeauthoryear {%
Kirisci%
, chi Hsu%
\BCBL {}\ \BBA {} Yu%
}{%
Kirisci%
\ \protect \BOthers {.}}{%
{\protect \APACyear {2001}}%
}]{%
kirisci2001robustness}
\APACinsertmetastar {%
kirisci2001robustness}%
\begin{APACrefauthors}%
Kirisci, L.%
, chi Hsu, T.%
\BCBL {}\ \BBA {} Yu, L.%
\end{APACrefauthors}%
\unskip\
\newblock
\APACrefYearMonthDay{2001}{}{}.
\newblock
{\BBOQ}\APACrefatitle {Robustness of Item Parameter Estimation Programs to
  Assumptions of Unidimensionality and Normality} {Robustness of item parameter
  estimation programs to assumptions of unidimensionality and
  normality}.{\BBCQ}
\newblock
\APACjournalVolNumPages{Applied Psychological Measurement}{25}{2}{146-162}.
\newblock
\begin{APACrefURL} \url{https://doi.org/10.1177/01466210122031975}
  \end{APACrefURL}
\newblock
\begin{APACrefDOI} \doi{10.1177/01466210122031975} \end{APACrefDOI}
\PrintBackRefs{\CurrentBib}

\bibitem [\protect \citeauthoryear {%
Laird%
}{%
Laird%
}{%
{\protect \APACyear {1978}}%
}]{%
laird78}
\APACinsertmetastar {%
laird78}%
\begin{APACrefauthors}%
Laird, N\BPBI M.%
\end{APACrefauthors}%
\unskip\
\newblock
\APACrefYearMonthDay{1978}{}{}.
\newblock
{\BBOQ}\APACrefatitle {Nonparametric maximum likelihood estimation of a mixing
  distribution} {Nonparametric maximum likelihood estimation of a mixing
  distribution}.{\BBCQ}
\newblock
\APACjournalVolNumPages{Journal of the American Statistical
  Association}{73}{}{805-807}.
\PrintBackRefs{\CurrentBib}

\bibitem [\protect \citeauthoryear {%
Li%
, M{\"u}ller%
\BCBL {}\ \BBA {} Lin%
}{%
Li%
\ \protect \BOthers {.}}{%
{\protect \APACyear {2011}}%
}]{%
li2011center}
\APACinsertmetastar {%
li2011center}%
\begin{APACrefauthors}%
Li, Y.%
, M{\"u}ller, P.%
\BCBL {}\ \BBA {} Lin, X.%
\end{APACrefauthors}%
\unskip\
\newblock
\APACrefYearMonthDay{2011}{}{}.
\newblock
{\BBOQ}\APACrefatitle {Center-adjusted inference for a nonparametric {B}ayesian
  random effect distribution} {Center-adjusted inference for a nonparametric
  {B}ayesian random effect distribution}.{\BBCQ}
\newblock
\APACjournalVolNumPages{Statistica Sinica}{21}{3}{1201--1223}.
\PrintBackRefs{\CurrentBib}

\bibitem [\protect \citeauthoryear {%
C.~Liu%
, Rubin%
\BCBL {}\ \BBA {} Wu%
}{%
C.~Liu%
\ \protect \BOthers {.}}{%
{\protect \APACyear {1998}}%
}]{%
liu1998parameter}
\APACinsertmetastar {%
liu1998parameter}%
\begin{APACrefauthors}%
Liu, C.%
, Rubin, D\BPBI B.%
\BCBL {}\ \BBA {} Wu, Y\BPBI N.%
\end{APACrefauthors}%
\unskip\
\newblock
\APACrefYearMonthDay{1998}{}{}.
\newblock
{\BBOQ}\APACrefatitle {Parameter expansion to accelerate {EM}: the {PX-EM}
  algorithm} {Parameter expansion to accelerate {EM}: the {PX-EM}
  algorithm}.{\BBCQ}
\newblock
\APACjournalVolNumPages{Biometrika}{85}{4}{755--770}.
\PrintBackRefs{\CurrentBib}

\bibitem [\protect \citeauthoryear {%
J\BPBI S.~Liu%
\ \BBA {} Wu%
}{%
J\BPBI S.~Liu%
\ \BBA {} Wu%
}{%
{\protect \APACyear {1999}}%
}]{%
liu1999parameter}
\APACinsertmetastar {%
liu1999parameter}%
\begin{APACrefauthors}%
Liu, J\BPBI S.%
\BCBT {}\ \BBA {} Wu, Y\BPBI N.%
\end{APACrefauthors}%
\unskip\
\newblock
\APACrefYearMonthDay{1999}{}{}.
\newblock
{\BBOQ}\APACrefatitle {Parameter expansion for data augmentation} {Parameter
  expansion for data augmentation}.{\BBCQ}
\newblock
\APACjournalVolNumPages{Journal of the American Statistical
  Association}{94}{448}{1264--1274}.
\PrintBackRefs{\CurrentBib}

\bibitem [\protect \citeauthoryear {%
S\BPBI J.~Liu%
}{%
S\BPBI J.~Liu%
}{%
{\protect \APACyear {1996}}%
}]{%
liu1996nonparametric}
\APACinsertmetastar {%
liu1996nonparametric}%
\begin{APACrefauthors}%
Liu, S\BPBI J.%
\end{APACrefauthors}%
\unskip\
\newblock
\APACrefYearMonthDay{1996}{}{}.
\newblock
{\BBOQ}\APACrefatitle {{Nonparametric hierarchical Bayes via sequential
  imputations}} {{Nonparametric hierarchical Bayes via sequential
  imputations}}.{\BBCQ}
\newblock
\APACjournalVolNumPages{The Annals of Statistics}{24}{3}{911 -- 930}.
\newblock
\begin{APACrefURL} \url{https://doi.org/10.1214/aos/1032526949}
  \end{APACrefURL}
\newblock
\begin{APACrefDOI} \doi{10.1214/aos/1032526949} \end{APACrefDOI}
\PrintBackRefs{\CurrentBib}

\bibitem [\protect \citeauthoryear {%
Lo%
}{%
Lo%
}{%
{\protect \APACyear {1984}}%
}]{%
lo1984class}
\APACinsertmetastar {%
lo1984class}%
\begin{APACrefauthors}%
Lo, A\BPBI Y.%
\end{APACrefauthors}%
\unskip\
\newblock
\APACrefYearMonthDay{1984}{}{}.
\newblock
{\BBOQ}\APACrefatitle {On a class of {B}ayesian nonparametric estimates: I.
  Density estimates} {On a class of {B}ayesian nonparametric estimates: I.
  density estimates}.{\BBCQ}
\newblock
\APACjournalVolNumPages{The Annals of Statistics}{}{}{351--357}.
\PrintBackRefs{\CurrentBib}

\bibitem [\protect \citeauthoryear {%
McHorney%
, Haley%
\BCBL {}\ \BBA {} Ware~Jr%
}{%
McHorney%
\ \protect \BOthers {.}}{%
{\protect \APACyear {1997}}%
}]{%
mchorney1997evaluation}
\APACinsertmetastar {%
mchorney1997evaluation}%
\begin{APACrefauthors}%
McHorney, C\BPBI A.%
, Haley, S\BPBI M.%
\BCBL {}\ \BBA {} Ware~Jr, J\BPBI E.%
\end{APACrefauthors}%
\unskip\
\newblock
\APACrefYearMonthDay{1997}{}{}.
\newblock
{\BBOQ}\APACrefatitle {Evaluation of the {MOS SF-36} physical functioning scale
  ({PF-40}): {II}. Comparison of relative precision using {L}ikert and {R}asch
  scoring methods} {Evaluation of the {MOS SF-36} physical functioning scale
  ({PF-40}): {II}. comparison of relative precision using {L}ikert and {R}asch
  scoring methods}.{\BBCQ}
\newblock
\APACjournalVolNumPages{Journal of Clinical Epidemiology}{50}{4}{451--461}.
\PrintBackRefs{\CurrentBib}

\bibitem [\protect \citeauthoryear {%
Micceri%
}{%
Micceri%
}{%
{\protect \APACyear {1989}}%
}]{%
micceri1989unicorn}
\APACinsertmetastar {%
micceri1989unicorn}%
\begin{APACrefauthors}%
Micceri, T.%
\end{APACrefauthors}%
\unskip\
\newblock
\APACrefYearMonthDay{1989}{}{}.
\newblock
{\BBOQ}\APACrefatitle {The unicorn, the normal curve, and other improbable
  creatures.} {The unicorn, the normal curve, and other improbable
  creatures.}{\BBCQ}
\newblock
\APACjournalVolNumPages{Psychological bulletin}{105}{1}{156}.
\PrintBackRefs{\CurrentBib}

\bibitem [\protect \citeauthoryear {%
Mislevy%
}{%
Mislevy%
}{%
{\protect \APACyear {1984}}%
}]{%
mislevy1984estimating}
\APACinsertmetastar {%
mislevy1984estimating}%
\begin{APACrefauthors}%
Mislevy, R\BPBI J.%
\end{APACrefauthors}%
\unskip\
\newblock
\APACrefYearMonthDay{1984}{}{}.
\newblock
{\BBOQ}\APACrefatitle {Estimating latent distributions} {Estimating latent
  distributions}.{\BBCQ}
\newblock
\APACjournalVolNumPages{Psychometrika}{49}{3}{359--381}.
\newblock
\begin{APACrefURL} \url{https://doi.org/10.1007/BF02306026} \end{APACrefURL}
\newblock
\begin{APACrefDOI} \doi{10.1007/BF02306026} \end{APACrefDOI}
\PrintBackRefs{\CurrentBib}

\bibitem [\protect \citeauthoryear {%
Miyazaki%
\ \BBA {} Hoshino%
}{%
Miyazaki%
\ \BBA {} Hoshino%
}{%
{\protect \APACyear {2009}}%
}]{%
miyazaki2009bayesian}
\APACinsertmetastar {%
miyazaki2009bayesian}%
\begin{APACrefauthors}%
Miyazaki, K.%
\BCBT {}\ \BBA {} Hoshino, T.%
\end{APACrefauthors}%
\unskip\
\newblock
\APACrefYearMonthDay{2009}{}{}.
\newblock
{\BBOQ}\APACrefatitle {A {B}ayesian semiparametric item response model with
  {D}irichlet {P}rocess priors} {A {B}ayesian semiparametric item response
  model with {D}irichlet {P}rocess priors}.{\BBCQ}
\newblock
\APACjournalVolNumPages{Psychometrika}{74}{3}{375--393}.
\PrintBackRefs{\CurrentBib}

\bibitem [\protect \citeauthoryear {%
Natesan%
, Nandakumar%
, Minka%
\BCBL {}\ \BBA {} Rubright%
}{%
Natesan%
\ \protect \BOthers {.}}{%
{\protect \APACyear {2016}}%
}]{%
natesan2016bayesian}
\APACinsertmetastar {%
natesan2016bayesian}%
\begin{APACrefauthors}%
Natesan, P.%
, Nandakumar, R.%
, Minka, T.%
\BCBL {}\ \BBA {} Rubright, J\BPBI D.%
\end{APACrefauthors}%
\unskip\
\newblock
\APACrefYearMonthDay{2016}{}{}.
\newblock
{\BBOQ}\APACrefatitle {{B}ayesian prior choice in {IRT} estimation using {MCMC}
  and variational {B}ayes} {{B}ayesian prior choice in {IRT} estimation using
  {MCMC} and variational {B}ayes}.{\BBCQ}
\newblock
\APACjournalVolNumPages{Frontiers in Psychology}{7}{}{1422}.
\PrintBackRefs{\CurrentBib}

\bibitem [\protect \citeauthoryear {%
Neal%
}{%
Neal%
}{%
{\protect \APACyear {2000}}%
}]{%
neal2000markov}
\APACinsertmetastar {%
neal2000markov}%
\begin{APACrefauthors}%
Neal, R\BPBI M.%
\end{APACrefauthors}%
\unskip\
\newblock
\APACrefYearMonthDay{2000}{}{}.
\newblock
{\BBOQ}\APACrefatitle {Markov chain sampling methods for {D}irichlet {P}rocess
  mixture models} {Markov chain sampling methods for {D}irichlet {P}rocess
  mixture models}.{\BBCQ}
\newblock
\APACjournalVolNumPages{Journal of Computational and Graphical
  Statistics}{9}{2}{249--265}.
\PrintBackRefs{\CurrentBib}

\bibitem [\protect \citeauthoryear {%
Nguyen%
\ \protect \BOthers {.}}{%
Nguyen%
\ \protect \BOthers {.}}{%
{\protect \APACyear {2020}}%
}]{%
nguyen2020}
\APACinsertmetastar {%
nguyen2020}%
\begin{APACrefauthors}%
Nguyen, D.%
, de Valpine, P.%
, Atchade, Y.%
, Turek, D.%
, Michaud, N.%
\BCBL {}\ \BBA {} Paciorek, C\BPBI J.%
\end{APACrefauthors}%
\unskip\
\newblock
\APACrefYearMonthDay{2020}{12}{}.
\newblock
{\BBOQ}\APACrefatitle {Nested Adaptation of MCMC Algorithms} {Nested adaptation
  of mcmc algorithms}.{\BBCQ}
\newblock
\APACjournalVolNumPages{Bayesian Analysis}{15}{4}{1323--1343}.
\newblock
\begin{APACrefURL} \url{https://doi.org/10.1214/19-BA1190} \end{APACrefURL}
\newblock
\begin{APACrefDOI} \doi{10.1214/19-BA1190} \end{APACrefDOI}
\PrintBackRefs{\CurrentBib}

\bibitem [\protect \citeauthoryear {%
Paulon%
, De~Iorio%
, Guglielmi%
\BCBL {}\ \BBA {} Ieva%
}{%
Paulon%
\ \protect \BOthers {.}}{%
{\protect \APACyear {2018}}%
}]{%
paulon2018}
\APACinsertmetastar {%
paulon2018}%
\begin{APACrefauthors}%
Paulon, G.%
, De~Iorio, M.%
, Guglielmi, A.%
\BCBL {}\ \BBA {} Ieva, F.%
\end{APACrefauthors}%
\unskip\
\newblock
\APACrefYearMonthDay{2018}{07}{}.
\newblock
{\BBOQ}\APACrefatitle {{Joint modeling of recurrent events and survival: a
  {B}ayesian non-parametric approach}} {{Joint modeling of recurrent events and
  survival: a {B}ayesian non-parametric approach}}.{\BBCQ}
\newblock
\APACjournalVolNumPages{Biostatistics}{21}{1}{1-14}.
\newblock
\begin{APACrefURL} \url{https://doi.org/10.1093/biostatistics/kxy026}
  \end{APACrefURL}
\newblock
\begin{APACrefDOI} \doi{10.1093/biostatistics/kxy026} \end{APACrefDOI}
\PrintBackRefs{\CurrentBib}

\bibitem [\protect \citeauthoryear {%
Pitman%
}{%
Pitman%
}{%
{\protect \APACyear {1996}}%
}]{%
pitman1996some}
\APACinsertmetastar {%
pitman1996some}%
\begin{APACrefauthors}%
Pitman, J.%
\end{APACrefauthors}%
\unskip\
\newblock
\APACrefYearMonthDay{1996}{}{}.
\newblock
{\BBOQ}\APACrefatitle {Some developments of the Blackwell-MacQueen urn scheme}
  {Some developments of the blackwell-macqueen urn scheme}.{\BBCQ}
\newblock
\BIn{} T\BPBI S.~Ferguson, L\BPBI S.~Shapley\BCBL {}\ \BBA {} J\BPBI
  B.~MacQueen\ (\BEDS), \APACrefbtitle {Statistics, probability and game
  theory} {Statistics, probability and game theory}\ (\BVOL\ Volume 30, \BPGS\
  245--267).
\newblock
\APACaddressPublisher{Hayward, CA}{Institute of Mathematical Statistics}.
\newblock
\begin{APACrefURL} \url{https://doi.org/10.1214/lnms/1215453576}
  \end{APACrefURL}
\newblock
\begin{APACrefDOI} \doi{10.1214/lnms/1215453576} \end{APACrefDOI}
\PrintBackRefs{\CurrentBib}

\bibitem [\protect \citeauthoryear {%
Qin%
}{%
Qin%
}{%
{\protect \APACyear {1998}}%
}]{%
qin1998nonparametric}
\APACinsertmetastar {%
qin1998nonparametric}%
\begin{APACrefauthors}%
Qin, L.%
\end{APACrefauthors}%
\unskip\
\newblock
\APACrefYear{1998}.
\unskip\
\newblock
\APACrefbtitle {Nonparametric {B}ayesian models for item response data}
  {Nonparametric {B}ayesian models for item response data}\
  \APACtypeAddressSchool {\BUPhD}{}{}.
\unskip\
\newblock
\APACaddressSchool {}{The Ohio State University}.
\PrintBackRefs{\CurrentBib}

\bibitem [\protect \citeauthoryear {%
Rasch%
}{%
Rasch%
}{%
{\protect \APACyear {1990}}%
}]{%
rasch1990probabilistic}
\APACinsertmetastar {%
rasch1990probabilistic}%
\begin{APACrefauthors}%
Rasch, G.%
\end{APACrefauthors}%
\unskip\
\newblock
\APACrefYear{1990}.
\newblock
\APACrefbtitle {Probabilistic Models for Some Intelligence and Attainment
  Tests.} {Probabilistic models for some intelligence and attainment tests.}
\newblock
\APACaddressPublisher{}{Copenhagen: Danish Institute for Educational Research.}
\PrintBackRefs{\CurrentBib}

\bibitem [\protect \citeauthoryear {%
Reise%
\ \BBA {} Rodriguez%
}{%
Reise%
\ \BBA {} Rodriguez%
}{%
{\protect \APACyear {2016}}%
}]{%
reise2016item}
\APACinsertmetastar {%
reise2016item}%
\begin{APACrefauthors}%
Reise, S.%
\BCBT {}\ \BBA {} Rodriguez, A.%
\end{APACrefauthors}%
\unskip\
\newblock
\APACrefYearMonthDay{2016}{}{}.
\newblock
{\BBOQ}\APACrefatitle {Item response theory and the measurement of psychiatric
  constructs: some empirical and conceptual issues and challenges} {Item
  response theory and the measurement of psychiatric constructs: some empirical
  and conceptual issues and challenges}.{\BBCQ}
\newblock
\APACjournalVolNumPages{Psychological Medicine}{46}{10}{2025--2039}.
\PrintBackRefs{\CurrentBib}

\bibitem [\protect \citeauthoryear {%
Rutkowski%
, Gonzalez%
, Joncas%
\BCBL {}\ \BBA {} von Davier%
}{%
Rutkowski%
\ \protect \BOthers {.}}{%
{\protect \APACyear {2010}}%
}]{%
rutkowski2010international}
\APACinsertmetastar {%
rutkowski2010international}%
\begin{APACrefauthors}%
Rutkowski, L.%
, Gonzalez, E.%
, Joncas, M.%
\BCBL {}\ \BBA {} von Davier, M.%
\end{APACrefauthors}%
\unskip\
\newblock
\APACrefYearMonthDay{2010}{}{}.
\newblock
{\BBOQ}\APACrefatitle {International large-scale assessment data: issues in
  secondary analysis and reporting} {International large-scale assessment data:
  issues in secondary analysis and reporting}.{\BBCQ}
\newblock
\APACjournalVolNumPages{Educational Researcher}{39}{2}{142--151}.
\PrintBackRefs{\CurrentBib}

\bibitem [\protect \citeauthoryear {%
Samejima%
}{%
Samejima%
}{%
{\protect \APACyear {1997}}%
}]{%
samejima1997departure}
\APACinsertmetastar {%
samejima1997departure}%
\begin{APACrefauthors}%
Samejima, F.%
\end{APACrefauthors}%
\unskip\
\newblock
\APACrefYearMonthDay{1997}{}{}.
\newblock
{\BBOQ}\APACrefatitle {Departure from normal assumptions: a promise for future
  psychometrics with substantive mathematical modeling} {Departure from normal
  assumptions: a promise for future psychometrics with substantive mathematical
  modeling}.{\BBCQ}
\newblock
\APACjournalVolNumPages{Psychometrika}{62}{4}{471--493}.
\PrintBackRefs{\CurrentBib}

\bibitem [\protect \citeauthoryear {%
San Mart{\'i}n%
, Jara%
, Rolin%
\BCBL {}\ \BBA {} Mouchart%
}{%
San Mart{\'i}n%
\ \protect \BOthers {.}}{%
{\protect \APACyear {2011}}%
}]{%
sanmartin2011}
\APACinsertmetastar {%
sanmartin2011}%
\begin{APACrefauthors}%
San Mart{\'i}n, E.%
, Jara, A.%
, Rolin, J\BHBI M.%
\BCBL {}\ \BBA {} Mouchart, M.%
\end{APACrefauthors}%
\unskip\
\newblock
\APACrefYearMonthDay{2011}{Jul}{01}.
\newblock
{\BBOQ}\APACrefatitle {On the {B}ayesian Nonparametric Generalization of
  {IRT}-Type Models} {On the {B}ayesian nonparametric generalization of
  {IRT}-type models}.{\BBCQ}
\newblock
\APACjournalVolNumPages{Psychometrika}{76}{3}{385--409}.
\newblock
\begin{APACrefURL} \url{https://doi.org/10.1007/s11336-011-9213-9}
  \end{APACrefURL}
\newblock
\begin{APACrefDOI} \doi{10.1007/s11336-011-9213-9} \end{APACrefDOI}
\PrintBackRefs{\CurrentBib}

\bibitem [\protect \citeauthoryear {%
Schmitt%
, Mehta%
, Aggen%
, Kubarych%
\BCBL {}\ \BBA {} Neale%
}{%
Schmitt%
\ \protect \BOthers {.}}{%
{\protect \APACyear {2006}}%
}]{%
schmitt2006semi}
\APACinsertmetastar {%
schmitt2006semi}%
\begin{APACrefauthors}%
Schmitt, J\BPBI E.%
, Mehta, P\BPBI D.%
, Aggen, S\BPBI H.%
, Kubarych, T\BPBI S.%
\BCBL {}\ \BBA {} Neale, M\BPBI C.%
\end{APACrefauthors}%
\unskip\
\newblock
\APACrefYearMonthDay{2006}{}{}.
\newblock
{\BBOQ}\APACrefatitle {Semi-nonparametric methods for detecting latent
  non-normality: A fusion of latent trait and ordered latent class modeling}
  {Semi-nonparametric methods for detecting latent non-normality: A fusion of
  latent trait and ordered latent class modeling}.{\BBCQ}
\newblock
\APACjournalVolNumPages{Multivariate Behavioral Research}{41}{4}{427--443}.
\PrintBackRefs{\CurrentBib}

\bibitem [\protect \citeauthoryear {%
Seong%
}{%
Seong%
}{%
{\protect \APACyear {1990}}%
}]{%
seong1990sensitivity}
\APACinsertmetastar {%
seong1990sensitivity}%
\begin{APACrefauthors}%
Seong, T.%
\end{APACrefauthors}%
\unskip\
\newblock
\APACrefYearMonthDay{1990}{}{}.
\newblock
{\BBOQ}\APACrefatitle {Sensitivity of marginal maximum likelihood estimation of
  item and ability parameters to the characteristics of the prior ability
  distributions} {Sensitivity of marginal maximum likelihood estimation of item
  and ability parameters to the characteristics of the prior ability
  distributions}.{\BBCQ}
\newblock
\APACjournalVolNumPages{Applied psychological measurement}{14}{3}{299--311}.
\PrintBackRefs{\CurrentBib}

\bibitem [\protect \citeauthoryear {%
Sethuraman%
}{%
Sethuraman%
}{%
{\protect \APACyear {1994}}%
}]{%
sethuraman1994}
\APACinsertmetastar {%
sethuraman1994}%
\begin{APACrefauthors}%
Sethuraman, J.%
\end{APACrefauthors}%
\unskip\
\newblock
\APACrefYearMonthDay{1994}{}{}.
\newblock
{\BBOQ}\APACrefatitle {A constructive definition of {D}irichlet priors} {A
  constructive definition of {D}irichlet priors}.{\BBCQ}
\newblock
\APACjournalVolNumPages{Statistica Sinica}{4}{2}{639--650}.
\PrintBackRefs{\CurrentBib}

\bibitem [\protect \citeauthoryear {%
Shaby%
\ \BBA {} Wells%
}{%
Shaby%
\ \BBA {} Wells%
}{%
{\protect \APACyear {2010}}%
}]{%
shaby2010}
\APACinsertmetastar {%
shaby2010}%
\begin{APACrefauthors}%
Shaby, B.%
\BCBT {}\ \BBA {} Wells, M.%
\end{APACrefauthors}%
\unskip\
\newblock
\APACrefYearMonthDay{2010}{}{}.
\newblock
\APACrefbtitle {Exploring an adaptive {M}etropolis algorithm} {Exploring an
  adaptive {M}etropolis algorithm}\ \APACbVolEdTR {}{Technical Report\ \BNUM\
  1011-14}.
\newblock
\APACaddressInstitution{}{Duke University Department of Statistical Science}.
\PrintBackRefs{\CurrentBib}

\bibitem [\protect \citeauthoryear {%
Sheng%
}{%
Sheng%
}{%
{\protect \APACyear {2010}}%
}]{%
sheng2010sensitivity}
\APACinsertmetastar {%
sheng2010sensitivity}%
\begin{APACrefauthors}%
Sheng, Y.%
\end{APACrefauthors}%
\unskip\
\newblock
\APACrefYearMonthDay{2010}{}{}.
\newblock
{\BBOQ}\APACrefatitle {A sensitivity analysis of Gibbs sampling for {3PNO IRT}
  models: Effects of prior specifications on parameter estimates} {A
  sensitivity analysis of gibbs sampling for {3PNO IRT} models: Effects of
  prior specifications on parameter estimates}.{\BBCQ}
\newblock
\APACjournalVolNumPages{Behaviormetrika}{37}{2}{87--110}.
\PrintBackRefs{\CurrentBib}

\bibitem [\protect \citeauthoryear {%
Smits%
, {\"O}{\u{g}}reden%
, Garnier-Villarreal%
, Terwee%
\BCBL {}\ \BBA {} Chalmers%
}{%
Smits%
\ \protect \BOthers {.}}{%
{\protect \APACyear {2020}}%
}]{%
smits2020study}
\APACinsertmetastar {%
smits2020study}%
\begin{APACrefauthors}%
Smits, N.%
, {\"O}{\u{g}}reden, O.%
, Garnier-Villarreal, M.%
, Terwee, C\BPBI B.%
\BCBL {}\ \BBA {} Chalmers, R\BPBI P.%
\end{APACrefauthors}%
\unskip\
\newblock
\APACrefYearMonthDay{2020}{}{}.
\newblock
{\BBOQ}\APACrefatitle {A study of alternative approaches to non-normal latent
  trait distributions in item response theory models used for health outcome
  measurement} {A study of alternative approaches to non-normal latent trait
  distributions in item response theory models used for health outcome
  measurement}.{\BBCQ}
\newblock
\APACjournalVolNumPages{Statistical Methods in Medical
  Research}{29}{4}{1030--1048}.
\PrintBackRefs{\CurrentBib}

\bibitem [\protect \citeauthoryear {%
Sosa%
\ \BBA {} Rodr{\`i}guez%
}{%
Sosa%
\ \BBA {} Rodr{\`i}guez%
}{%
{\protect \APACyear {2021}}%
}]{%
sosa2021latent}
\APACinsertmetastar {%
sosa2021latent}%
\begin{APACrefauthors}%
Sosa, J.%
\BCBT {}\ \BBA {} Rodr{\`i}guez, A.%
\end{APACrefauthors}%
\unskip\
\newblock
\APACrefYearMonthDay{2021}{}{}.
\newblock
{\BBOQ}\APACrefatitle {A latent space model for cognitive social structures
  data} {A latent space model for cognitive social structures data}.{\BBCQ}
\newblock
\APACjournalVolNumPages{Social Networks}{65}{}{85 - 97}.
\PrintBackRefs{\CurrentBib}

\bibitem [\protect \citeauthoryear {%
{Stan Development Team}%
}{%
{Stan Development Team}%
}{%
{\protect \APACyear {2018}}%
}]{%
stan2018}
\APACinsertmetastar {%
stan2018}%
\begin{APACrefauthors}%
{Stan Development Team}.%
\end{APACrefauthors}%
\unskip\
\newblock
\APACrefYearMonthDay{2018}{}{}.
\newblock
\APACrefbtitle {Stan Modeling Language Users Guide and Reference Manual,
  Version 2.18.0.} {Stan modeling language users guide and reference manual,
  version 2.18.0.}
\newblock
\begin{APACrefURL} \url{http://mc-stan.org} \end{APACrefURL}
\PrintBackRefs{\CurrentBib}

\bibitem [\protect \citeauthoryear {%
Vats%
, Flegal%
\BCBL {}\ \BBA {} Jones%
}{%
Vats%
\ \protect \BOthers {.}}{%
{\protect \APACyear {2019}}%
}]{%
vats2019multivariate}
\APACinsertmetastar {%
vats2019multivariate}%
\begin{APACrefauthors}%
Vats, D.%
, Flegal, J\BPBI M.%
\BCBL {}\ \BBA {} Jones, G\BPBI L.%
\end{APACrefauthors}%
\unskip\
\newblock
\APACrefYearMonthDay{2019}{}{}.
\newblock
{\BBOQ}\APACrefatitle {Multivariate output analysis for {M}arkov chain {M}onte
  {C}arlo} {Multivariate output analysis for {M}arkov chain {M}onte
  {C}arlo}.{\BBCQ}
\newblock
\APACjournalVolNumPages{Biometrika}{106}{2}{321--337}.
\PrintBackRefs{\CurrentBib}

\bibitem [\protect \citeauthoryear {%
Ware%
}{%
Ware%
}{%
{\protect \APACyear {2003}}%
}]{%
ware2003sf36}
\APACinsertmetastar {%
ware2003sf36}%
\begin{APACrefauthors}%
Ware, J\BPBI E.%
\end{APACrefauthors}%
\unskip\
\newblock
\APACrefYearMonthDay{2003}{}{}.
\newblock
{\BBOQ}\APACrefatitle {SF-36 health survey: Manual and interpretation guide}
  {Sf-36 health survey: Manual and interpretation guide}.{\BBCQ}.
\PrintBackRefs{\CurrentBib}

\bibitem [\protect \citeauthoryear {%
Watanabe%
\ \BBA {} Opper%
}{%
Watanabe%
\ \BBA {} Opper%
}{%
{\protect \APACyear {2010}}%
}]{%
watanabe2010asymptotic}
\APACinsertmetastar {%
watanabe2010asymptotic}%
\begin{APACrefauthors}%
Watanabe, S.%
\BCBT {}\ \BBA {} Opper, M.%
\end{APACrefauthors}%
\unskip\
\newblock
\APACrefYearMonthDay{2010}{}{}.
\newblock
{\BBOQ}\APACrefatitle {Asymptotic equivalence of {B}ayes cross validation and
  widely applicable information criterion in singular learning theory.}
  {Asymptotic equivalence of {B}ayes cross validation and widely applicable
  information criterion in singular learning theory.}{\BBCQ}
\newblock
\APACjournalVolNumPages{Journal of Machine Learning Research}{11}{12}{}.
\PrintBackRefs{\CurrentBib}

\bibitem [\protect \citeauthoryear {%
Woods%
}{%
Woods%
}{%
{\protect \APACyear {2007}}%
}]{%
woods2007empirical}
\APACinsertmetastar {%
woods2007empirical}%
\begin{APACrefauthors}%
Woods, C\BPBI M.%
\end{APACrefauthors}%
\unskip\
\newblock
\APACrefYearMonthDay{2007}{}{}.
\newblock
{\BBOQ}\APACrefatitle {Empirical histograms in item response theory with
  ordinal data} {Empirical histograms in item response theory with ordinal
  data}.{\BBCQ}
\newblock
\APACjournalVolNumPages{Educational and Psychological
  Measurement}{67}{1}{73--87}.
\PrintBackRefs{\CurrentBib}

\bibitem [\protect \citeauthoryear {%
Woods%
\ \BBA {} Thissen%
}{%
Woods%
\ \BBA {} Thissen%
}{%
{\protect \APACyear {2006}}%
}]{%
woods2006item}
\APACinsertmetastar {%
woods2006item}%
\begin{APACrefauthors}%
Woods, C\BPBI M.%
\BCBT {}\ \BBA {} Thissen, D.%
\end{APACrefauthors}%
\unskip\
\newblock
\APACrefYearMonthDay{2006}{}{}.
\newblock
{\BBOQ}\APACrefatitle {Item response theory with estimation of the latent
  population distribution using spline-based densities} {Item response theory
  with estimation of the latent population distribution using spline-based
  densities}.{\BBCQ}
\newblock
\APACjournalVolNumPages{Psychometrika}{71}{2}{281}.
\PrintBackRefs{\CurrentBib}

\bibitem [\protect \citeauthoryear {%
Yang%
\ \BBA {} Dunson%
}{%
Yang%
\ \BBA {} Dunson%
}{%
{\protect \APACyear {2010}}%
}]{%
yang2010bayesian}
\APACinsertmetastar {%
yang2010bayesian}%
\begin{APACrefauthors}%
Yang, M.%
\BCBT {}\ \BBA {} Dunson, D\BPBI B.%
\end{APACrefauthors}%
\unskip\
\newblock
\APACrefYearMonthDay{2010}{}{}.
\newblock
{\BBOQ}\APACrefatitle {{B}ayesian semiparametric structural equation models
  with latent variables} {{B}ayesian semiparametric structural equation models
  with latent variables}.{\BBCQ}
\newblock
\APACjournalVolNumPages{Psychometrika}{75}{4}{675--693}.
\PrintBackRefs{\CurrentBib}

\bibitem [\protect \citeauthoryear {%
Yang%
, Dunson%
\BCBL {}\ \BBA {} Baird%
}{%
Yang%
\ \protect \BOthers {.}}{%
{\protect \APACyear {2010}}%
}]{%
yang2010semiparametric}
\APACinsertmetastar {%
yang2010semiparametric}%
\begin{APACrefauthors}%
Yang, M.%
, Dunson, D\BPBI B.%
\BCBL {}\ \BBA {} Baird, D.%
\end{APACrefauthors}%
\unskip\
\newblock
\APACrefYearMonthDay{2010}{}{}.
\newblock
{\BBOQ}\APACrefatitle {Semiparametric {B}ayes hierarchical models with mean and
  variance constraints} {Semiparametric {B}ayes hierarchical models with mean
  and variance constraints}.{\BBCQ}
\newblock
\APACjournalVolNumPages{Computational Statistics \& Data
  Analysis}{54}{9}{2172--2186}.
\PrintBackRefs{\CurrentBib}

\end{thebibliography}

%% ITEM 11 [See the "howto.tex" file.]
%%%% You can put your Figures and Tables here
%%%% after the Reference Section.
%%%% BE SURE TO MARK IN THE TEXT WHERE
%%%% YOU WANT EACH FIGURE AND TABLE TO BE PLACED.
%%%% If you prefer, you can integrate your figures and tables into the text of your paper,
%%%% PROVIDED you will provide camera-ready copies of each figure.
%\vspace{\fill}\pagebreak
%\linespacing{1}

%\section*{Figures}
%
%\begin{figure}[h]
%\centerline{\includegraphics{figure01.eps}}
%\caption{Your figure caption goes here.}
%\end{figure}
%\vskip6pt

%\vspace{\fill}\pagebreak

%\section*{Tables}

%\vspace{\fill}\pagebreak

\end{document}